\documentclass[12pt]{article}
\usepackage{a4wide}
\usepackage{epsfig}

\newcommand{\ssigma}{\hbox{$\kern2.5pt\vrule height4pt\kern-2.5pt\sigma$}}
\newcommand{\oone}{\hbox{$1\kern-2pt\vrule height8pt\kern2pt$}}
\newcommand{\as}{{\alpha_s}}

\newcommand{\GeV}{\mbox{\sl\,GeV}}
\newcommand{\real}{\mbox{\sl Re\,}}
\newcommand{\imag}{\mbox{\it Im\,}}

\newcommand\sxi{{\sqrt\xi}}
\newcommand\pfrac[2]{\left(\frac{#1}{#2}\right)}
\newcommand{\Li}{{\rm Li}}
\newcommand{\proj}{{\mit\Pi}}
\newcommand{\eps}{\varepsilon}
\newcommand{\cz}{\chi_{\scriptscriptstyle Z}}
\newcommand{\pbreak}{\break\vbox to0pt{\hbox to0pt{}}\vskip-12pt}
\begin{document}
\thispagestyle{empty}

\begin{flushright}
MZ-TH/08-39\\
arXiv:0811.2728 [hep-ph]\\
November 2008\\
\end{flushright}

\vspace{0.5cm}
\begin{center}
{\Large\bf Analytical results for \boldmath{$O(\as)$} radiative 
  corrections}\\[.2cm]
{\Large\bf to \boldmath{$e^+e^-\to \bar tt^\uparrow$} up to a given gluon 
energy cut}\\
\vspace{1.3cm}
{\large S.~Groote$^{1,2}$ and J.G.~K\"orner$^2$}\\[1cm]
$^1$ Loodus- ja Tehnoloogiateaduskond, F\"u\"usika Instituut,\\[.2cm]
  Tartu \"Ulikool, T\"ahe 4, 51010 Tartu, Estonia\\[.5truecm]
$^2$ Institut f\"ur Physik, Johannes-Gutenberg-Universit\"at,\\[.2cm]
Staudinger Weg 7, 55099 Mainz, Germany\\
\end{center}
\vspace{1cm}
\begin{abstract}\noindent
We determine the $O(\as)$ radiative corrections to polarized top quark pair
production in $e^+e^-$ annihilations with a specified gluon energy cut. We
write down fully analytical results for the unpolarized and polarized $O(\as)$
cross sections $e^+e^-\rightarrow\bar tt(G)$ and
$e^+e^-\rightarrow\bar tt^\uparrow(G)$ including their polar orientation
dependence relative to the beam direction. In the soft-gluon limit we recover
the usual factorizing form known from the soft-gluon approximation. In the
limit when the gluon energy cut takes its maximum value we recover the totally
inclusive unpolarized and polarized cross sections calculated previously. We
provide some numerical results on the cut-off dependence of the various
polarized and unpolarized cross sections and discuss how the exact results
numerically differ from the approximate soft-gluon results.
\end{abstract}

\vspace{0.5cm}

\newpage

\section{Introduction}
After the discovery of the heavy top quark at the Tevatron in 1995 there has
been much interest in the use of the proposed high energy linear $e^+e^-$
collider as a copious source of top quark pairs. When the proposed linear
collider ILC comes into operation it is necessary to have available detailed
radiative corrections to the production and the decay of top quark pairs.
Concerning production there are a number of unpolarized and single spin
polarized structure functions that describe the $e^+e^-$ production process of
massive top quark pairs. In the unpolarized case one has the three structure
functions $H_U$ (unpolarized transverse), $H_L$ (longitudinal), and $H_F$
(forward-backward) which determine the polar angle orientation of the top pair
relative to the beam axis. Partial results on the full $O(\as)$ radiative
corrections to the unpolarized structure functions $H_U$, $H_L$ and $H_F$ had
been written down in Refs.~\cite{Grunberg:1979ru,Jersak:1981sp} starting with
the early work on the $O(\alpha)$ QED radiative corrections to the vector
current $(\gamma_{V} e^+e^-)$ vertex function~\cite{Jost:1950}. Complete
results on the $O(\as)$ unpolarized structure functions have been first given
in Refs.~\cite{Stav:1992gj,Stav:1994se}. All of the unpolarized $O(\as)$
structure functions were recalculated in the course of computing the top
quark's $O(\as)$ polarization asymmetries where the unpolarized structure
functions were needed to normalize the polarization asymmetries
\cite{Tung:1996dq,Korner:1993dy,Groote:1995yc,Groote:1995ky,Groote:1996nc}.
The numerators of the polarization asymmetries are expressed in terms of
polarized structure functions. In the case of the longitudinal polarization of
the top one has the three structure functions $H_U^l$, $H_L^l$ and $H_F^l$ for
which the full $O(\as)$ radiative corrections were given in
Refs.~\cite{Korner:1993dy,Groote:1995yc,Groote:1996nc}. In the case of a top
quark polarized transverse or normal to the event plane, one has two structure
functions in each case which are $H_I^T$ and $H_A^T$, and $H_I^N$ and
$H_A^N$, respectively (see e.g.\ Ref.~\cite{Groote:1995ky}). These were
calculated in Refs.~\cite{Groote:1995ky,Ravindran:2000rz}. 

When doing the full $O(\as)$ radiative corrections one integrates over the
full (hard and soft) gluon phase space. For some applications it is also
interesting to consider radiative corrections where one integrates over gluon
phase space up to a given gluon energy cut $E_c$.\footnote{Technically, this
means that one is dealing with a three-scale problem with the scales $q^2$,
$m_t$ and $E_c$.} Such radiative corrections may be dictated by experimental
considerations when soft gluons accompanying the top quark pair cannot be
resolved by the detector. Alternatively one could attempt to measure the cross
section for top--antitop--gluon production with a given gluon energy cut $E_c$
and compare the energy cut dependence of the cross section with the
predictions of QCD. Finally, one could define a hard gluon region by
introducing a lower gluon energy cut and compare experiment with QCD in the
hard gluon region.

In this paper we provide analytical results for the $O(\as)$ radiative
corrections to the three unpolarized structure functions $H_U$, $H_L$, and
$H_F$ as well as for the seven polarized structure functions $H_U^\ell$,
$H_L^\ell$, $H_F^\ell$, $H_I^{T,N}$, and $H_A^{T,N}$ for polarized top
quarks where we integrate over the gluon energy phase space up to a given
energy cut $E_c$. We mention that radiative corrections with a gluon energy 
cut have been treated before in the unpolarized
case~\cite{Stav:1996ep,Arbuzov:1991pr}. 

We emphasize that we are not using the soft-gluon approximation (SGA) in the
present calculation but integrate over the full $O(\as)$ matrix element tree
graph structure. However, we will compare our results with the soft-gluon
approximation. The soft-gluon approximation consists of the factorization of
the tree graph contribution into the Born term contribution and a universal
soft-gluon piece which can be easily integrated. An $O(\as)$ calculation of
some of the structure functions appearing in polarized top pair production
using variants of the soft-gluon approximation has been done before in
Refs.~\cite{Groote:1996nc,Akatsu:1997tq}. 

One of the further aims of the present investigation is to find out to what 
extent one can pin down a new non-SM (Standard Model) coupling structure in
top quark pair production in the presence of $O(\as)$ radiative corrections
with an exact treatment of gluon emission rather than soft-gluon emission. In
the latter approximation the tree graph contribution is Born term like and
thus polarization-type observables would not be affected by the radiative tree
graph corrections but only by the non-Born term structure of the one-loop
contributions. Deviations from SM predictions for the polarization-type
observables could result from new non-SM coupling structure or from an exact
treatment of radiative corrections. As an example we will introduce an
anomalous $CP$-odd axial current and compare the results of our exact
next-to-leading order (NLO) calculation with the contributions of the
anomalous axial current for some relevant observables and structure functions.

\section{Unpolarized and polarized structure functions}
\begin{figure}
\epsfig{figure=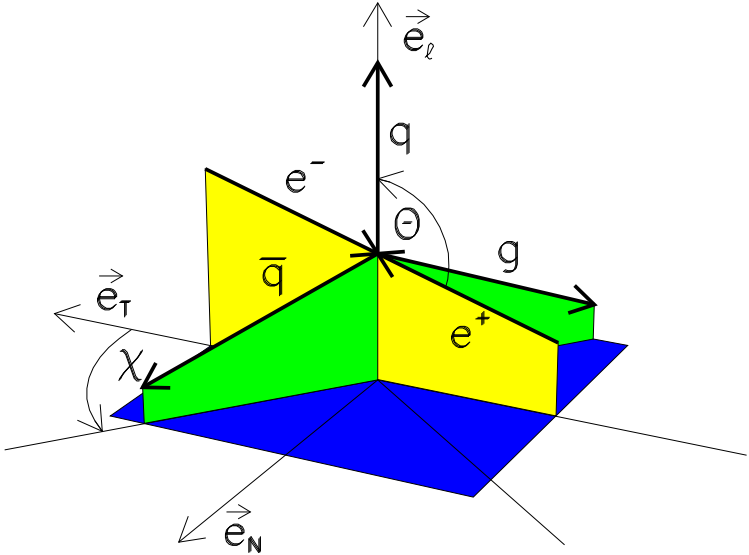, scale=0.8}
\caption{\label{tsys} Orthonormal spin basis $\vec e_T$,
  $\vec e_N$ and $\vec e_\ell$ for the top quark. Also shown are the beam 
plane (light gray, respectively, yellow) and the event plane (dark gray,
respectively, green).}
\end{figure}

In order to acquaint the reader with our notation, we use this section to
outline the main structure of the cross section calculation and to introduce
the various unpolarized and polarized structure functions that come into play.
To start with, we define a polarized hadron tensor for the three-body process
$(\gamma_V,Z)\rightarrow q(p_1)+\bar q(p_2)+G(p_3)$ according to
\begin{equation}\label{hadten}
H_{\mu\nu}(q,p_1,p_2,s)=\sum_{\bar q,G\ {\rm spins}}
  \langle\bar qq(s)G|j_\mu|0\rangle\langle 0|j_\nu^\dagger|\bar qq(s)G\rangle
\end{equation}
where $p_1$, $p_2$ and $p_3$ are the four-momenta of the quark, antiquark and
gluon, respectively, and $q=p_1+p_2+p_3$ is the four-momentum of the
intermediate gauge boson. The spin vector of the quark is denoted by $s$. A
similar definition holds for the Born case
$(\gamma_V,Z)\rightarrow q(p_1)+\bar q(p_2)$. The hadron tensor defined in
Eq.~(\ref{hadten}) depends on the vector ($V$: $\gamma_\mu$) and axial-vector
($A$: $\gamma_\mu\gamma_5$) composition of the product of currents $j_\mu$ and
$j_\nu$. It is convenient to introduce the four independent hadron tensor
components
$H_{\mu\nu}^i$ ($i=1,2,3,4$) defined according to
\begin{eqnarray}\label{H^{i}}
H_{\mu\nu}^1&=&\frac12(H_{\mu\nu}^{VV}+H_{\mu\nu}^{AA}),\qquad
H_{\mu\nu}^2\ =\ \frac12(H_{\mu\nu}^{VV}-H_{\mu\nu}^{AA}),\nonumber\\
H_{\mu\nu}^3&=&\frac{i}2(H_{\mu\nu}^{VA}-H_{\mu\nu}^{AV}),\qquad
H_{\mu\nu}^4\ =\ \frac12(H_{\mu\nu}^{VA}+H_{\mu\nu}^{AV}).
\end{eqnarray}
For notational convenience we have omitted all arguments in the hadron tensor
components in Eqs.~(\ref{H^{i}}). In the following we will use explicit
arguments only when they are needed. For example, we include the spin vector
argument when we define unpolarized and polarized structure functions
$H_{\mu\nu}^i$ and $H_{\mu\nu}^{i,m}$ ($i=1,2,3,4$, $m=\ell,T,N$) according to
\begin{equation}
\label{spinprojection}
H_{\mu\nu}^i=H_{\mu\nu}^i(s^m)+H_{\mu\nu}^i(-s^m),\qquad
H_{\mu\nu}^{i,m}=H_{\mu\nu}^i(s^m)-H_{\mu\nu}^i(-s^m)
\end{equation}
where $s^m$ is the spin vector corresponding to longitudinal ($m=\ell$),
transverse ($m=T$) and normal ($m=N$) polarization of the top quark. Our
choices of the three orthonormal spin directions
$(\vec e_T,\vec e_N,\vec e_\ell)$ are given by
\begin{equation}\label{basistop}
\vec e_T=\frac{(\vec p_{e^-}\times\vec p_1)\times\vec p_1}
{|(\vec p_{e^-}\times\vec p_1)\times\vec p_1|},\qquad
\vec e_N=\frac{\vec p_{e^-}\times\vec p_1}
{|\vec p_{e^-}\times\vec p_1|},\qquad 
\vec e_\ell=\frac{\vec p_1}{|\vec p_1|}
\end{equation}
(cf.\ Fig.~\ref{tsys}).
For the hadron tensor components we introduce the compact notation
$H_{\mu\nu}^{i(m)}$ where the round brackets indicate that, in the unpolarized
case, the index $m$ and the round bracket is omitted. We use this compact
notation to display the general features common to the unpolarized and
polarized parts.

For the process $e^+e^-\rightarrow\bar qq(G)$, the cross section can be
written in modular form consisting of the hadron tensor, the lepton tensor and
the model dependent coupling coefficients $g_{ij}$. The Standard Model (SM)
values of the coupling coefficients $g_{ij}$ are listed in Appendix~A. The
unpolarized and polarized cross sections read 
\begin{equation}
d\sigma^{(m)}=\frac{e^4}{2q^6}\sum_{i,j=1}^4
  g_{ij}L^{i\,\mu\nu}H_{\mu\nu}^{j(m)}dPS
\end{equation}
where $dPS$ is the phase-space factor. The lepton tensor components
$L^i_{\mu\nu}$ ($i=1,2,3,4$) are defined in the same way as in
Eq.~(\ref{H^{i}}). The process $e^+e^-\rightarrow\bar qq(G)$ can be described
either in the {\em beam plane\/} spanned by the electron and positron beam and
the outgoing quark, or the {\it event plane\/} spanned by the quark, the
antiquark, and the gluon. In the Born case where no gluon is emitted, both
planes coincide by convention. The polar angle between the quark momentum and
the electron momentum is denoted by $\theta$ (or by $\theta_{te^-}$), and 
the azimuthal angle between
the two planes is denoted by $\chi$. In order to determine directions, we
define different frames with the $(x,z)$ plane lying in the corresponding
plane. For the beam plane we define a {\it lepton frame\/} with the $z$ axis
determined by the momentum direction of the electron, and a {\it beam frame\/}
with the $z$ axis determined by the momentum direction of the quark. For the
event plane we define an {\it event frame\/} with the $z$ direction determined
again by the momentum direction of the quark. The transition from one frame to
the other is performed by using the two Euler angles $\theta$ and $\chi$.

The natural frame for describing the hadron tensor is the event frame which
makes no reference to the beam plane. On the other hand, the lepton tensor is
most naturally described in the lepton frame. In this frame the lepton
tensor component $L^{3\,\mu\nu}$ vanishes identically and $L^{2\,\mu\nu}$
vanishes for zero lepton masses (which we assume). The remaining two
components have the simple form
\begin{equation}
L^{1\,\mu\nu}=\frac{q^2}2\pmatrix{0&0&0&0\cr 0&1&0&0\cr 0&0&1&0
\cr 0&0&0&0\cr},\qquad
L^{4\,\mu\nu}=\frac{q^2}2\pmatrix{0&0&0&0\cr 0&0&-i&0\cr 0&i&0&0\cr 0&0&0&0
\cr}.
\end{equation}
The contraction of the lepton and hadron tensor has to be done in one
particular frame for which we choose the event frame. We therefore have to
rotate the lepton tensor into the event frame. In doing so a variety of
angular dependences appear. In fact we can decompose the lepton tensors
according to
\begin{eqnarray}\label{decomp}
L^{1\,\mu\nu}&=&\frac{q^2}2\left\{\frac12(1+\cos^2\theta)\proj_U^{\mu\nu}
+\sin^2\theta\,\proj_L^{\mu\nu}
  -2\sqrt2\sin\theta\cos\theta\,\proj_I^{\mu\nu}\right\},\nonumber\\
L^{4\,\mu\nu}&=&\frac{q^2}2\left\{\cos\theta\,\proj_F^{\mu\nu}
-2\sqrt2\sin\theta\,\proj_A^{\mu\nu}\right\}
\end{eqnarray}
where $\proj_I$ and $\proj_A$ contain an implicit linear dependence on 
$\sin\chi$ and $\cos\chi$. The matrices $\proj_U$, $\proj_L$, $\proj_I$,
$\proj_F$ and $\proj_A$ are called {\it projectors\/} because when contracting
the lepton tensor with the hadron tensor they project out the relevant
coefficients of the hadron tensor that give rise to the various angular
dependences. The decomposition in Eq.~(\ref{decomp}) describes the complete
angular dependence of unpolarized and polarized top production in the process
$e^+e^-\rightarrow\bar qq(G)$. It gives rise to the decomposition of the
differential cross section according to
\begin{equation}\label{polar}
\frac{d\sigma^{(m)}}{d\cos\theta}=\frac38(1+\cos^2\theta)\sigma_U^{(m)}
  +\frac34\sin^2\theta\,\sigma_L^{(m)}+\frac34\cos\theta\,\sigma_F^{(m)}
  -\frac3{\sqrt2}\sin\theta\cos\theta\,\sigma_I^{(m)}
  -\frac3{\sqrt2}\sin\theta\,\sigma_A^{(m)}
\end{equation}
where
\begin{eqnarray}\label{sigmaa}
\sigma_a^{(m)}&=&\frac{(4\pi\alpha)^2}{3q^4}
  \sum_{j=1}^4g_{ij}\int H_a^{j(m)}\frac{dPS}{d\cos\theta},\quad
  H_a^{j(m)}=\proj_a^{\mu\nu}H_{\mu\nu}^{j(m)}.
\end{eqnarray}
Without beam polarization effects one finds the following pattern. For $i=1$
one has contributions from $a=U,L,I$ and for $i=4$ one has contributions from
$a=F,A$ as written out in Eq.~(\ref{decomp}). More details about the coupling
pattern including transverse and longitudinal beam polarization effects can be
found in~\cite{Groote:1996nc}. In Eqs.~(\ref{sigmaa}) we have divided out the
$d\cos\theta$ differential which has already been taken into account in the
polar distribution (\ref{polar}). For the two-particle final states (Born term
and loop contribution) one has the phase-space factor
\begin{equation}\label{dps2}
dPS_2=\frac{v}{8(2\pi)^2}\ d\cos\theta\,d\chi\rightarrow
  \frac{v}{16\pi}\ d\cos\theta.
\end{equation}
where $v=\sqrt{1-4m^2/q^2}$ is the velocity of the outgoing quark. The
transition to the rightmost form in Eq.~(\ref{dps2}) marked by an arrow
expresses the fact that the azimuthal integration over $\chi$ is always
implied throughout this paper. As we shall see, the transverse and normal spin
dependence drop out for the components $a=U,L,F$ in $H^{j(m)}_a$ but are
retained for the components $a=I,A$ after the azimuthal integration over
$\chi$.\footnote{It is important to keep in mind that the transverse and
normal spin components are defined w.r.t.\ the beam frame. When defined w.r.t.\
the event frame the transverse and normal spin components average to zero
after azimuthal averaging.} Just the opposite happens to the spin independent
and longitudinal spin components. For the two-particle final state one obtains
\begin{equation}\label{sigmab}
\sigma_a^{(m)}({\it Born,loop\/})=\frac{\pi\alpha^2v}{3q^4}
  \sum_{j=1}^4g_{ij}H_a^{j(m)}({\it Born,loop\/})
\end{equation}
with $a=U,L,I$ for $i=1$ and $a=F,A$ for $i=4$ as above.

Next we turn to the $O(\as)$ tree graph contributions. The relevant three
particle final state phase-space is given by
\begin{equation}
dPS_3=\frac{v}{8(2\pi)^2}\ d\cos\theta\,d\chi\ \frac{q^2}{16\pi^2v}\ dy\,dz
  \rightarrow\frac{v}{16\pi}\ d\cos\theta\ \frac{q^2}{16\pi^2v}\ dy\,dz
\end{equation}
where the transition to the last expression is again due to the azimuthal
integration. We have introduced two phase-space variables
$y=1-2p_1\cdot q/q^2$ and $z=1-2p_2\cdot q/q^2$. The $O(\as)$ tree graph
contributions to the various cross sections $\sigma_a^{(m)}$ are written as
\begin{eqnarray}\label{sigmac}
\sigma_a^{(m)}({\it tree\/})&=&\frac{\pi\alpha^2v}{3q^4}
  \left(\frac{q^2}{16\pi^2v}\sum_{j=1}^4g_{ij}\int H_a^{j(m)}(y,z)
  dy\,dz\right)
\end{eqnarray}
with $a=U,L,I$ for $i=1$ and $a=F,A$ for $i=4$, as before. It is convenient to
introduce the tree graph helicity structure functions
$H_a^{j(m)}({\it tree\/})$ by defining
\begin{equation}
\label{integmeasure}
H_a^{j(m)}({\it tree\/})=\frac{q^2}{16\pi^2v}\int H_a^{j(m)}(y,z)dy\,dz.
\end{equation}
The Born term and the $O(\as)$ corrections $H_a^{j(m)}({\it Born\/})$ and 
$H_a^{j(m)}(\as)=H_a^{j(m)}({\it tree\/})+H_a^{j(m)}({\it loop\/})$ will be
referred to as the {\it unpolarized\/} and {\it polarized structure
functions\/} to leading (LO) and next-to-leading (NLO) order, respectively,
while the sum of the LO and NLO contributions will be referred to as the
$O(\as)$ results.

In summary, one has three unpolarized and seven polarized hadronic helicity
structure functions. It is instructive to list them together including a
specification of whether they are fed by the parity conserving ($pc$) or by the
parity violating ($pv$) part of the product of hadronic currents and to which
of the two classes of the so-called $T$-even and $T$-odd structure functions
they belong to. One has  
\begin{eqnarray}
\mbox{unpolarized:}             &&H_U(pc),\,H_L(pc),\,H_F(pv)
\qquad\mbox{$T$-even}\\
\mbox{longitudinally polarized:}&&H_U^\ell(pv),\,H_L^\ell(pv),\,H_F^\ell(pc)
\qquad\mbox{$T$-even}\\ 
\mbox{transversely polarized:}  &&H_A^T(pc),\,H_I^T(pv)
\hspace{2.3cm}\mbox{$T$-even}\\
\label{class4}
\mbox{normal polarization:}     &&H_I^N(pc),\,H_A^N(pv)
\hspace{2.2cm}\mbox{$T$-odd}
\end{eqnarray}
If one neglects contributions proportional to the imaginary part $\imag\cz$ of
the Breit--Wigner line shape of the $Z$-boson (see Appendix A) the $T$-odd
helicity structure functions $H_A^N(pv)$ and $H_I^N(pc)$ are contributed to by
the imaginary parts of the one-loop amplitudes leading to nonvanishing triple
product correlations of the type $\vec s_t\cdot(\vec l\times\vec p_t)$,
whereas the $T$-even structure functions obtain contributions from the Born
term, the $O(\as)$ tree graph contributions and the real part of the one-loop
contributions.

If one includes the contributions proportional to the imaginary part
$\imag\cz$ the structure functions $H_F(pv)$, $H_U^\ell(pv)$ and $H_I^T(pv)$
are also contributed to by the imaginary parts of the one-loop contributions,
and, vice versa, $H_A^N(pv)$ obtains also contributions from the Born term,
the $O(\as)$ tree graph contributions and the real part of the one-loop
contributions. All the latter contributions originate from the $(VA-AV)$ part
of the product of hadron currents and thus belong to the class of helicity
structure functions $H_a^{3(m)}$ according to the classification of
Eq.~(\ref{H^{i}}). The latter contributions can only be probed through the
imaginary part of the Breit--Wigner resonance shape which is strongly
suppressed for $(t\bar t)$ production. In fact, the contributions coming from
the imaginary part of the Breit--Wigner resonance shape are of order
$O(\imag\cz(q^{2})/\real\cz(q^{2}))$ and can thus safely be neglected for top
quark pair production. For example, in the threshold region of top quark pair
production $\imag\cz/\real\cz$ is approximately $0.1\%$ and decreases further
with a $1/q^2$ power fall-off behaviour. We shall nevertheless include all
$H_a^{3(m)}$ contributions for completeness and for possible applications in
$(b\bar b)$ production where the $H_a^{3(m)}$ contributions cannot be
neglected in the $Z$ resonance region.

\section{Covariant expressions for the projectors}
The projectors $\proj_a$ will be written in covariant form. We go to the rest
frame of the gauge boson such that $q=(\sqrt{q^2};0,0,0)$. The $z$ axis is
defined by the momentum direction of the top quark. For the top quark momentum
one has
\begin{equation}
p_1=\frac12\sqrt{q^2}\left(1-y;0,0,\sqrt{(1-y)^2-\xi}\right)
\end{equation}
($y=0$ for two-body decays) with $\xi=1-v^2=4m^2/q^2$. We construct a
four-transverse quark momentum and a four-transverse metric tensor
\begin{equation}
\hat g_{\mu\nu}=g_{\mu\nu}-\frac{q_\mu q_\nu}{q^2},\qquad
\hat p_{1\mu}=\hat g_{\mu\nu}p_1^\nu=p_{1\mu}-\frac{p_1\cdot q}{q^2}q_\mu
\end{equation}
and use $q$ and $\hat p_1$ to build up two elements of a coordinate basis,
\begin{eqnarray}
e_0^\mu&=&\left(q^\mu/\sqrt{q^2}\right)\  
  \bigg(\,=(1;0,0,0)\,\mbox{in the gauge boson rest system}\bigg),\\
e_3^\mu&=&\left(\hat p_1^\mu/\sqrt{(p_1\cdot q)^2/q^2-m^2}\right)
\ \bigg(\,=(0;0,0,1)\,\mbox{in the gauge boson rest system}\bigg).\qquad
\end{eqnarray}
In covariant form the longitudinal spin vector of the top quark 
reads (see e.g.~\cite{Fischer:2001gp})
\begin{equation}\label{covspin}
s^{\ell\mu}=-(q^\mu-\frac{p_1q}{m^2}p_1^\mu)/\sqrt{(p_1q)^2/m^2-q^2}.
\end{equation}
In the gauge boson rest system Eq.~(\ref{covspin}) turns into
\begin{equation}
s^\ell=\frac1{\sxi}\left(\sqrt{(1-y)^2-\xi};0,0,1-y\right),
\end{equation}
while in the top quark rest system one has $s^\ell=(0;0,0,1)$. The
longitudinal spin vector $s^\ell$ can be seen to be a linear combination of
the two basis vectors $e_0$ and $e_3$ and does not provide a new direction in 
our vierbein basis. The projectors that can be constructed with the help of
$e_0$ and $e_3$ are limited to the three projectors
\begin{eqnarray}\label{project1}
\proj_U^{\mu\nu}&=&-\hat g^{\mu\nu}-e_3^\mu e_3^\nu,\nonumber\\
\proj_L^{\mu\nu}&=&e_3^\mu e_3^\nu,\nonumber\\
\proj_F^{\mu\nu}&=&i\eps_{\mu\nu\rho\sigma}e_3^\rho e_0^\sigma
\end{eqnarray}
where $\eps_{\mu\nu\rho\sigma}$ is the totally antisymmetric Levi-Civit\`a
tensor with $\eps_{0123}=1$. They project out the three unpolarized and three
longitudinally polarized helicity structure functions where, according to
Eq.~(\ref{spinprojection}), the polarized structure functions $H^{i\ell}_a$
($a=U,L,F$) are obtained from $H^{i\ell}_a=\proj_a^{\mu\nu}
\left(H^i_{\mu\nu}(s^\ell)-H^{i}_{\mu\nu}(-s^\ell)\right)$.

The transverse and normal polarization vectors of the top quark are defined 
in the beam frame. Viewed from the event frame they are given by
\begin{equation}
e_T=(0;\cos\chi,-\sin\chi,0),\qquad
e_N=(0;\sin\chi,\cos\chi,0).
\end{equation}
These two vectors therefore allow one to span the beam plane and a plane
perpendicular to the beam plane in event frame coordinates. With these new
elements it is possible to construct the remaining additional projectors. They
read ($m=T,N$)
\begin{eqnarray}\label{project2}
\proj_I^{\mu\nu}(e_m)=\frac{-1}{2\sqrt2}(s^\mu e_3^\nu+e_3^\mu e_m^\nu),&&
\proj_I^{\prime\,\mu\nu}(e_m)=\frac{-1}{2\sqrt2}\left(\eps_{\mu\rho\sigma\tau}
  e_3^\nu+\eps_{\nu\rho\sigma\tau}e_3^\mu\right)e_0^\rho e_3^\sigma e_m^\tau,
  \nonumber\\
\proj_A^{\mu\nu}(e_m)=\frac{-i}{2\sqrt2}\eps_{\mu\nu\rho\sigma}e_0^\rho
  e_m^\sigma,&&
\proj_A^{\prime\,\mu\nu}(e_m)=\frac{i}{2\sqrt2}(e_m^\mu e_3^\nu
  -e_3^\mu e_m^\nu).
\end{eqnarray}
For example, according to Eq.~(\ref{spinprojection}), one obtains the
structure function $H_{I}^{4T}$ by calculating
$H^{4T}_{I}=\proj_I^{\mu\nu}(e_T)
\left(H^4_{\mu\nu}(s^T)-H^i_{\mu\nu}(-s^T)\right)$. Note that since
$\proj_I(e_N)=\proj'_I(e_T)$, $\proj_I(e_T)=-\proj'_I(e_N)$,
$\proj_A(e_N)=-\proj'_A(e_T)$ and $\proj_A(e_T)=\proj'_A(e_N)$ the primed
projectors are redundant. This set of four (Eq.~(\ref{project2})) and six
(Eq.~(\ref{project1})) covariant projectors allows one to calculate the
complete set of ten helicity structure functions from the hadron tensor.

In the following we list the Born term and loop contributions calculated
already in previous papers~\cite{Groote:1995yc,Groote:1995ky,Groote:1996nc}.
The nonvanishing unpolarized Born term contributions are given by
\begin{eqnarray}
H_U^1({\it Born\/})=2N_cq^2(1+v^2),&&
H_L^1({\it Born\/})=N_cq^2(1-v^2)\ =\ H_L^2({\it Born\/}),\nonumber\\[3pt]
H_U^2({\it Born\/})=2N_cq^2(1-v^2),&&
H_F^4({\it Born\/})=4N_cq^2v.
\end{eqnarray}
The longitudinally polarized contributions read
\begin{eqnarray}
H_U^{4\ell}({\it Born\/})=4N_cq^2v,&&
H_F^{1\ell}({\it Born\/})=2N_cq^2(1+v^2),\nonumber\\[3pt]
H_L^{4\ell}({\it Born\/})=0,&&
H_F^{2\ell}({\it Born\/})=2N_cq^2(1-v^2).
\end{eqnarray}
For the transverse and normal polarization components one
has~\cite{Groote:1995ky}
\begin{eqnarray}
H_I^{4T}({\it Born\/})=N_cq^2v\sqrt{\frac\xi2},&&
H_A^{1T}({\it Born\/})=N_cq^2\sqrt{\frac\xi2}=H_A^{2T}({\it Born\/}),
  \nonumber\\
&&H_A^{3N}({\it Born\/})=N_cq^2v\sqrt{\frac\xi2}.
\end{eqnarray}
Note that one has $H_L^1=H_L^2$, $H_U^1=H_F^{1\ell}$, $H_U^2=H_F^{2\ell}$, 
$H_F^4=H_U^{4\ell}$, $H_A^{1T}=H_A^{2T}$, and $H_I^{4T}=H_A^{3N}$ at the Born
term level. We will return to these relations when we discuss the $ O(\as)$
tree graph contributions.

Note that the transverse and normal spin components $T$ and $N$ are
proportional to $\sqrt\xi=2m/\sqrt{q^{2}}$. The origin of this suppression
factor is a helicity flip suppression factor at the $\gamma/Z-t\bar t$ vertex.
The same suppression factor also occurs in the $O(\alpha_s)$ one-loop and tree
graph radiative corrections to be treated later on. It is clear that this
overall suppression factor is not important for $(t\bar t)$ production in the
threshold region and not very significant in the range of beam energies
considered in this paper. Altogether this means that the transverse and normal
spin components of the top quark are non-negligible in the present
application~\cite{Groote:1995ky,Ravindran:2000rz}.
  
Most of the nonvanishing one-loop contributions have already been given 
in~\cite{Groote:1995yc,Groote:1995ky,Groote:1996nc}
\begin{eqnarray}
H_U^1({\it loop\/})&=&4N_cq^2\left((1+v^2)\real A-2v^2\real B\right),
  \nonumber\\
H_U^2({\it loop\/})&=&4N_cq^2\left((1-v^2)\real A+2v^2\real B\right),
  \nonumber\\
H_L^1({\it loop\/})&=&2N_cq^2\left((1-v^2)\real A+v^2\real B\right)
  \ =\ H_L^2({\it loop\/}),\nonumber\\[3pt]
H_F^3({\it loop\/})&=&-8N_cq^2v\imag B,\nonumber\\[3pt]
H_F^4({\it loop\/})&=&8N_cq^2v\left(\real A-\real B\right),\nonumber\\[7pt]
H_U^{3\ell}({\it loop\/})&=&-8N_cq^2v\imag B,\nonumber\\[3pt]
H_U^{4\ell}({\it loop\/})&=&8N_cq^2v\left(\real A-\real B\right),
  \nonumber\\[3pt]
H_L^{3\ell}({\it loop\/})&=&0\ =\ H_L^{4\ell}({\it loop\/}),\nonumber\\[3pt]
H_F^{1\ell}({\it loop\/})&=&4N_cq^2\left((1+v^2)\real A-2v^2\real B\right),
  \nonumber\\
H_F^{2\ell}({\it loop\/})&=&4N_cq^2\left((1-v^2)\real A+2v^2\real B\right),
  \nonumber\\[7pt]
H_I^{3T}({\it loop\/})&=&-N_cq^2v\sqrt{\frac\xi2}(1+\xi)\imag B/\xi,
  \nonumber\\
H_I^{4T}({\it loop\/})&=&N_cq^2v\sqrt{\frac\xi2}
  \left(2\real A+(1-3\xi)\real B/\xi\right),\nonumber\\
H_A^{1T}({\it loop\/})&=&N_cq^2\sqrt{\frac\xi2}
  \left(\real A+v^2\real B/\xi\right)
  \ =\ H_A^{2T}({\it loop\/}),\nonumber\\[7pt]
H_I^{1N}({\it loop\/})&=&N_cq^2\sqrt{\frac\xi2}(1-\xi)\imag B/\xi
  \ =\ H_I^{2N}({\it loop\/}),\nonumber\\
H_A^{3N}({\it loop\/})&=&N_cq^2v\sqrt{\frac\xi2}
  \left(2\real A+(1-3\xi)\real B/\xi\right),\nonumber\\
H_A^{4N}({\it loop\/})&=&N_cq^2v\sqrt{\frac\xi2}(1+\xi)\imag B/\xi.
\end{eqnarray}
where the real part of the form factor $A$ and the real and imaginary parts
of the form factor $B$ read
\begin{eqnarray}
\real A&=&-\frac{\alpha_sC_F}{4\pi}
  \Bigg\{\left(2+\frac{1+v^2}v\ln\pfrac{1-v}{1+v}\right)
  \ln\pfrac{\Lambda q^2}{m^2}+3v\ln\pfrac{1-v}{1+v}+4\nonumber\\&&\qquad
  +\frac{1+v^2}v\left(\Li_2\pfrac{2v}{1+v}+\frac14\ln^2\pfrac{1-v}{1+v}
  -\frac{\pi^2}2\right)\Bigg\},\nonumber\\
\real B&=&\frac{\alpha_sC_F}{4\pi}\ \frac{1-v^2}v\ln\pfrac{1-v}{1+v},\qquad
\imag B\ =\ \frac{\alpha_sC_F}{4\pi}\ \frac{1-v^2}v\pi
\end{eqnarray}
The imaginary contributions $H_F^3({\it loop\/})$ and
$H_U^{3\ell}({\it loop\/})$ complete the list of one-loop contributions given
in~\cite{Groote:1995yc,Groote:1995ky,Groote:1996nc}. We are now in full
agreement with the one-loop contributions given in~\cite{Ravindran:2000rz}.  
$\imag B$ contributes to the $T$-odd structure functions $H_I^{4T}$ and
$H_A^{4N}$ as mentioned after Eq.~(\ref{class4}). The infrared singularity has
been regularized by the introduction of a gluon mass $m_G$ via
$m_G^2=\Lambda q^2$. The loop induced infrared singularities in the real part
of the one-loop contributions can be seen to cancel against the corresponding
infrared singularities in the tree graph contributions to be treated later on.

In the next section we will present our results on the cut-off dependent
helicity structure functions. They must coincide with the fully integrated
results written down in Refs.~\cite{Groote:1995yc,Groote:1995ky,Groote:1996nc}
when the cut-off is taken to its maximal value
$E_G({\rm max})=(q^2-4m^2)/(2\sqrt{q^2})$. This will be verified in Sec.~5.

\section{Exact result up to a given gluon energy cut}
\stepcounter{figure}\noindent%
\parbox[b]{7truecm}{In this section we present the results of our calculations
for the $O(\as)$ corrections to the helicity structure functions with a
given cut on the gluon energy. We define a scaled gluon energy cut
$\lambda=E_G/\sqrt{q^2}$ and do the phase-space integration in the region
$0\le E_G\le\lambda_{{\rm max}}\sqrt{q^2}$. The maximal value that the cut 
parameter $\lambda$ can take is $\lambda_{{\rm max}}=(1-\xi)/2$. In terms of
our phase-space variables $y$ and $z$ the cut phase-space is defined by
$0\le y+z\le2\lambda$. In Fig.~\arabic{figure} we have drawn a $(y,z)$
phase-space plot choosing a specific value for $\xi=0.1$ for illustrative
purposes. The shaded area corresponds to the integration region with the
specific choice of cut value $\lambda=0.3$. The upper boundary of the
integration region is given by the straight line $z=-y+2\lambda$.\pbreak}\hfill
\parbox[b]{9truecm}{\quad\epsfig{figure=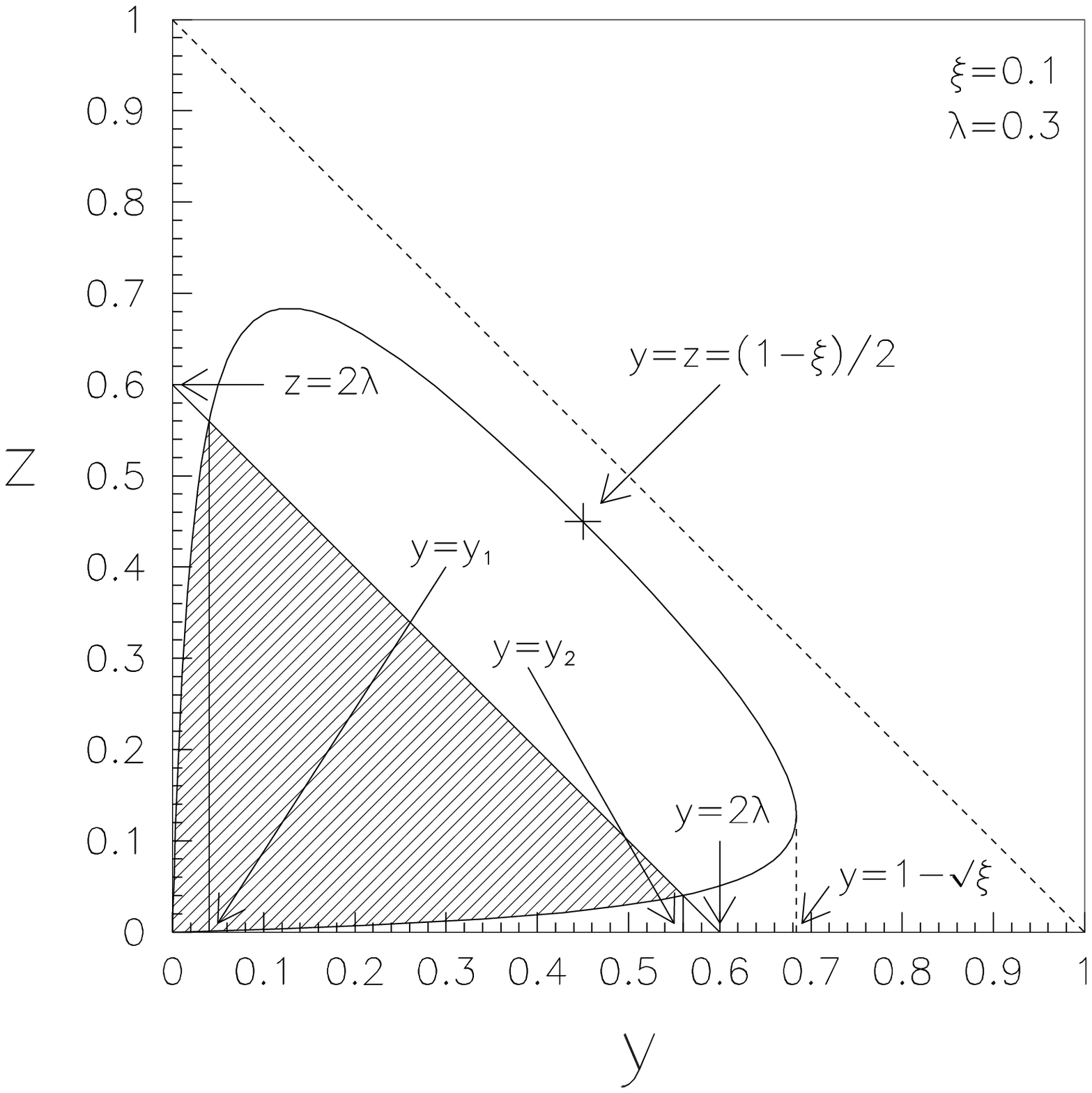, scale=0.5}\\[7pt]\strut
\qquad Figure~\arabic{figure}: Phase-space diagram with gluon cut
\vspace{7pt}}

\vspace{12pt}
The full phase-space is bounded from above and below by the two functions
$z_+$ and $z_-$ where
\begin{eqnarray}\label{boundz}
z_\pm=\frac{2y-2y^2-\xi y\pm 2y\sqrt{(1-y)^2-\xi}}{4y+\xi}.
\end{eqnarray}
The upper gluon cut given by $z=-y+2\lambda$ intersects the two boundary
curves (\ref{boundz}) at
\begin{equation}
y_1=\lambda\left(1-\sqrt{\frac{1-2\lambda-\xi}{1-2\lambda}}\right),\qquad
y_2=\lambda\left(1+\sqrt{\frac{1-2\lambda-\xi}{1-2\lambda}}\right).
\end{equation}
Since the phase-space is symmetric with respect to reflections along the
diagonal, the corresponding $z$-values are $z_1=y_2$ and $z_2=y_1$. 

From a visual inspection of the phase-space plot Fig.~\arabic{figure} one can
see that one has to discuss two cases when integrating the cut phase-space
depending on whether (case~A) $\lambda\le\lambda_{\rm trans}$ or (case~B)
$\lambda>\lambda_{\rm trans}$. The transition value
$\lambda_{\rm trans}=(1-\sxi)/(2-\sxi)$ is defined by the $\lambda$ value at
which the straight boundary line of the cut intersects the phase-space
boundary at the point $(y_2,z_2)=(1-\sxi,\sxi(1-\sxi)/(2-\sxi))$. At this
point the tangent of the full phase-space boundary is vertical. From an
inspection of the phase-space plot Fig.~\arabic{figure} one concludes that in
case~A the integration region is divided into two parts, whereas one has to
consider three integration regions in case~B. The specific example shown in
Fig.~\arabic{figure} corresponds to case~A.

Let us denote the general $y$- and $z$-dependent tree graph integrand in
case~A by $I(y,z)$. One has to do the two integrations
\begin{equation}\label{caseA}
\int_0^{y_1}\int_{z_-}^{z_+} I(y,z)dy\,dz
  +\int_{y_1}^{y_2}\int_{z_-}^{2\lambda-y}I(y,z)dy\,dz,
\end{equation}
while in case~B one has an additional integration, viz.\
\begin{equation}\label{caseB}          
\int_0^{y_1}\int_{z_-}^{z_+}I(y,z)dy\,dz
  +\int_{y_1}^{y_2}\int_{z_-}^{2\lambda-y}I(y,z)dy\,dz
  +\int_{y_2}^{1-\sqrt{\xi}}\int_{z_-}^{z_+}I(y,z)dy\,dz.
\end{equation}
It is clear that one should recover the fully integrated results listed in
Sec.~5 when setting $\lambda$ to its maximal value
$\lambda_{\rm max}=(1-\xi)/2$. When comparing to the fully integrated result
one has to discuss case~B with $\lambda=\lambda_{\rm max}=y_1=y_2=(1-\xi)/2$.
In this case the second integral in Eq.~(\ref{caseB}) vanishes and the
remaining two integrals can be merged to give
\begin{equation}
\int_0^{1-\sxi}\int_{z_-}^{z_+}I(y,z)dy\,dz
\end{equation}
which corresponds to the fully integrated tree graph contribution entering the
full NLO result given in Sec.~5.

Let us return to case~A involving the two integrations in Eq.~(\ref{caseA}).
For most practical applications case~A will be the relevant case since the
ratio
\begin{equation}
\frac{\lambda_{\rm trans}}{\lambda_{\rm max}}=\frac2{(1+\sxi)(2-\sxi)}
\end{equation}
remains close to $1$ over most of the range of $\xi$ values. The integration
over $z$ is straightforward. The second integration over $y$ is done by using
the Euler substitution
\begin{equation}\label{Eulersub}
y=1-\sxi\frac{1+w^2}{1-w^2}.
\end{equation}
Eq.~(\ref{Eulersub}) is easily inverted. The $y$-integration limits $y=0$,
$y_1$, $y_2$, $2\lambda$, and $1-\sxi$ translate into $w=w_0$, $w_1$, $w_2$,
$w_\lambda $, and $0$, where
\begin{equation}
w_0=\sqrt{\frac{1-\sxi}{1+\sxi}},\quad
w_{1,2}=\sqrt{\frac{1-y_{1,2}-\sxi}{1-y_{1,2}+\sxi}},\quad
w_\lambda=\sqrt{\frac{1-2\lambda-\sxi}{1-2\lambda+\sxi}}.
\end{equation}
The value $w_\lambda$ corresponds to the intersection of the upper gluon cut
boundary with any of the two axes. In addition to the velocity parameter
$v=\sqrt{1-\xi}$ we introduce modified velocity parameters
$v_i=\sqrt{(1-y_i)^2-\xi}$ and $v_\lambda=\sqrt{(1-2\lambda)^2-\xi}$. We shall
also use the abbreviations $a=2+\sxi$ and $b=2-\sxi$. Our results for case~A
read ($N=\as N_cC_Fq^2/(4\pi v)$)
\begin{eqnarray}
\lefteqn{H^1_U=N\Bigg\{2(2-\xi)^2(t_{0-}-t_{0+})
  -(8-10\xi-\xi^2)(t_{1-}-t_{1+})+\strut}\nonumber\\&&\strut
  +\sxi(1-\sxi)(2+4\sxi-3\xi)t_w-(6-11\xi)v+2(2-\xi)\ell_{4+}
  -16\xi\ell_{5+}+\strut\nonumber\\&&\strut-\frac14(8+12\xi-\xi^2)\ell_{6+}
  +\left(16\lambda-8\lambda^2-2\xi+\xi^2-\frac{4\lambda^2\xi^2}{v_\lambda^2}
  \right)\ell_{7+}+\strut\nonumber\\&&\strut
  -\left(16\lambda-6\lambda^2-2(1+4\lambda+\lambda^2)\xi+\xi^2
  -\frac{2\lambda^2}{v^2}\right)\ell_{8+}+2y_1(4-y_1)\ell_1-2y_2(4-y_2)\ell_2
  +\strut\nonumber\\&&\strut
  -\frac{2(1-2\lambda-(1-\lambda)\sxi)}{(1-\sxi)\sxi}
  \Bigg((1-2\lambda)(2-\xi)+\nonumber\\&&\qquad\qquad\qquad\qquad\qquad\strut
  +2\lambda\sxi-(4+3\lambda)\xi\sxi+3\xi^2
  -\frac{4\lambda^2\xi\sxi}{1-2\lambda-\sxi}\Bigg)\ell_3
  +\strut\nonumber\\&&\strut
  -\frac14(24+5\xi)v_1+\frac{b\xi\sxi}{2(b-aw_1)}
  -\frac14(48+5\xi)y_1+6y_1^2+\strut\nonumber\\&&\strut
  -\frac14(24+5\xi)v_2-\frac{b\xi\sxi}{2(b+aw_2)}
  +\frac14(48+5\xi)y_2-6y_2^2\Bigg\}\\[12pt]
\lefteqn{H^2_U=\xi N\Bigg\{2(2-\xi)(t_{0-}-t_{0+})
  -(4-\xi)(t_{1-}-t_{1+})+\sxi(1-\sxi)t_w+2v+2\ell_{4+}+\strut}
  \nonumber\\&&\strut
  -\frac32\xi\ell_{6+}+(8\lambda-\xi)(\ell_{7+}-\ell_{8+})
  +\frac{2\lambda^2\xi}{v^2}\ell_{8+}+4y_1\ell_1-4y_2\ell_2
  +\strut\nonumber\\&&\strut
  -2\frac{1-2\lambda-(1-\lambda)\sxi}{(1-\sxi)\sxi}
    (1-2\lambda-\xi+\lambda\sxi)\ell_3-5v_1-5y_1-5v_2+5y_2\Bigg\}\\[12pt]
\lefteqn{H^{4\ell}_U=N\Bigg\{4(2-\xi)v(t_{0-}+t_{0+})
  -(8+2\xi+3\xi^2)(t_{1-}+t_{1+})-\frac12(1+\sxi)(2-\sxi)^2+\strut}
  \nonumber\\&&\strut
  +4v\ell_{4-}-3(2-5\xi)v\ell_{5-}+\frac\xi4(28-17\xi)\ell_{6-}
  +\left((6-16\lambda-13\xi)v+\frac{8\lambda^2}v\right)\ell_{8-}
  +\strut\nonumber\\&&\strut
  -v_\lambda\left(6-4\lambda-13\xi-2\lambda\xi
  +\frac{4\lambda\xi}{v_\lambda^2}(7\lambda+2\lambda^2+\xi)
  +\frac{8\lambda^3\xi^2}{v_\lambda^4}\right)\ell_{7-}
  +\strut\nonumber\\&&\strut
  +v_1\left(2\left(y_1-3(1-\xi)\right)
  +\frac{3(1+\sxi)^2(2+\sxi)\sxi}{2(1-y_1+\sxi)}
  -\frac{3(1-\sxi)^2(2-\sxi)\sxi}{2(1-y_1-\sxi)}\right)\ell_1
  +\strut\nonumber\\&&\strut
  -v_2\left(2\left(y_2-3(1-\xi)\right)
  +\frac{3(1+\sxi)^2(2+\sxi)\sxi}{2(1-y_2+\sxi)}
  -\frac{3(1-\sxi)^2(2-\sxi)\sxi}{2(1-y_2-\sxi)}\right)\ell_2
  +\strut\nonumber\\&&\strut
  -\frac{1-2\lambda-(1-\lambda)\sxi}{(1-2\lambda-\sxi)\sxi}
    \Bigg\{\,2-8\lambda+8\lambda^2-7\xi+5\lambda\xi+2\lambda^2\xi-3\xi^2
  +\strut\nonumber\\&&\qquad\qquad\qquad\qquad\qquad\strut
  -(1-6\lambda+8\lambda^2-9\xi+3\lambda\xi)\sxi\Bigg\}
  \left(\frac1{w_1}-\frac1{w_2}\right)+\strut\nonumber\\&&\strut
  +\frac{1-2\lambda+(1-\lambda)\sxi}{(1-2\lambda+\sxi)\sxi}
    \Bigg\{\,2-8\lambda+8\lambda^2-7\xi+5\lambda\xi+2\lambda^2\xi-3\xi^2
  +\strut\nonumber\\&&\qquad\qquad\qquad\qquad\qquad\strut
  +(1-6\lambda+8\lambda^2-9\xi+3\lambda\xi)\sxi\Bigg\}(w_1-w_2)
  +\strut\nonumber\\&&\strut
  +\frac14(40-48\lambda-33\xi)(v_1-v_2)
  -4\lambda\xi\left(\frac{v_1}{y_1}-\frac{v_2}{y_2}\right)
  +\frac{b\xi\sxi}{2(b-aw_1)}+\frac{b\xi\sxi}{2(b+aw_2)}
  +\strut\nonumber\\&&\strut
  +\frac14(24-33\xi-8v_1)y_1-2y_1^2
  +\frac14(24-33\xi+8v_2)y_2-2y_2^2\Bigg\}\\[12pt]
\lefteqn{H^1_L=N\Bigg\{\xi(2-\xi)(t_{0-}-t_{0+})-2\xi(2+\xi)(t_{1-}-t_{1+})
  -\sxi(1-\sxi)(2+4\sxi-3\xi)t_w+\strut}\nonumber\\&&\strut
  +\frac14(16-54\xi+3\xi^2)v+\xi\ell_{4+}+16\xi\ell_{5+}
  +\frac\xi{16}(8+8\xi-3\xi^2)\ell_{6+}+\strut\nonumber\\&&\strut
  -\frac\xi{2v_\lambda^2}(8\lambda-28\lambda^2+16\lambda^3+16\lambda^4
    +\xi-12\lambda\xi-8\lambda^2\xi-\xi^2)\ell_{7+}+\strut\nonumber\\&&\strut
  -\frac\xi{2v^2}(8\lambda+4\lambda^2-\xi-8\lambda\xi+\xi^2)\ell_{8+}
  -\xi\left(2y_1+\frac12y_1^2\right)\ell_1
  +\xi\left(2y_2+\frac12y_2^2\right)\ell_2+\strut\nonumber\\&&\strut
  +\frac{2(1-2\lambda-(1-\lambda)\sxi)}{(1-2\lambda-\sxi)(1-\sxi)\sxi}
    (2-8\lambda+8\lambda^2-\xi+2\lambda\xi-4\lambda^2\xi+7\xi^2
    -3\lambda\xi^2+\strut\nonumber\\&&\qquad\qquad\qquad\qquad\strut
    -(2-6\lambda+4\lambda^2+3\xi-3\lambda\xi-2\lambda^2\xi+3\xi^2)\sxi)
    \ell_3+\strut\nonumber\\&&\strut
  -\frac1{16}(32-72\xi+5\xi^2-8\xi y_1)v_1+\frac{b\xi^2\sxi}{8(b-aw_1)}
  -\frac1{16}(32-72\xi+5\xi^2)y_1+\strut\nonumber\\&&\strut
  -\frac1{16}(32-72\xi+5\xi^2-8\xi y_2)v_2-\frac{b\xi^2\sxi}{8(b+aw_2)}
  +\frac1{16}(32-72\xi+5\xi^2)y_2\Bigg\}
  \nonumber\\\\
\lefteqn{H^2_L=\xi N\Bigg\{(2-\xi)(t_{0-}-t_{0+})-2v^2(t_{1-}-t_{1+})
  -\sxi(1-\sxi)t_w+\frac14(22-3\xi)v+\strut}\nonumber\\&&\strut
  +\ell_{4+}+\frac1{16}(8-8\xi+3\xi^2)\ell_{6+}
  +\frac12(8\lambda+4\lambda^2-\xi)(\ell_{7+}-\ell_{8+})
  -\frac{2\lambda^2\xi}{v^2}\ell_{8+}+\strut\nonumber\\&&\strut
  +\left(2y_1+\frac12y_1^2\right)\ell_1
  -\left(2y_2+\frac12y_2^2\right)\ell_2
  +2\frac{1-2\lambda-(1-\lambda)\sxi}{(1-\sxi)\sxi}
    (1-2\lambda-\xi+\lambda\sxi)\ell_3+\strut\nonumber\\&&\strut
  -\frac1{16}(72-5\xi+8y_1)v_1-\frac{b\xi\sxi}{8(b-aw_1)}
  -\frac1{16}(72-5\xi)y_1+\strut\nonumber\\&&\strut
  -\frac1{16}(72-5\xi+8y_2)v_2+\frac{b\xi\sxi}{8(b+aw_2)}
  +\frac1{16}(72-5\xi)y_2\Bigg\}\\[12pt]
\lefteqn{H^{4\ell}_L=N\Bigg\{\xi(10+3\xi)(t_{1-}+t_{1+})
  -\frac\xi2(24-7\xi)\ell_{6-}-13v\xi(\ell_{5-}-\ell_{8-})
  -\frac{4\lambda^2\xi}v\ell_{8-}+\strut}\nonumber\\&&\strut
  -v_\lambda\left(4\xi-\frac{8\lambda^3\xi^2}{v_\lambda^4}
    +\frac{3(1+\sxi)^2(2+\sxi)\sxi}{2(1-2\lambda+\sxi)}
    -\frac{3(1-\sxi)^2(2-\sxi)\sxi}{2(1-2\lambda-\sxi)}\right)\ell_{7-}
  +\strut\nonumber\\&&\strut
  -v_1\left(4\xi+\frac{3(1+\sxi)^2(2+\sxi)\sxi}{2(1-y_1+\sxi)}
    -\frac{3(1-\sxi)^2(2-\sxi)\sxi}{2(1-y_1-\sxi)}\right)\ell_1
  +\strut\nonumber\\&&\strut
  +v_2\left(4\xi+\frac{3(1+\sxi)^2(2+\sxi)\sxi}{2(1-y_2+\sxi)}
    -\frac{3(1-\sxi)^2(2-\sxi)\sxi}{2(1-y_2-\sxi)}\right)\ell_2
  +\strut\nonumber\\&&\strut
  +\frac{1-2\lambda-(1-\lambda)\sxi}{(1-2\lambda-\sxi)\sxi}
    \Bigg\{\,2-8\lambda+8\lambda^2-7\xi+5\lambda\xi+2\lambda^2\xi-3\xi^2
  +\strut\nonumber\\&&\qquad\qquad\qquad\qquad\qquad\strut
  -(1-6\lambda+8\lambda^2-9\xi+3\lambda\xi)\sxi\Bigg\}
  \left(\frac1{w_1}-\frac1{w_2}\right)+\strut\nonumber\\&&\strut
  -\frac{1-2\lambda+(1-\lambda)\sxi}{(1-2\lambda+\sxi)\sxi}
    \Bigg\{\,2-8\lambda+8\lambda^2-7\xi+5\lambda\xi+2\lambda^2\xi-3\xi^2
  +\strut\nonumber\\&&\qquad\qquad\qquad\qquad\qquad\strut
  +(1-6\lambda+8\lambda^2-9\xi+3\lambda\xi)\sxi\Bigg\}(w_1-w_2)
  +\strut\nonumber\\&&\strut
  +(2+7\xi)v_1+(2+7\xi)y_1-(2+7\xi)v_2+(2+7\xi)y_2\Bigg\}\\[12pt]
\lefteqn{H^{1\ell}_F=N\Bigg\{2(2-\xi)^2(t_{0-}-t_{0+})
  -(8+2\xi+\xi^2)(t_{1-}-t_{1+})+\strut}\nonumber\\[7pt]&&\qquad\strut
  +\sxi(1-\sxi)(2-\sxi)(4+\sxi)t_w+\strut\nonumber\\&&\strut
  -2(6+\xi)v+2(2-\xi)\ell_{4+}+8\xi\ell_{5+}
  -\frac12(4+6\xi-3\xi^2)\ell_{6+}+\strut\nonumber\\&&\strut
  +\left(16\lambda(1-\lambda^2)-2(1+6\lambda)\xi+\xi^2
  -\frac{8\lambda^2(1-2\lambda)^3}{v_\lambda^2}\right)\ell_{7+}
  +\strut\nonumber\\&&\strut
  -\left(2\lambda(8-3\lambda)-2(1+4\lambda+\lambda^2)\xi+\xi^2
  -\frac{2\lambda^2}{v^2}\right)\ell_{8+}+\strut\nonumber\\&&\strut
  +4(1-2\lambda)(3-2\lambda)\ell_{9+}
  +2y_1(4-3\xi-y_1)\ell_1-2y_2(4-3\xi-y_2)\ell_2+\strut\nonumber\\&&\strut
  -\frac{2(1-2\lambda-(1-\lambda)\sxi)}{(1-2\lambda-\sxi)(1-\sxi)}
    \Big(6-16\lambda+8\lambda^2+11\xi-5\lambda\xi+2\lambda^2\xi-\xi^2
    +\strut\nonumber\\&&\qquad\qquad\qquad\qquad\qquad\strut
    -(15-22\lambda+8\lambda^2+\xi+\lambda\xi)\sxi\Big)\ell_3
  +\strut\nonumber\\&&\strut
  -\frac34(4-7\xi-4y_1)v_1+\frac{b\xi\sxi}{2(b-aw_1)}
  -\frac14(16-21\xi)y_1+y_1^2+\strut\nonumber\\&&\strut
  -\frac34(4-7\xi-4y_2)v_2-\frac{b\xi\sxi}{2(b+aw_2)}
  +\frac14(16-21\xi)y_2-y_2^2\Bigg\}\\[12pt]
\lefteqn{H^{2\ell}_F=\xi N\Bigg\{2(2-\xi)(t_{0-}-t_{0+})
  -(4-\xi)(t_{1-}-t_{1+})+\sxi(1-\sxi)t_w+2v+2\ell_{4+}+\strut}
  \nonumber\\&&\strut
  -\frac32\xi\ell_{6+}+(8\lambda-\xi)(\ell_{7+}-\ell_{8+})
  +\frac{2\lambda^2\xi}{v^2}\ell_{8+}+4y_1\ell_1-4y_2\ell_2
  +\strut\nonumber\\&&\strut
  -2\frac{1-2\lambda-(1-\lambda)\sxi}{(1-\sxi)\sxi}
    (1-2\lambda-\xi+\lambda\sxi)\ell_3-5v_1-5y_1-5v_2+5y_2\Bigg\}\\[12pt]
\lefteqn{H^4_F=N\Bigg\{4v(2-\xi)(t_{0-}+t_{0+})-2(4-5\xi)(t_{1-}+t_{1+})
  -\frac12(1+\sxi)(2-\sxi)^2+\strut}\nonumber\\&&\strut
  +4v\ell_{4-}-6v\ell_{5-}-8\xi\ell_{6-}
  -\frac2{v_\lambda}(3-14\lambda+20\lambda^2-8\lambda^3-2\xi-\xi^2)
  \ell_{7-}+\strut\nonumber\\&&\strut
  +\frac2v(3-8\lambda+4\lambda^2-2\xi+8\lambda\xi-\xi^2)\ell_{8-}
  +4(1-2\lambda)(3-2\lambda)\ell_{9-}+\strut\nonumber\\[3pt]&&\strut
  -2(3-y_1)v_1\ell_1+2(3-y_2)v_2\ell_2+\strut\nonumber\\[3pt]&&\strut
  +\frac14(12+16\lambda+\xi-4y_1)v_1+\frac{b\xi\sxi}{2(b-aw_1)}
  -\frac{4\lambda\xi v_1}{y_1}+\frac14(24+\xi)y_1-y_1^2
  +\strut\nonumber\\&&\strut
  -\frac14(12+16\lambda+\xi-4y_2)v_2+\frac{b\xi\sxi}{2(b+aw_2)}
  +\frac{4\lambda\xi v_2}{y_2}+\frac14(24+\xi)y_2-y_2^2\Bigg\}
  \label{H4Fcut}\\[12pt]
\lefteqn{H^{4T}_I=\frac12\sqrt{\frac\xi2}N\Bigg\{2v(2-\xi)(t_{0-}+t_{0+})
  -\frac12(16+7\xi)(t_{1-}+t_{1+})+\strut}\nonumber\\&&\strut
  -\frac14(1+\sxi)(2-\sxi)^2+2v\ell_{4-}
  +\frac18(72-30\xi-3\xi^2)\ell_{6-}+\strut\nonumber\\&&\strut
  +v_\lambda\Bigg(\frac\xi2-\frac\xi{v_\lambda^4}(1-2\lambda-\xi)
  (1-4\lambda-\xi)+\strut\nonumber\\&&\qquad\strut
  +\frac{3(1+\sxi)^2(2+\sxi)}{2(1-2\lambda+\sxi)}
  +\frac{3(1-\sxi)^2(2-\sxi)}{2(1-2\lambda-\sxi)}\Bigg)\ell_{7-}
  +\strut\nonumber\\&&\strut
  -\frac12(12+7\xi)v(\ell_{5-}-\ell_{8-})
  -\frac1v(8\lambda-4\lambda^2-\xi-8\lambda\xi-\lambda^2\xi+\xi^2)\ell_{8-}
  +2(1-2\lambda)\ell_{9-}+\strut\nonumber\\&&\strut
  +v_1\left(\frac\xi2+\frac{3(1+\sxi)^2(2+\sxi)}{2(1-y_1+\sxi)}
  +\frac{3(1-\sxi)^2(2-\sxi)}{2(1-y_1-\sxi)}\right)\ell_1
  +\strut\nonumber\\&&\strut
  -v_2\left(\frac\xi2+\frac{3(1+\sxi)^2(2+\sxi)}{2(1-y_2+\sxi)}
  +\frac{3(1-\sxi)^2(2-\sxi)}{2(1-y_2-\sxi)}\right)\ell_2
  +\strut\nonumber\\&&\strut
  +\frac{1-2\lambda-(1-\lambda)\sxi}{(1-2\lambda-\sxi)\xi}
    \Bigg\{\,2-8\lambda+8\lambda^2-7\xi+5\lambda\xi+2\lambda^2\xi-3\xi^2
  +\strut\nonumber\\&&\qquad\qquad\qquad\qquad\qquad\strut
  -(1-6\lambda+8\lambda^2-9\xi+3\lambda\xi)\sxi\Bigg\}
  \left(\frac1{w_1}-\frac1{w_2}\right)+\strut\nonumber\\&&\strut
  +\frac{1-2\lambda+(1-\lambda)\sxi}{(1-2\lambda+\sxi)\xi}
    \Bigg\{\,2-8\lambda+8\lambda^2-7\xi+5\lambda\xi+2\lambda^2\xi-3\xi^2
  +\strut\nonumber\\&&\qquad\qquad\qquad\qquad\qquad\strut
  +(1-6\lambda+8\lambda^2-9\xi+3\lambda\xi)\sxi\Bigg\}(w_1-w_2)
  +\strut\nonumber\\&&\strut
  -\frac18(28+5\xi)v_1-\frac{2\lambda\xi v_1}{y_1}
  +\frac{b\xi\sxi}{4(b-aw_1)}-\frac18(28+5\xi)y_1+\strut\nonumber\\&&\strut
  +\frac18(28+5\xi)v_2+\frac{2\lambda\xi v_2}{y_2}
  +\frac{b\xi\sxi}{4(b+aw_2)}-\frac18(28+5\xi)y_2\Bigg\}\\[12pt]
\lefteqn{H^{1T}_A=\frac12\sqrt{\frac\xi2}N\Bigg\{2(2-\xi)(t_{0-}-t_{0+})
  -\frac12(16-3\xi)(t_{1-}-t_{1+})+\strut}\nonumber\\&&\strut
  -\frac12(1-\sxi)(2-\sxi)(4+\sxi)t_w-\frac12(16-3\xi)v
  +\strut\nonumber\\&&\strut
  +2\ell_{4+}+2(7-\xi)\ell_{5+}-\frac18(8-6\xi+3\xi^2)\ell_{6+}
  +2(1-2\lambda)\ell_{9+}+\frac\xi2y_1\ell_1-\frac\xi2y_2\ell_2
  +\strut\nonumber\\&&\strut
  -\xi\left(1-\lambda-\frac{4\lambda^2}{v_\lambda^2}\right)\ell_{7+}
  -\left(4\lambda(2-\lambda)-\xi+\frac{\lambda^2\xi}v\right)\ell_{8+}
  +\strut\nonumber\\&&\strut
  +\frac{1-2\lambda-(1-\lambda)\sxi}{(1-\sxi)\sxi}\Big(6-4\lambda
  -(9-\xi)\sxi+(2-\lambda)\xi-\frac{8\lambda^2\sxi}{1-2\lambda-\sxi}\Big)
  \ell_3+\strut\nonumber\\&&\strut
  -\frac18(4+5\xi)v_1+\frac{b\xi\sxi}{4(b-aw_1)}
  -\frac18(4+5\xi)y_1+\strut\nonumber\\&&\strut
  -\frac18(4+5\xi)v_2-\frac{b\xi\sxi}{4(b+aw_2)}+\frac18(4+5\xi)y_2
  \Bigg\}\\[12pt]
\lefteqn{H^{2T}_A=\frac12\sqrt{\frac\xi2}N\Bigg\{2(2-\xi)(t_{0-}-t_{0+})
  -\frac12(8-3\xi)(t_{1-}-t_{1+})-\frac\xi2(1-\sxi)t_w+\strut}
  \nonumber\\&&\strut
  +\frac32(4-\xi)v+2\ell_{4+}+2(1+\xi)\ell_{5+}+\frac18(2-\xi)(4-3\xi)\ell_{6+}
  +(8\lambda-\xi-\lambda\xi)\ell_{7+}+\strut\nonumber\\&&\strut
  -\left(4\lambda(2+\lambda)-\xi+\frac{\lambda^2\xi}v\right)\ell_{8+}
  -2(1-2\lambda)\ell_{9+}+\strut\nonumber\\&&\strut
  +\frac12(8-\xi)y_1\ell_1-\frac12(8-\xi)y_2\ell_2
  +\frac{1-2\lambda-(1-\lambda)\sxi}{1-\sxi}(1-2\lambda+\lambda\sxi-\xi)
  \ell_3+\strut\nonumber\\&&\strut
  -\frac18(52-5\xi)v_1-\frac{b\xi\sxi}{4(b-aw_1)}
  -\frac18(52-5\xi)y_1+\strut\nonumber\\&&\strut
  -\frac18(52-5\xi)v_2+\frac{b\xi\sxi}{4(b+aw_2)}
  +\frac18(52-5\xi)y_2\Bigg\}\\[12pt]
\lefteqn{H^{3N}_A=\frac12\sqrt{\frac\xi2}N\Bigg\{2v(2-\xi)(t_{0-}+t_{0+})
  -\frac12(8-13\xi)(t_{1-}+t_{1+})+\strut}\nonumber\\&&\strut  
  +\frac14(2-\sxi)^2(1+\sxi)+2v\ell_{4-}-\frac12(8+\xi)v\ell_{5-}
  +\frac18(8-30\xi+3\xi^2)\ell_{6-}+\strut\nonumber\\&&\strut
  -\left(\frac{8+\xi}2v_\lambda-\xi\frac{1-2\lambda-\xi}{v_\lambda}\right)
  \ell_{7-}+v\left(4(1-\lambda)^2+\frac32\xi-\frac{3\lambda^2\xi}{v^2}\right)
  \ell_{8-}+\strut\nonumber\\&&\strut
  -2(1-2\lambda)\ell_{9-}
  -\frac12(8+\xi)v_1\ell_1+\frac12(8+\xi)v_2\ell_2+\strut\nonumber\\&&\strut
  +\frac18(52+5\xi)v_1-\frac{2\lambda\xi v_1}{y_1}
  -\frac{b\xi\sxi}{4(b-aw_1)}+\frac18(52+5\xi)y_1+\strut\nonumber\\&&\strut
  -\frac18(52+5\xi)v_2+\frac{2\lambda\xi v_2}{y_2}
  -\frac{b\xi\sxi}{4(b+aw_2)}+\frac18(52+5\xi)y_2\Bigg\}
\end{eqnarray}
The logarithmic rate terms $\ell_i$ and the double and dilogarithmic rate
terms $t_{0-}$, $t_{0+}$, $t_{1-}$, $t_{1+}$, and $t_w$ are listed in
Appendix~C. Note the exact $O(\as)$ tree graph relation $H_F^{2\ell}=H_U^2$
which was also noticed in~\cite{Ravindran:2000rz}. We have not been able to
derive this relation from general principles.

We shall not dwell on the technical details of how the finite integrals have
been calculated but rather concentrate on the class of IR-divergent integrals.
For instance, the integral
\begin{equation}
\tilde I_z(-1,-1)=\int_0^{y_1}\int_{z_-}^{z_+}\frac{dy\,dz}{yz}
  =\int_0^{y_1}\ln\pfrac{z_+(y)}{z_-(y)}\frac{dy}y
\end{equation}
is IR-divergent and will be regularized by a gluon mass
$m_G=\sqrt{\Lambda q^2}$. The introduction of a gluon mass changes the lower
$y$ limit from $0$ to $y_-=\Lambda+\sqrt{\Lambda\xi}$, and the $z$ limits to
\begin{equation}
z_\pm(y)=\frac1{4y+\xi}\left(2y-2y^2-\xi y+2\Lambda y+2\Lambda
  \pm2\sqrt{(y-\Lambda)^2-\Lambda\xi}\sqrt{(1-y)^2-\xi}\right).
\end{equation}
Therefore, the integration over $z$ gives rise to
\begin{equation}
\tilde I_z(-1,-1)=\int_{y_-}^{y_1}
  \ln\pfrac{2y-2y^2-\xi y+2\Lambda y+2\Lambda+2\sqrt{(y-\Lambda)^2-\Lambda\xi}
  \sqrt{(1-y)^2-\xi}}{2y-2y^2-\xi y+2\Lambda y+2\Lambda
  -2\sqrt{(y-\Lambda)^2-\Lambda\xi}\sqrt{(1-y)^2-\xi}}\frac{dy}y.
\end{equation}
This integral is not analytically calculable for general values of $\Lambda$.
However, we can divide the integral into a divergent and a convergent part
which are separately calculable as long as $\Lambda$ is a small parameter. The
residue of the divergent part should coincide with the residue of the original
integrand at the IR-singular pole at $y=0$. A simplified IR-divergent part can
be constructed from the full integrand by neglecting higher powers in $y$
whenever possible. Before this approximation we shift the integration by
$-\Lambda$ in order to facilitate the expansion around the lower boundary. We
obtain
\begin{equation}
\tilde I_z^D(-1,-1)=\int_{\sqrt{\Lambda\xi}}^{y_1}
  \ln\pfrac{(1+v^2)y+2v\sqrt{y^2-\Lambda\xi}}{(1+v^2)y
  -2v\sqrt{y^2-\Lambda\xi}}\frac{dy}y.
\end{equation}
This integral can be calculated analytically and one obtains
\begin{eqnarray}\label{divpart}
\lefteqn{\tilde I_z^D(-1,-1)\ =\ \ln\pfrac{1+v}{1-v}
  \ln\pfrac{y_1^2}{\Lambda\xi}
  -\Li_2\pfrac{2v}{(1+v)^2}+\Li_2\pfrac{-2v}{(1-v)^2}\,+}\nonumber\\&&
  -\frac12\Li_2\left(-\frac{(1+v)^2}{(1-v)^2}\right)
  +\frac12\Li_2\left(-\frac{(1-v)^2}{(1+v)^2}\right)
  \ =:\ t_p-\ln\pfrac{1+v}{1-v}\ln\Lambda.\qquad\quad
\end{eqnarray}
In the case $\Lambda\rightarrow 0$ we have the limiting value (we write
$\eps=\sqrt{\Lambda \xi}$)
\begin{equation}
\tilde I_z^D(-1,-1)\rightarrow 2\ln\pfrac{1+v}{1-v}
  \lim_{\eps\rightarrow0}\int_\eps^{y_1}\frac{dy}y
\end{equation}
which is an ill-defined quantity for $\eps=0$. However, we can subtract the
singular piece from the original integral also taken in the limit
$\Lambda\rightarrow 0$. As a result the divergences cancel and one obtains
\begin{equation}
\tilde I_z^C(-1,-1)=\lim_{\eps\rightarrow 0}
  \left\{\int_\eps^{y_1}\ln\pfrac{2-2y-\xi+2\sqrt{(1-y)^2-\xi}}{2-2y-\xi
  -2\sqrt{(1-y)^2-\xi}}\frac{dy}y-2\ln\pfrac{1+v}{1-v}
  \int_\eps^{y_1}\frac{dy}y\right\}
\end{equation}
or symbolically
\begin{equation}\label{conpart}
\tilde I_z^C(-1,-1)=\lim_{\eps\rightarrow 0}
  \left\{\hat I_z^{ba\prime}(-1)-2\ln\pfrac{1+v}{1-v}\hat I'_z(-1)\right\}
\end{equation}
where the primes indicates that the lower limit is $\eps$. With the Euler
substitution Eq.~(\ref{Eulersub}), and after partial fractioning according to
\begin{equation}
\frac{dy}y=-\frac{dw}{w_0-w}+\frac{dw}{w_0+w}+\frac{dw}{1-w}-\frac{dw}{1+w}
\end{equation}
one obtains
\begin{eqnarray}
\hat I_z^{ba\prime}(-1)&=&\int_\eps^{y_1}\ln\pfrac{2-2y-\xi
  +2\sqrt{(1-y)^2-\xi}}{2-2y-\xi-2\sqrt{(1-y)^2-\xi}}\frac{dy}y\ =\nonumber\\
  &=&I^{ba}_{0-}(w'_0)-I^{ba}_{0-}(w_1)-I^{ba}_{0+}(w'_0)+I^{ba}_{0+}(w_1)
  +\strut\nonumber\\[7pt]&&
  -I^{ba}_{1-}(w'_0)+I^{ba}_{1-}(w_1)+I^{ba}_{1+}(w'_0)-I^{ba}_{1+}(w_1),
  \label{Izbaprime}\\[7pt]
\hat I'_z(-1)\ =\ \int_\eps^{y_1}\frac{dy}y
  &=&I_{0-}(w'_0)-I_{0-}(w_1)-I_{0+}(w'_0)+I_{0+}(w_1)\,+\nonumber\\&&
  -I_{1-}(w'_0)+I_{1-}(w_1)+I_{1+}(w'_0)-I_{1+}(w_1)\label{Izprime}\qquad\quad
\end{eqnarray}
where
\begin{equation}
w'_0=\sqrt{\frac{1-\eps-\sxi}{1-\eps+\sxi}} =
  w_0\left(1-\frac{\eps\sxi}{1-\xi}+\ldots\right).
\end{equation}
The variable $w'_0$ tends to $w_0$ for $\eps\to 0$. It is instructive to note 
that the divergences now reside in the terms $I^{ba}_{0-}(w'_0)$ and
$I_{0-}(w'_0)$ which contain the integrand factor $(w_0-w)^{-1}$. We obtain
\begin{equation}\label{Iba0m}
I^{ba}_{0-}(w)=t^l_p(w)-2\ln\pfrac{1+v}{1-v}\ln(w_0-w),\qquad
I_{0-}(w)=-\ln(w_0-w)
\end{equation}
where $t^l_p$ is a decay rate term which vanishes in the limit $w\to w_0$. For
this reason the two expressions in Eq.~(\ref{Iba0m}) do not contribute to the
convergent part at all. Using Eqs.~(\ref{Izbaprime}) and~(\ref{Izprime}) we can
calculate the convergent part in Eq.~(\ref{conpart}) and add the divergent part
in Eq.~(\ref{divpart}) to obtain
\begin{eqnarray}\label{tilde}
\tilde I_z(-1,-1)&=&t_p-\ln\pfrac{1+v}{1-v}\ln\Lambda
  -\left\{I^{ba}_{0-}(w_1)-2\ln\pfrac{1+v}{1-v}I_{0-}(w_1)\right\}
  +\ldots\ =\nonumber\\
  &=&t^{ba}_{0-}(w_0)-t^{ba}_{0-}(w_1)-t^{ba}_{0+}(w_0)+t^{ba}_{0+}(w_1)
  +\ldots-\ln\pfrac{1+v}{1-v}\ln\Lambda.\qquad\quad
\end{eqnarray}
The dots indicate further contributions according to Eqs.~(\ref{Izbaprime})
and~(\ref{Izprime}) where we can replace $w'_0$ by $w_0$. The decay rate terms
$t^{ba}_{0-}(w)$, $t_{0+}^{ba}(w)$, \dots\ are listed in Appendix~C. It is
obvious that $t^{ba}_{0-}(w_0)$ in Eq.~(\ref{tilde}) has to be replaced by the
special value $t_p$ defined in Eq.~(\ref{divpart}).

We now turn to case~B when $\lambda>\lambda_{\rm trans}$. As discussed in the
beginning of this section this entails the calculation of the second integral
in Eq.~(\ref{caseB}) which has to be added to the first and the third integral
in Eq.~(\ref{caseB}). The latter two integrals are already known from case~A.
Using some additional decay rate terms listed in Appendix C the results for
this additional phase-space portion are given by
\begin{eqnarray}
\lefteqn{H^1_U\ =\ N\Bigg\{2(2-\xi)^2(t_{0-}^c-t_{0+}^c)-(8-10\xi-\xi^2)
  (t_{1-}^c-t_{1+}^c)+\strut}\nonumber\\&&\strut
  +\sxi(1-\sxi)(2+4\sxi-3\xi)t_w^c-8(2-\xi)v\ell_{4+}^c-16\xi\ell_{5+}^c
  +\strut\nonumber\\&&\strut
  +\left(\frac14(8+12\xi-\xi^2)+8y_2-2y_2^2\right)\ell_2^c
  +\frac12(24+5\xi)v_2-\frac{\xi(4-\xi)v_2}{2(4y_2+\xi)}\Bigg\},\\
\lefteqn{H^2_U\ =\ \xi N\Bigg\{2(2-\xi)(t_{0-}^c-t_{0+}^c)-(4-\xi)
  (t_{1-}^c-t_{1+}^c)+\strut}\nonumber\\&&\strut
  +\sxi(1-\sxi)t_w^c-8v\ell_{4+}^c+\left(\frac32\xi+4y_2\right)\ell_2^c+10v_2
  \Bigg\},\\
\lefteqn{H^{4\ell}_U\ =\ N\Bigg\{4(2-\xi)v(t_{0-}^c+t_{0+}^c)-(8+2\xi+3\xi^2)
  (t_{1-}^c+t_{1+}^c)-8(1-\xi)\ell_{4-}^c+\strut}\nonumber\\&&\strut
  +16(1-\xi)\ell_{5-}^c-\frac\xi4(28-17\xi)(2\ell_{5-}^c+\ell_{6-}^c)
  -16(1-\xi)\ell_{7-}^c+\strut\nonumber\\&&\strut
  +v_2\left(2\left(y_2-3(1-\xi)\right)
  +\frac{3(1+\sxi)^2(2+\sxi)\sxi}{2(1-y_2+\sxi)}
  -\frac{3(1-\sxi)^2(2-\sxi)\sxi}{2(1-y_2-\sxi)}\right)\ell_2^c
  +\strut\nonumber\\&&\strut
  +\frac18(64+64\sxi-352\xi+232\xi\sxi+\xi^2)-\frac32(8-11\xi)y_2+4y_2^2
  -\frac{\xi(4-\xi)^2}{8(4y_2+\xi)}\Bigg\},\\[12pt]
\lefteqn{H^1_L\ =\ N\Bigg\{\xi(2-\xi)(t_{0-}^c-t_{0+}^c)-2\xi(2+\xi)
  (t_{1-}^c-t_{1+}^c)-\sxi(2+2\sxi-7\xi+3\xi\sxi)t_w^c+\strut}
  \nonumber\\&&\strut
  -4\xi v\ell_{4+}^c+16\xi\ell_{5+}^c-\xi\left(\frac1{16}(8+8\xi-3\xi^2)
  +2y_2+\frac12y_2^2\right)\ell_2^c+\strut\nonumber\\&&\strut
  +\frac18(32-72\xi+5\xi^2-8\xi y_2)v_2
  -\frac{\xi^2(4-\xi)v_2}{8(4y_2+\xi)}\Bigg\},\\
\lefteqn{H^2_L\ =\ \xi N\Bigg\{(2-\xi)(t_{0-}^c-t_{0+}^c)
  -2(1-\xi)(t_{1-}^c-t_{1+}^c)-\sxi(1-\sxi)t_w^c-4v\ell_{4+}^c+\strut}
  \nonumber\\&&\strut
  -\left(\frac1{16}(8-8\xi+3\xi^2)-2y_2-\frac12y_2^2\right)\ell_2^c
  +\frac18(72-5\xi+8y_2)v_2+\frac{\xi(4-\xi)v_2}{8(4y_2+\xi)}\Bigg\},\\
\lefteqn{H^{4\ell}_L\ =\ N\Bigg\{\xi(10+3\xi)(t_{1-}^c+t_{1+}^c)
  +\xi(24-7\xi)\ell_{5-}^c+\frac12\xi(24-7\xi)\ell_{6-}^c+\strut}
  \nonumber\\&&\strut
  +\left(\frac{3\sxi}{2w_2}(2-\sxi)(1-\sxi)^2
  -\frac{3\sxi w_2}2(2+\sxi)(1+\sxi)^2-4\xi v_2\right)\ell_2^c
  +\strut\nonumber\\&&\strut
  +2(1-\sxi)(2-6\sxi+13\xi)-2(2+7\xi)y_2\Bigg\},\\[12pt]
\lefteqn{H^{1\ell}_F\ =\ N\Bigg\{2(2-\xi)^2(t_{0-}^c-t_{0+}^c)
  -(8+2\xi+\xi^2)(t_{1-}^c-t_{1+}^c)+\strut}\nonumber\\&&\strut
  +\sxi(8-10\sxi+\xi+\xi\sxi)t_w^c-8(2-\xi)v\ell_{4+}^c+8\xi\ell_{5+}^c+\strut
  \nonumber\\&&\strut
  +\left(\frac12(4+6\xi-3\xi^2)+2(4-3\xi)y_2-2y_2^2\right)\ell_2^c
  +\frac32(4-7\xi-4y_2)v_2-\frac{\xi(4-\xi)v_2}{2(4y_2+\xi)}\Bigg\},\qquad\\
\lefteqn{H^{2\ell}_F\ =\ \xi N\Bigg\{2(2-\xi)(t_{0-}^c-t_{0+}^c)
  -(4-\xi)(t_{1-}^c-t_{1+}^c)+\strut}\nonumber\\&&\strut
  +\sxi(1-\sxi)t_w^c-8v\ell_{4+}+\left(\frac32\xi+4y_2\right)\ell_2^c
  +10v_2\Bigg\},\\
\lefteqn{H^4_F\ =\ N\Bigg\{4(2-\xi)v(t_{0-}^c+t_{0+}^c)
  -2(4-5\xi)(t_{1-}^c+t_{1+}^c)-8(1-\xi)\ell_{4-}^c+\strut}\nonumber\\&&\strut
  +16\ell_{5-}^c+8\xi\ell_{6-}^c-16(1-\xi)\ell_{7-}^c-2(3-y_2)v_2\ell_2^c
  +\strut\nonumber\\&&\strut
  +\frac18(80-64\sxi-8\xi+\xi^2)-\frac12(24+\xi)y_2+2y_2^2
  -\frac{\xi(4-\xi)^2}{8(4y_2+\xi)}\Bigg\},\label{H4Fadd}\\[12pt]
\lefteqn{H^{4T}_I\ =\ \frac12\sqrt{\frac\xi2}N
  \Bigg\{2(2-\xi)v(t_{0-}^c+t_{0+}^c)
  -\frac12(16+7\xi)(t_{1-}^c+t_{1+}^c)-4(1-\xi)\ell_{4-}^c+\strut}
  \nonumber\\&&\strut
  -\frac14(4-\xi)(10+3\xi)\ell_{5-}^c-\frac18(72-30\xi-3\xi^2)\ell_{6-}^c
  +\strut\nonumber\\&&\strut
  -8(1-\xi)\ell_{7-}^c+\left(\frac{\xi v_2}2+\frac3{2w_2}(2-\sxi)(1-\sxi)^2
  +\frac{3w_2}2(2+\sxi)(1+\sxi)^2\right)\ell_2^c+\strut\nonumber\\&&\strut
  -\frac1{16}(304-496\sxi+208\xi-24\xi\sxi-\xi^2)+\frac14(28+5\xi)y_2
  -\frac{\xi(4-\xi)^2}{16(4y_2+\xi)}\Bigg\},\\[12pt]
\lefteqn{H^{1T}_A\ =\ \frac12\sqrt{\frac\xi2}N
  \Bigg\{2(2-\xi)(t_{0-}^c-t_{0+}^c)
  -\frac12(16-3\xi)(t_{1-}^c-t_{1+}^c)+\strut}\nonumber\\&&\strut
  -\frac12(8-10\sxi+\xi+\xi\sxi)t_w^c-8v\ell_{4+}^c
  +2(7-\xi)\ell_{5+}^c+\strut\nonumber\\&&\strut
  +\left(\frac18(8-6\xi+3\xi^2)+\frac12\xi y_2\right)\ell_2^c
  +\frac14(4+5\xi)v_2-\frac{\xi(4-\xi)v_2}{4(4y_2+\xi)}\Bigg\},\\
\lefteqn{H^{2T}_A\ =\ \frac12\sqrt{\frac\xi2}N
  \Bigg\{2(2-\xi)(t_{0-}^c-t_{0+}^c)-\frac12(8-3\xi)(t_{1-}^c-t_{1+}^c)
  -\frac12\xi(1-\sxi)t_w^c-8v\ell_{4+}^c+\strut}\nonumber\\&&\strut
  +2(1+\xi)\ell_{5+}^c-\left(\frac18(2-\xi)(4-3\xi)
  -\frac12(8-\xi)y_2\right)\ell_2^c+\frac14(52-5\xi)v_2
  +\frac{\xi(4-\xi)v_2}{4(4y_2+\xi)}\Bigg\},\nonumber\\\\
\lefteqn{H^{3N}_A\ =\ \frac12\sqrt{\frac\xi2}N
  \Bigg\{2(2-\xi)v(t_{0-}^c+t_{0+}^c)
  -\frac12(8-13\xi)(t_{1-}^c+t_{1+}^c)-4(1-\xi)\ell_{4-}^c+\strut}
  \nonumber\\&&\strut
  +\frac14(24-2\xi-3\xi^2)\ell_{5-}^c-\frac18(8-30\xi+3\xi^2)\ell_{6-}^c
  -8(1-\xi)\ell_{7-}^c-\frac12(8+\xi)v_2\ell_2^c+\strut\nonumber\\&&\strut
  +\frac1{16}(208-208\sxi+16\xi-24\xi\sxi-\xi^2)-\frac14(52+5\xi)y_2
  +\frac{\xi(4-\xi)^2}{4(4y_2+\xi)}\Bigg\}. 
\end{eqnarray}
Note again that one has the remarkable relation $H_F^{2\ell}=H_U^2$. We
mention that, differing from Ref.~\cite{Stav:1996ep}, we have been able to
obtain a closed form result for the cut-dependent structure function $H^4_F$
(see Eqs.~(\ref{H4Fcut}) and~(\ref{H4Fadd})).

Numerically, the contribution of the second integral in Eq.~(\ref{caseB})
calculated above is quite small. This is because the relevant integration
region is far away from the IR region where the rate is largest. Nevertheless,
this contribution is needed if one wants to check on the consistency of our
case~B result with the fully integrated results in Refs.~\cite{Korner:1993dy,%
Groote:1995yc,Groote:1995ky,Groote:1996nc,Ravindran:2000rz}. In fact, we have
performed an explicit check that for each of the unpolarized and polarized
rate functions the sum of the three integrals in Eq.~(\ref{caseB}) reproduce
the full phase-space result calculated previously in
Refs.~\cite{Korner:1993dy,Groote:1995yc,Groote:1995ky,Groote:1996nc,%
Ravindran:2000rz} when the gluon energy cut is set to its maximal value
$\lambda_{\rm max}=(1-\xi)/2$ (which corresponds to setting $y_1$ and $y_2$ to
$(1-\xi)/2$ in Eq.~(\ref{caseB})). We have also checked that our exact result
converges to the soft-gluon expression to be derived in Sec.~6 when
$\lambda\rightarrow 0$.

\section{Fully integrated $O(\as)$ results}
The cut-off dependent helicity structure functions calculated in the previous
section must coincide with the fully integrated results written down in
Refs.~\cite{Groote:1995yc,Groote:1995ky,Groote:1996nc} when the cut-off is
taken to its maximal value. For the convenience of the reader we collect the
fully integrated results of
\cite{Groote:1995yc,Groote:1995ky,Groote:1996nc} and list them in terms of the
sum $H_a^{j(m)}(\as)=H_a^{j(m)}({\it tree\/})+H_a^{j(m)}({\it loop\/})$.
As before we define ($N=\as N_cC_Fq^2/(4\pi v)$). One has
\begin{eqnarray}
H_U^1(\as)&=&N\bigg\{(2+7\xi)v+\frac12(48-48\xi+7\xi^2)t_3
  +\sxi(2-7\xi)t_4+\strut\nonumber\\&&\strut
  +\xi(2+3\xi)(t_4-t_5)-2(2-\xi)\left((2-\xi)(t_8-t_9)+2v(t_{10}+2t_{12})
  \right)
  \bigg\},\nonumber\\
H_U^2(\as)&=&\xi N\bigg\{6v+(6-\xi)t_3+\sxi t_4-\xi(t_4-t_5)
  +\strut\nonumber\\&&\strut
  -2\left((2-\xi)(t_8-t_9)+2v(t_{10}+2t_{12})\right)\bigg\},\nonumber\\
H_L^1(\as)&=&\!\!N\bigg\{\frac14(16-46\xi+3\xi^2)v
  +\frac\xi8(88-32\xi+3\xi^2)t_3-\sxi(2-7\xi)t_4+\strut\nonumber\\&&\strut
  -\xi(2+3\xi)(t_4-t_5)-\xi\left((2-\xi)(t_8-t_9)+2v(t_{10}+2t_{12})\right)
  \bigg\},\nonumber\\
H_L^2(\as)&=&\xi N\bigg\{\frac34(10-\xi)v+\frac18(24-16\xi-3\xi^2)t_3
  -\sxi t_4+\strut\nonumber\\&&\strut+\xi(t_4-t_5)
  -\left((2-\xi)(t_8-t_9)+2v(t_{10}+2t_{12})\right)\bigg\},\nonumber\\
H_F^3(\as)&=&-4\xi N v\pi,\nonumber\\[3pt]
H_F^4(\as)&=&N\bigg\{-8\sxi(1-\sxi)-8(t_1-t_2)+4(2-3\xi)vt_3
  +\strut\nonumber\\&&\strut-2(4-5\xi)t_6
  -4v\left((2-\xi)(t_8-t_7)+2v(t_{10}+t_{11})\right)\bigg\},\\[7pt]
H_U^{3\ell}(\as)&=&-4\xi Nv\pi,\nonumber\\[3pt]
H_U^{4\ell}(\as)&=&N\bigg\{-(2+35\xi)+\sxi(8+29\xi)
  -\frac14(32-60\xi+17\xi^2)(t_1-t_2)+\strut\nonumber\\&&\strut
  +2(4+9\xi)vt_3-(8+2\xi+3\xi^2)t_6
  -4v\left((2-\xi)(t_8-t_7)+v(t_{10}+t_{11})\right)\bigg\},\nonumber\\
H_L^{3\ell}(\as)&=&0,\nonumber\\[3pt]
H_L^{4\ell}(\as)&=&N\bigg\{2(2+19\xi)-2\sxi(8+13\xi)
  +\strut\nonumber\\&&\strut
  -\frac12\xi(24-7\xi)(t_1-t_2)-26\xi vt_3+\xi(10+3\xi)t_6\bigg\},\nonumber\\
H_F^{1\ell}(\as)&=&N\bigg\{-2(2+3\xi)v+(24-12\xi+\xi^2)t_3
  +\sxi(8+\xi)t_4+\strut\nonumber\\&&\strut-\xi(10-\xi)(t_4-t_5)
  -2(2-\xi)\left((2-\xi)(t_8-t_9)+2v(t_{10}+2t_{12})\right)\bigg\},\nonumber\\
H_F^{2\ell}(\as)&=&\xi N\bigg\{6v+(6-\xi)t_3+\sxi t_4
  -\xi(t_4-t_5)+\strut\nonumber\\&&\strut
  -2\left((2-\xi)(t_8-t_9)+2v(t_{10}+2t_{12})\right)\bigg\},\\[7pt]
H_I^{3T}(\as)&=&-\frac12\sqrt{\frac\xi2}N(1+\xi)v\pi,\nonumber\\
H_I^{4T}(\as)&=&-\frac14\sqrt{\frac\xi2}N\bigg\{48+17\xi-\sxi(62+3\xi)
  -\frac14(4-\xi)(10+3\xi)(t_1-t_2)+\strut\nonumber\\&&\strut
  -2(21+2\xi)vt_3+(16+7\xi)t_6
  +4v\left((2-\xi)(t_8-t_7)+2v(t_{10}+t_{11})\right)\bigg\},\nonumber\\
H_A^{1T}(\as)&=&-\frac14\sqrt{\frac\xi2}N\bigg\{(8-3\xi)v
  -\frac12(72-38\xi+3\xi^2)t_3-\sxi(10-\xi)t_4+\strut\nonumber\\&&\strut
  +(8+\xi)(t_4-t_5)+4\left((2-\xi)(t_8-t_9)+2v(t_{10}+2t_{12})\right)\bigg\},
  \nonumber\\
H_A^{2T}(\as)&=&-\frac14\sqrt{\frac\xi2}N\bigg\{-(20-3\xi)v
  -\frac12(32-14\xi-3\xi^2)t_3-\xi\sxi t_4+\strut\nonumber\\&&\strut
  +\xi(t_4-t_5)+4\left((2-\xi)(t_8-t_9)+2v(t_{10}+2t_{12})\right)\bigg\},
  \\[7pt]
H_I^{1N}(\as)&=&\frac12\sqrt{\frac\xi2}Nv^2\pi
  \ =\ H_I^{2N}(\as),\nonumber\\
H_A^{3N}(\as)&=&\frac14\sqrt{\frac\xi2}N\bigg\{20+9\xi-\sxi(26+3\xi)
  -\frac14(24-2\xi-3\xi^2)(t_1-t_2)+\strut\nonumber\\&&\strut
  +2(1-6\xi)vt_3-(8-13\xi)t_6-4v\left((2-\xi)(t_8-t_7)+2v(t_{10}+t_{11})\right)
  \bigg\},\nonumber\\
H_A^{4N}(\as)&=&\frac12\sqrt{\frac\xi2}N(1+\xi)v\pi.
\end{eqnarray}
The fully integrated $O(\as)$ results are given in terms of the rate
functions $t_1$ to $t_{12}$ which are listed in Appendix~B. It is clear that 
one again has the relation $H_F^{2\ell}(\as)=H_U^2(\as)$ because
both loop and tree contributions satisfy this identity.
 
\section{The soft-gluon approximation}
The basic ingredient of the soft-gluon approximation (SGA) for the tree graph
matrix elements is the eikonal approximation where the gluon momentum is
neglected in the numerators of Feynman diagram contributions. In the eikonal
approximation the hadron tensor is proportional to the Born term. In the
present case one has
\begin{equation}\label{eikonal}
H_{\mu\nu}^i({\it soft\/})=g_s^2C_F\left(\frac{p_1^2}{(p_1p_3)^2}
  -\frac{2(p_1p_2)}{(p_1p_3)(p_2p_3)}+\frac{p_2^2}{(p_2p_3)^2}\right)
  H_{\mu\nu}^i({\it Born\/})
\end{equation}
where $H_{\mu\nu}^i({\it Born\/})$ refers to the Born term tensor in the
two-body case where $q=p_1+p_2$. On the other hand, the eikonal factor
multiplying $H_{\mu\nu}^i({\it Born\/})$ refers to the three-body case where
$q=p_1+p_2+p_3$ and depends on the dimensionless three-body phase-space
variables $x=E_G/\sqrt{q^2}=p_3q/q^2$ and $u=(p_1-p_2)q/q^2$. When integrating
$H_{\mu\nu}^i({\it soft\/})$ over the three-body phase-space the Born term
contribution $H_{\mu\nu}^i({\it Born\/})$ can be taken outside of the
integral. In this sense the integration on the soft-gluon factor in
Eq.~(\ref{eikonal}) is universal in the sense that it is process and
polarization independent. 

When projecting the eikonal contribution in Eq.~(\ref{eikonal}) onto the
various helicity structure functions one recovers the various Born term
contributions $H_a^i({\it Born\/})$ listed in Sec.~3. Referring to the
integration measure in Eq.~(\ref{integmeasure}) and using $dy\,dz=2dx\,du$ one
obtains
\begin{equation}
\frac{2q^2}{16\pi^2v}\int_{\sqrt\Lambda}^\lambda
  \int_{-u_+}^{u_+}H_{\mu\nu}^i({\it soft\/})dx\,du
  =H_{\mu\nu}^i({\it Born\/})\frac{\alpha_sC_F}{4\pi v}
  \int_{\sqrt\Lambda}^\lambda\int_{-u_+}^{u_+}h(x,u)dx\,du
\end{equation}
where
\begin{equation}
h(x,u)=8\frac{(1-2x+\Lambda)(u^2-(x-\Lambda)^2)
  +\xi(x-\Lambda)^2}{(u^2-(x-\Lambda)^2)^2}.
\end{equation}
The limits of the $u$ integration are given by $\pm u_+$ where
\begin{equation}
u_+(x)=\left((x^2-\Lambda)\frac{1-2x+\Lambda-\xi}{1-2x+\Lambda}\right)^{1/2}
\end{equation}
After integration over $u$ one obtains 
\begin{equation}
h(x)=-4\left(\frac{-2u_+\xi}{(x-\Lambda)^2-u_+^2}
  +\frac{2-4x-2\Lambda-\xi}{x-\Lambda}\ln\pfrac{x-\Lambda+u_+}{x-\Lambda-u_+}
  \right).
\end{equation}
Further integrating over the scaled gluon energy $x$ from $\Lambda$ to 
$\lambda$ one finally has
\begin{eqnarray}\label{sga1}
h_{\rm eik}\!\!\!&=&\!\!\!-\frac{\alpha_sC_F}{\pi v}\Bigg\{\left(2v-(2-\xi)
  \ln\pfrac{1+v}{1-v}\right)\ln\pfrac{2\lambda}{\sqrt\Lambda}
  +4\left(\sqrt{\strut1-2\lambda}\sqrt{1-2\lambda-\xi}
  -v\right)\,+\nonumber\\&&\qquad
  +2v\left(\ln\pfrac{z_\lambda}{z_0}+2\ln\pfrac{z_0^2-1}{z_\lambda z_0-1}
  \right)-\ln z_0+4\lambda\ln z_\lambda\,+\nonumber\\&&\qquad
  +(2-\xi)\Bigg(\frac12\ln^2\pfrac{z_\lambda}{z_0}+2\ln z_0
  \ln\pfrac{z_\lambda z_0-1}{z_0^2-1}+\frac14\ln^2z_0
  \,+\nonumber\\&&\qquad\qquad
  +\Li_2\pfrac{2v}{1+v}+\Li_2\left(1-\frac{z_\lambda}{z_0}\right)
  +\Li_2(1-z_\lambda z_0)-\Li_2(1-z_0^2)\Bigg)\Bigg\}
\end{eqnarray}
where
\begin{equation}
z_0=\frac{1+v}{1-v},\qquad z_\lambda=\frac{\sqrt{1-2\lambda}
  +\sqrt{1-2\lambda-\xi}}{\sqrt{1-2\lambda}-\sqrt{1-2\lambda-\xi}}.
\end{equation}
The function $h_{\rm eik}$ will be referred to as the eikonal form of the
SGA factor.

For $\lambda\rightarrow 0$ one obtains 
\begin{eqnarray}\label{limfac}
h_{\rm SGA}&=&-\frac{\alpha_sC_F}{\pi v}\Bigg\{\left(2v-(2-\xi)
  \ln\pfrac{1+v}{1-v}\right)\ln\pfrac{2\lambda}{\sqrt\Lambda}
  +\strut\nonumber\\&&\qquad
  -\ln\frac{1+v}{1-v}+(2-\xi)\left(\frac14\ln^2\pfrac{1+v}{1-v}
  +\Li_2\pfrac{2v}{1+v}\right)\Bigg\}.
\end{eqnarray}
Following the literature \cite{Akatsu:1997tq,Kodaira:1998gt} we shall refer to
the SGA factor (\ref{limfac}) as the soft-gluon approximation of
Eq.~(\ref{sga1}).

In addition to the check on our case~A results discussed in Sec.~4 we have
performed a second and independent check by taking the $\lambda\rightarrow 0$
limit in the relevant exact expressions in Sec.~4. In this limit the exact
result can be seen to factor into a Born term contribution times the
soft-gluon factor given in Eq.~(\ref{limfac}). This proves that the exact
results given in Sec.~4 have the correct soft-gluon limiting behaviour.

\begin{figure}\begin{center}
\epsfig{figure=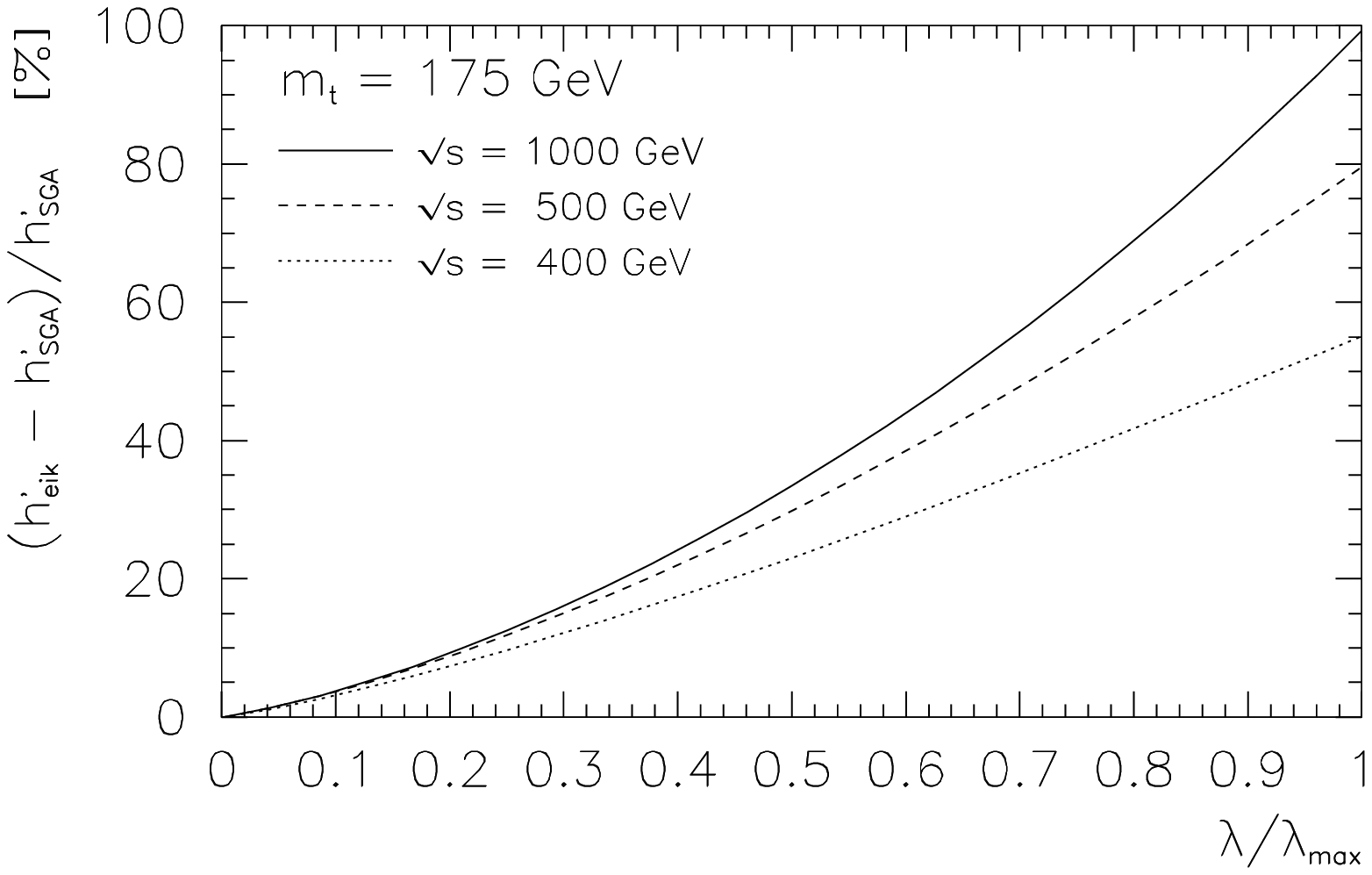, scale=0.7}\end{center}
\caption{\label{intewsga}
Dependence of the relative fraction $(h'_{\rm eik}-h'_{\rm SGA})/h'_{\rm SGA}$
on the scaled gluon energy cut-off parameter $\lambda/\lambda_{\rm max}$ where
$\lambda_{\rm max}=(1-\xi)/2$. Curves are shown for the three center-of-mass
energies $\sqrt s=400$ (dotted), $500$ (dashed), and $1000\GeV$ (full
line).}
\end{figure}

In order to be able to compare the eikonal SGA factor Eq.~(\ref{sga1}) and its
approximate version Eq.~(\ref{limfac}) we (minimally) subtract the
IR-divergent piece $h_{\it IR}$ from both expressions where
\begin{equation}
h_{\it IR}=-\frac{\alpha_sC_F}{\pi v}\Bigg\{\left(2v+(2-\xi)
  \ln\pfrac{1-v}{1+v}\right)\ln\pfrac1{\sqrt\Lambda}\Bigg\}.
\end{equation}
The remaining IR-finite pieces are then $h'_{\rm eik}= h_{\rm eik}-h_{\it IR}$
and $h'_{\rm SGA}=h_{\rm SGA}-h_{\it IR}$. In Fig.~\ref{intewsga} we show a
plot of the relative fraction $(h'_{\rm eik}-h'_{\rm SGA})/h'_{\rm SGA}$ as a
function of the cut-off parameter $\lambda/\lambda_{\rm max}$.
Fig.~\ref{intewsga} shows that $|h'_{\rm eik}|>|h'_{\rm SGA}|$ since both
functions $h'_{\rm eik}$ and $h'_{\rm SGA}$ are negative over the whole range
of $\lambda$. The SGA Eq.~(\ref{limfac}) is a poor approximation to the eikonal
approximation Eq.~(\ref{sga1}) except for the region very close to the
soft-gluon point. For $\sqrt{s}=1000\GeV$ the fractional deviation can become
as large as $100\%$ at the maximal cut value.

\begin{figure}\begin{center}
\epsfig{figure=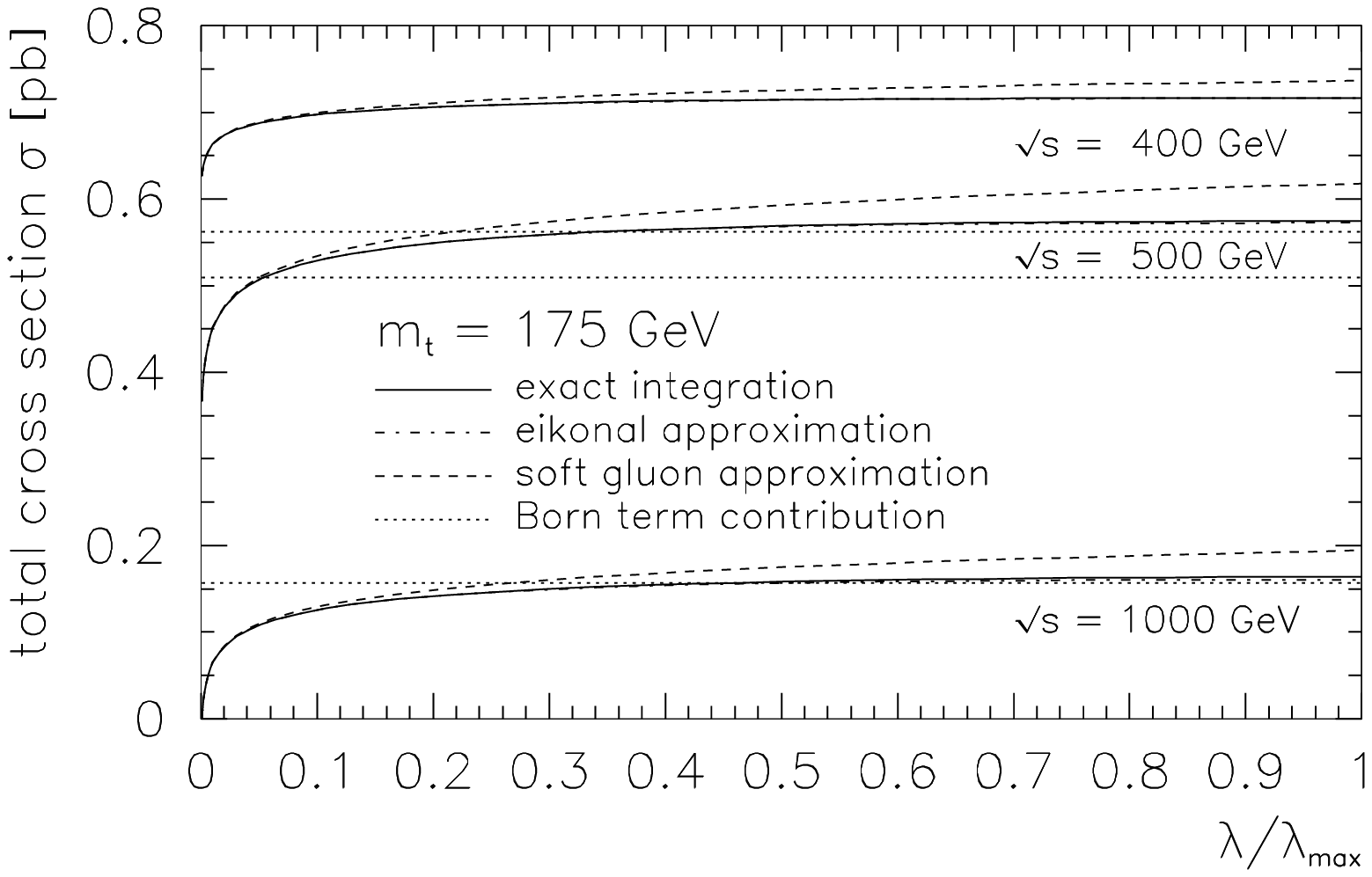, scale=0.7}\end{center}
\caption{\label{intewfir}
Dependence of the total rate (solid line: exact NLO; dash-dotted line:
eikonal, dashed line: SGA) on the scaled gluon energy cut-off parameter
$\lambda/\lambda_{\rm max}$ where $\lambda_{\rm max}=(1-\xi)/2$. Also
shown are the respective cut-off independent LO Born term contributions
(horizontal dotted lines). Curves are shown for the three center-of-mass
energies $\sqrt s=400$, $500$, and $1000\GeV$.}
\end{figure}

As it turns out the eikonal approximation with the eikonal factor (\ref{sga1})
approximates the exact result rather well numerically even up to the hard end
of the gluon spectrum. In Fig.~\ref{intewfir} we show a plot of the total rate
$\sigma$ (=$\sigma_{U+L}$) as a function of the cut-off parameter
$\lambda/\lambda_{\rm max}$ for the three center-of-mass energies
$\sqrt s=400$, $500$ and $1000\GeV$ where we take $m_t=175\GeV$ and
$\as=0.0964$, $0.0941$ and $0.0875$, respectively, for the above three
energies. The rates rise very quickly from the soft region to values close to
the total rates showing that the contributions from the soft region dominate
the total rates. The quality of the eikonal approximation becomes marginally
weaker when the hard gluon region becomes larger with the increase of the
center-of-mass energy. The exact result is hardly discernible from the eikonal
result at the scale of the figure even for the highest c.m.\ energy. The SGA
approximation can be seen to be quite poor. Also shown are the respective LO
Born term contributions which appear as dotted horizontal lines in
Fig.~\ref{intewfir}. The radiative corrections can be seen to be quite large.
At the point where the $O(\as)$ rate intersects the LO Born term rate the
$\as$ corrections go to zero. This can be seen to happen at
$\lambda/\lambda_{\rm max}=2\times 10^{-6}$, $0.014$, and $0.200$ for the above
three c.m.\ energies. At even smaller cut values the total $O(\alpha_{s})$ rate
goes to zero altogether. This happens at $\lambda/\lambda_{\rm max}=10^{-21}$,
$10^{-8}$, and $5\times 10^{-4}$ for the same three above c.m.\ energies. It is
clear that perturbation theory should not be used for such small values of
$\lambda$. This holds, in particular, for the polarization-type observables to
be discussed later on since they are normalized to the total rate and are thus
very sensitive to the vanishing of the total rate. It is important to keep in
mind that the NLO rate goes to $-\infty$ when $\lambda\to 0$ even if this is
not apparent in Fig.~\ref{intewfir}.  

\begin{figure}\begin{center}
\epsfig{figure=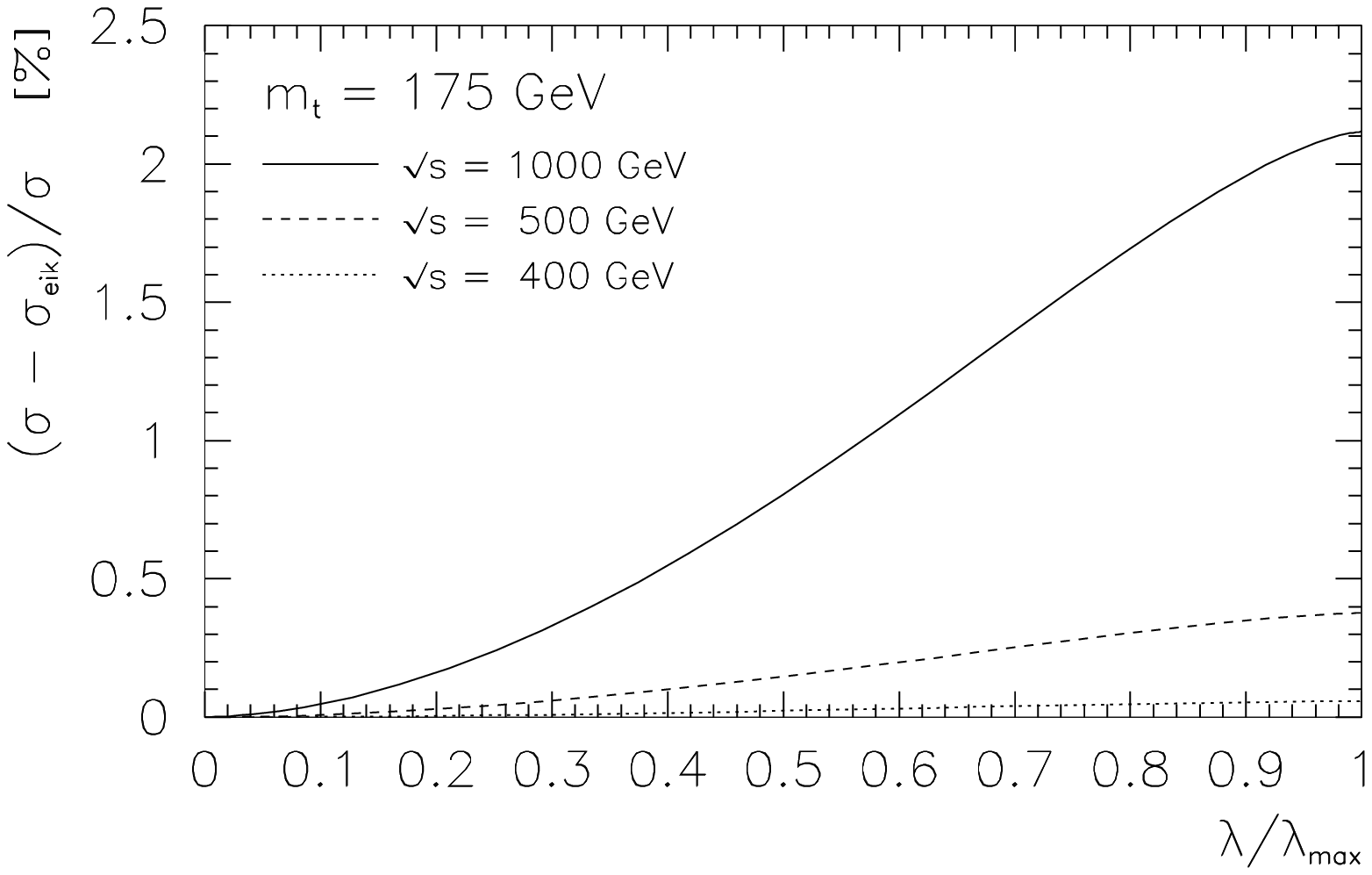, scale=0.7}\end{center}
\caption{\label{intewrel}
Dependence of the relative difference of the exact cross section and the
eikonal approximation on $\lambda/\lambda_{\rm max}$ where
$\lambda_{\rm max}=(1-\xi)/2$ for center-of-mass energies $\sqrt s=400$
(dotted line), $500$ (dashed line), and $1000\GeV$ (full line).}
\end{figure}

In order to show the quality of the eikonal approximation in
Fig.~\ref{intewrel} we show a plot of the cut-off dependence of the relative
difference of the exact cross section and the eikonal approximation
$(\sigma-\sigma_{\rm eik})/\sigma$ for the same three center-of-mass energies.
For $\sqrt s=400\GeV$ the relative difference is very small and remains below
$0.1\%$ over the whole gluon energy spectrum. For the largest energy shown
($\sqrt s=1000\GeV$), where the hard gluon region is the largest, the relative
difference rises from zero at the soft end of the spectrum to about $2\%$ at
the hard end of the spectrum.

\section{Numerical results}                                               
Let us begin the numerical section by the statement that we shall, as in the
previous section, always use a top quark mass of $175\GeV$ in our numerical
results. Since all our results are given in analytical form the corresponding
results for other values of the top quark mass can be readily calculated. For
the strong coupling constant we take the same values as described at the end
of the previous section.

We shall divide our numerical results into two subsections according to
whether the observables or structure functions have a nonvanishing or
vanishing Born term contribution.

\subsection{NLO corrections to nonvanishing LO observables}             
We shall use a terminology where the NLO results are partitioned into a soft
and a hard region by a cut-off value for the gluon energy $E_c$. The soft and
hard regions are defined by their respective integration regions. In the soft
region one integrates from zero gluon energy up to the gluon energy cut $E_c$
including, of course, the one-loop results. In the hard region, one integrates
from the (lower) gluon energy cut $E_c$ to the maximal gluon energy
$E=(1-\xi)\sqrt{q^2}/2$. We use this terminology to differentiate between
choosing an upper cut-off (soft region) and a lower cut-off (hard region) even
if the respective integrations extend into regions with maximal and minimal
gluon energy. The hard gluon contribution can be obtained by subtraction.
Thus, for example, $\sigma({\it hard\/})=\sigma-\sigma({\it soft\/})$. The
definition of the two regions holds irrespective of the actual value of the
cut-off energy.

\begin{figure}\begin{center}
\epsfig{figure=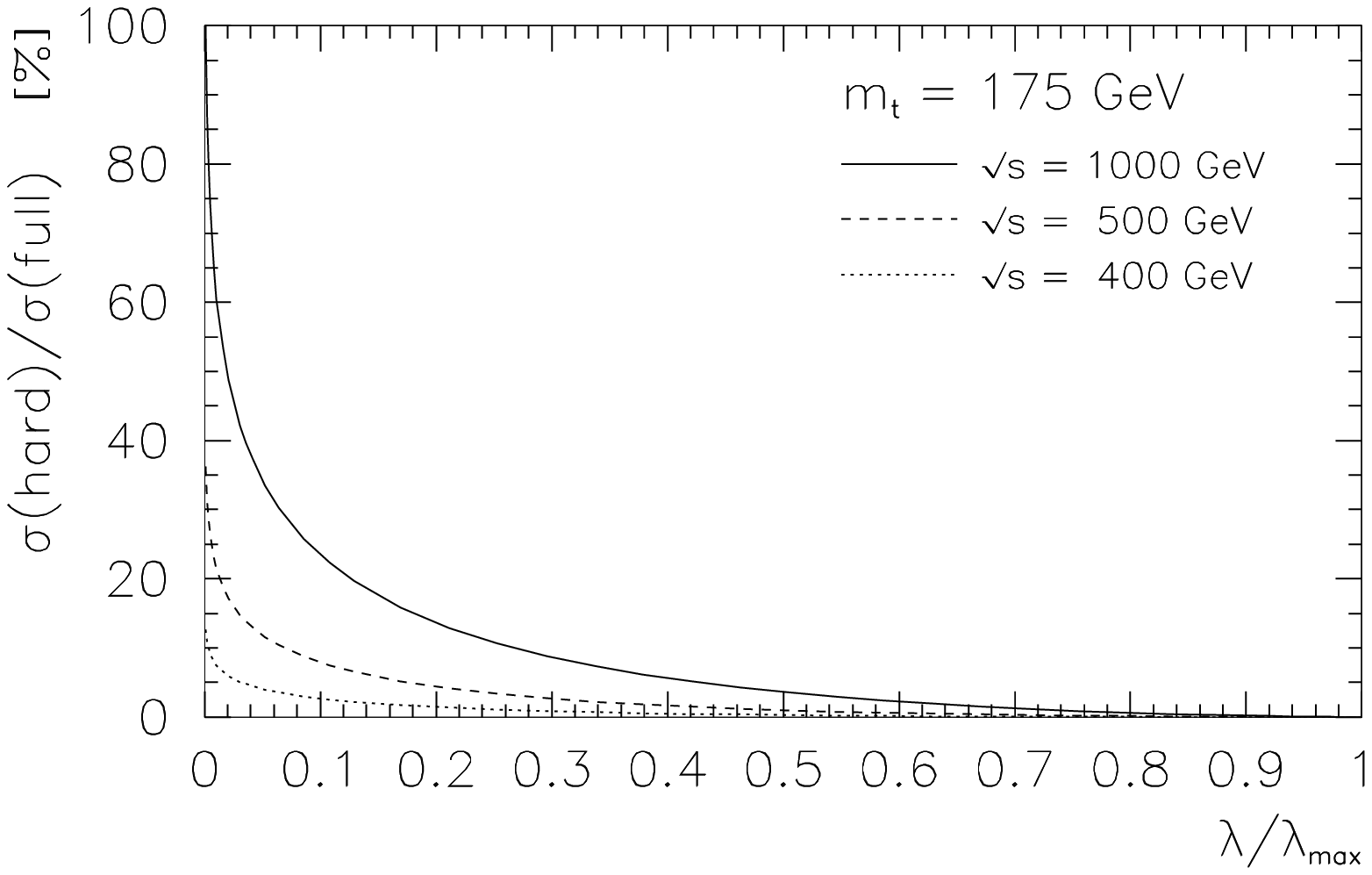, scale=0.7}\end{center}
\caption{\label{intewhar}
Dependence of the ratio $\sigma({\it hard\/})/\sigma({\it full\/})$ on
$\lambda/\lambda_{\rm max}$ in the hard region where $\lambda$ denotes a lower
cut-off. Curves are shown for the three center-of-mass energies
$\sqrt s=400\GeV$ (dotted), $500\GeV$ (dashed), and $1000\GeV$ (full line).}
\end{figure}

In Fig.~\ref{intewhar} we show a plot of the ratio
$\sigma({\it hard\/})/\sigma({\it full\/})$ ($\sigma({\it full\/})=\sigma$) as
a function of the cut-off parameter $\lambda/\lambda_{\rm max}$ for the three
c.m.\ energies $\sqrt s=400$, $500$ and $1000\GeV$. Note that the hard gluon
fraction is proportional to $\alpha_s$. The hard gluon fraction is generally
quite small. As the lower cut-off tends to zero $\sigma({\it hard\/})$ and
thereby $\sigma({\it hard\/})/\sigma({\it full\/})$ tends to $+\infty$ (due to
the positive $-\log\lambda$ singularity). Away from $\lambda=0$ the hard gluon
fraction then drops very quickly as the lower cut-off is raised and reaches
zero at $\lambda/\lambda_{\rm max}=1$ where there is no phase-space left. The
hard gluon fraction becomes larger as the energy increases. For example, at
$\lambda/\lambda_{\rm max}=0.2$ the hard gluon fraction is $1.5$, $4.4$,
and $13.6\%$ for $\sqrt{s}=400$, $500$ and $1000\GeV$, respectively. The
corresponding soft-gluon fractions can be obtained by subtraction as mentioned
above.

We do not show corresponding plots for the other partial unpolarized and
polarized rates $\sigma_i^{(m)}$ because they do not differ much from those
shown in Fig.~\ref{intewhar}. This can be understood from the discussion in
Sec.~6 where we demonstrated that the real gluon emission contributions are
very well approximated by the eikonal approximation which in turn is
proportional to the Born term contribution. This implies that all ratios
$\sigma_i^{(m)}({\it hard\/})/\sigma_i^{(m)}({\it full\/})$ are approximately
equal to one another as well as approximately equal to
$\sigma({\it hard\/})/\sigma({\it full\/})$. An exception is $\sigma_L^\ell$
where the Born term contribution is zero.  This case will be discussed in more
detail later on.

\begin{figure}\begin{center}
\epsfig{figure=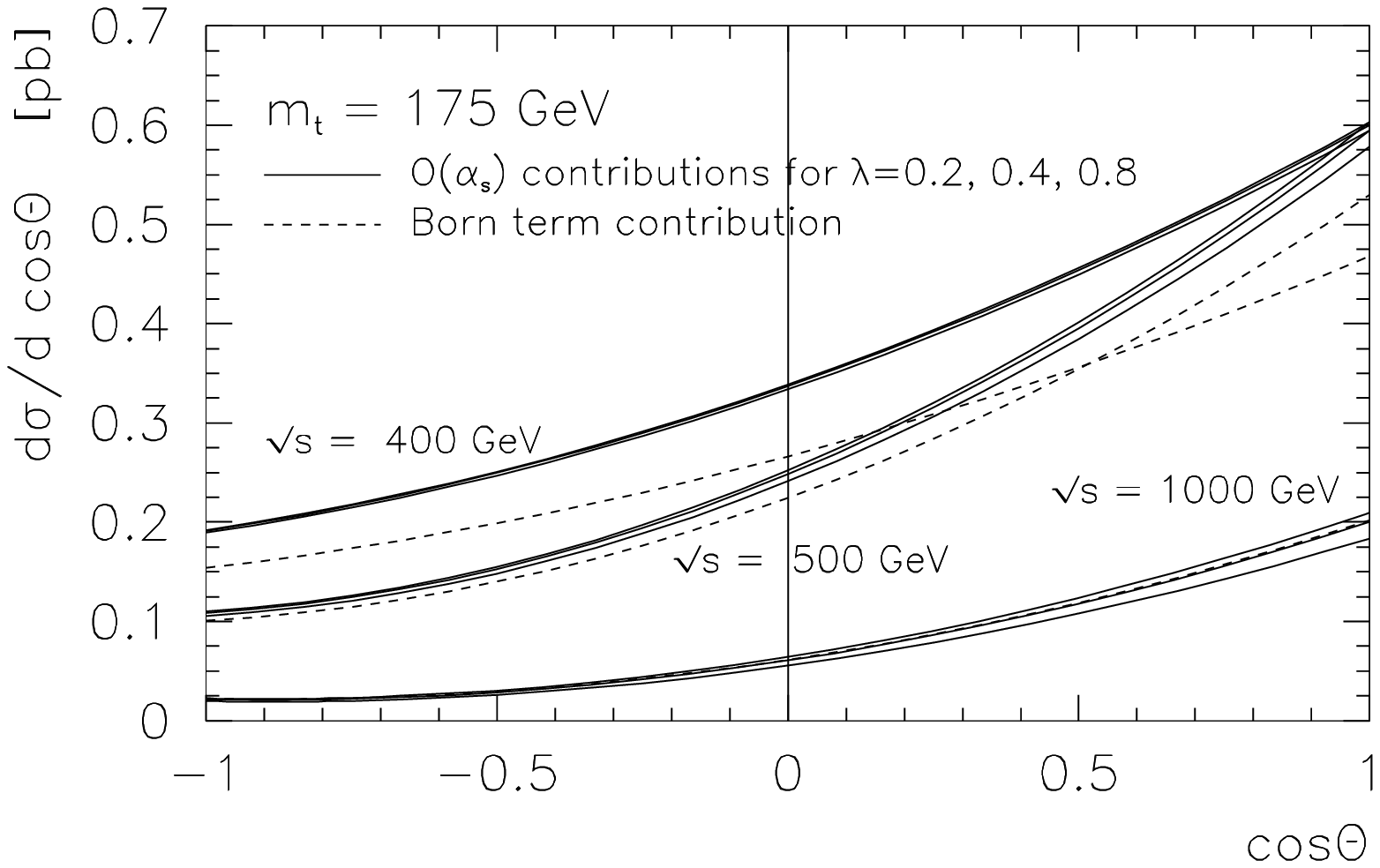, scale=0.7}\end{center}
\caption{\label{intewsit}
Dependence of the differential rate $d\sigma/d\cos\theta$ on $\cos\theta$
in the soft region. Curves are shown for the three center-of-mass energies
$\sqrt s=400$ (dotted line), $500$ (dashed line), and $1000\GeV$ (full line)
and three upper cut-off values $\lambda/\lambda_{\rm max}=0.2$, $0.4$ and
$0.8$ (from bottom to top).}
\end{figure}

In Fig.~\ref{intewsit} we show a plot of $d\sigma/d\cos\theta$ as a function
of $\cos\theta$ for the three c.m.\ energies $\sqrt{s}=400$, $500$ and
$1000\GeV$ and for three respective cut-off parameter values of
$\lambda/\lambda_{\rm max}=0.2$, $0.4$ and $0.8$. The $\cos\theta$ dependence
is marked and strongest for $\sqrt s=500\GeV$ showing that the forward-backward
contribution $\sigma_{F}$ is non-neglible. The radiative corrections are large
for $\sqrt s=400\GeV$ and $\sqrt s=500\GeV$ similar to the total rate plotted
in Fig.~\ref{intewfir}. The cut-off dependence is generally quite weak showing
that the bulk of the different partial rates comes from the region close to
the soft-gluon point $\lambda=0$.

\begin{figure}\begin{center}
\epsfig{figure=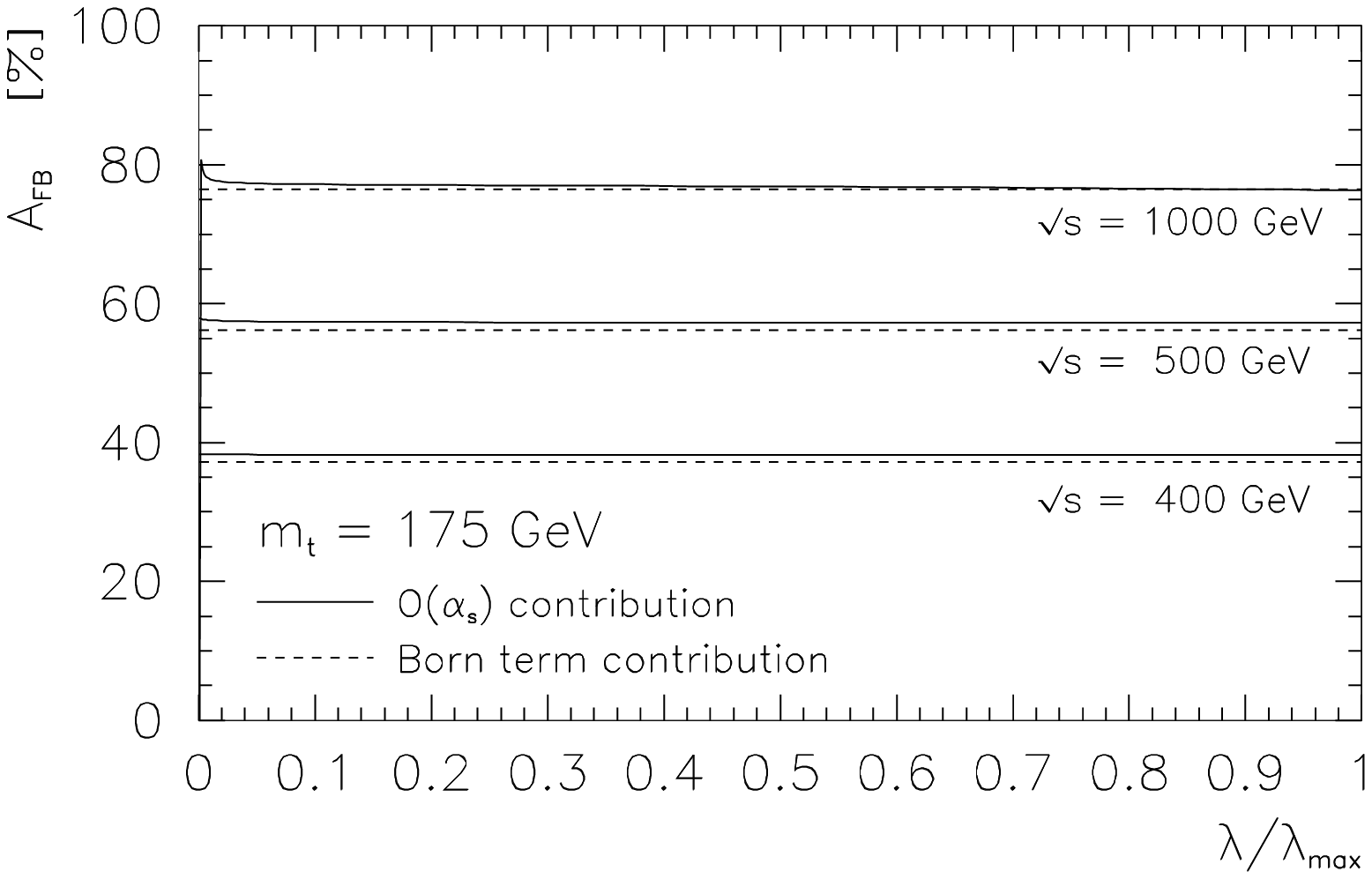, scale=0.7}\end{center}
\caption{\label{intewafb}
Dependence of the forward-backward asymmetry $A_{FB}$ on the (upper) cut-off
$\lambda/\lambda_{\rm max}$ in the soft region (full line). Curves are shown
for the three center-of-mass energies $\sqrt s=400$, $500$, and $1000\GeV$.
Also shown are the respective cut-off independent LO Born term contributions
(horizontal dashed lines).}
\end{figure}

In Fig.~\ref{intewafb} we show a plot of $A_{FB}$ as a function of the upper
cut-off $\lambda/\lambda_{\rm max}$ again for the three c.m.\ energies
$\sqrt{s}=400$, $500$ and $1000\GeV$ where we have defined the
forward-backward asymmetry by
\begin{equation}\label{FB1}
A_{FB}=\frac{\sigma({\it forward\/})-\sigma({\it backward\/})}
{\sigma({\it forward\/})+\sigma({\it backward\/})}.
\end{equation}
Note that one has to separately integrate the numerator and denominator of
Eq.~(\ref{FB1}) over the gluon energy when calculating $A_{FB}$. The radiative
corrections are generally small and the dependence on the cut-off $\lambda$ is
quite weak. $A_{FB}$ is largest for $\sqrt s=1000\GeV$ as can also be
appreciated by looking at Fig.~\ref{intewsit}. The radiative corrections are
largest for $\sqrt s=400\GeV$. For example, for an upper cut-off of
$\lambda/\lambda_{\rm max}=0.2$ they amount to $2.7\%$.

The radiative corrections to polarization-type observables $P_i^{(m)}$ are in
general quite small even if the radiative corrections to the polarized rates
themselves are large. The reason is that polarization-type observables
correspond to normalized density matrix elements defined by the ratio of a
polarized rate and the total rate. The radiative corrections to the numerator
and the denominator tend to go in the same direction and thus tend to cancel
out in the ratio. Take, for example, a generic polarization observable
$P_i^{(m)}$ which, at $O(\as)$, is defined by\footnote{The forward-backward
asymmetry $A_{FB}$ defined in Eq.~(\ref{FB1}) is such a  polarization-type
observable with $\sigma_i^{(m)}=\sigma_F$.} 
\begin{eqnarray}\label{approx}
P_i^{(m)}(O(\as);\lambda)&=&\frac{\sigma_i^{(m)}({\it Born})
  +\sigma_i^{(m)}(\as;\lambda)}{\sigma({\it Born})+\sigma(\as;\lambda)}
  \nonumber\\
  &\approx&\frac{\sigma_i^{(m)}({\it Born})(1+h'_{\rm eik}(\as;\lambda))}
  {\sigma({\it Born})(1+h'_{\rm eik}(\as;\lambda))}
  \ =\ P_i^{(m)}({\it Born}). 
\end{eqnarray} 
Thus $P_i^{(m)}(O(\as);\lambda)=P_i^{(m)}({\it Born})$ as long as one can
neglect non-Born term like structures in the radiative $\alpha_s$-corrections
resulting either from the one-loop or the $\lambda$-dependent hard gluon
corrections. As it turns out the non-Born term like $\alpha_s$ corrections are
in general small but can amount to several percent. The above reasoning breaks
down when either the numerator or the denominator in Eq.~(\ref{approx})
approaches zero which can happen for very small values of $\lambda$. As has
been argued before such small cut values are not acceptable from the physics
point of view.

\begin{figure}\begin{center}
\epsfig{figure=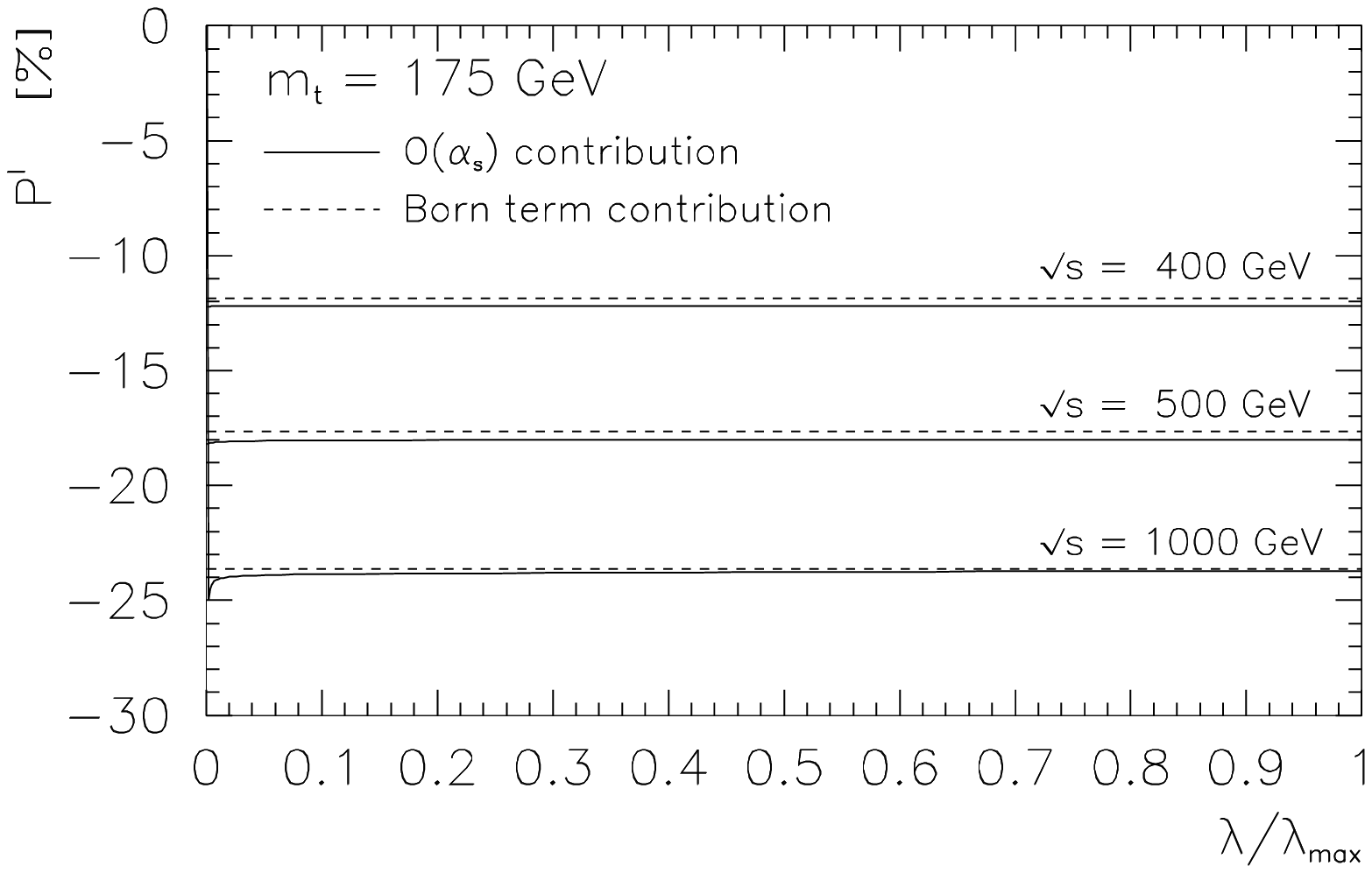, scale=0.7}\end{center}
\caption{\label{intewpol}
Dependence of the longitudinal polarization $P^\ell$ on the (upper) cut-off
$\lambda/\lambda_{\rm max}$ in the soft region (full line). Curves are shown
for the three center-of-mass energies $\sqrt s=400$, $500$ and $1000\GeV$.
Also shown are the respective cut-off independent LO Born term contributions
(horizontal dashed lines).}
\end{figure}

In Fig.~\ref{intewpol} we show a plot of $P^\ell$ as a function of
$\lambda/\lambda_{\rm max}$ again for the three c.m.\ energies $\sqrt{s}=400$,
$500$ and $1000\GeV$ where $P^\ell$ is the longitudinal polarization of the
top quark $P^\ell=\sigma^\ell/\sigma$. Note that again one has to separately
integrate the numerator and denominator over the gluon energy when calculating
$P^\ell$, i.e.\ $P^\ell(\lambda)=\sigma^\ell(\lambda)/\sigma(\lambda)$. As in
Fig.~\ref{intewafb} the radiative corrections and the dependence on $\lambda$
can be seen to be quite small. The longitudinal polarization $P^\ell$ is
largest for $\sqrt s= 1000\GeV$. 

\begin{figure}\begin{center}
\epsfig{figure=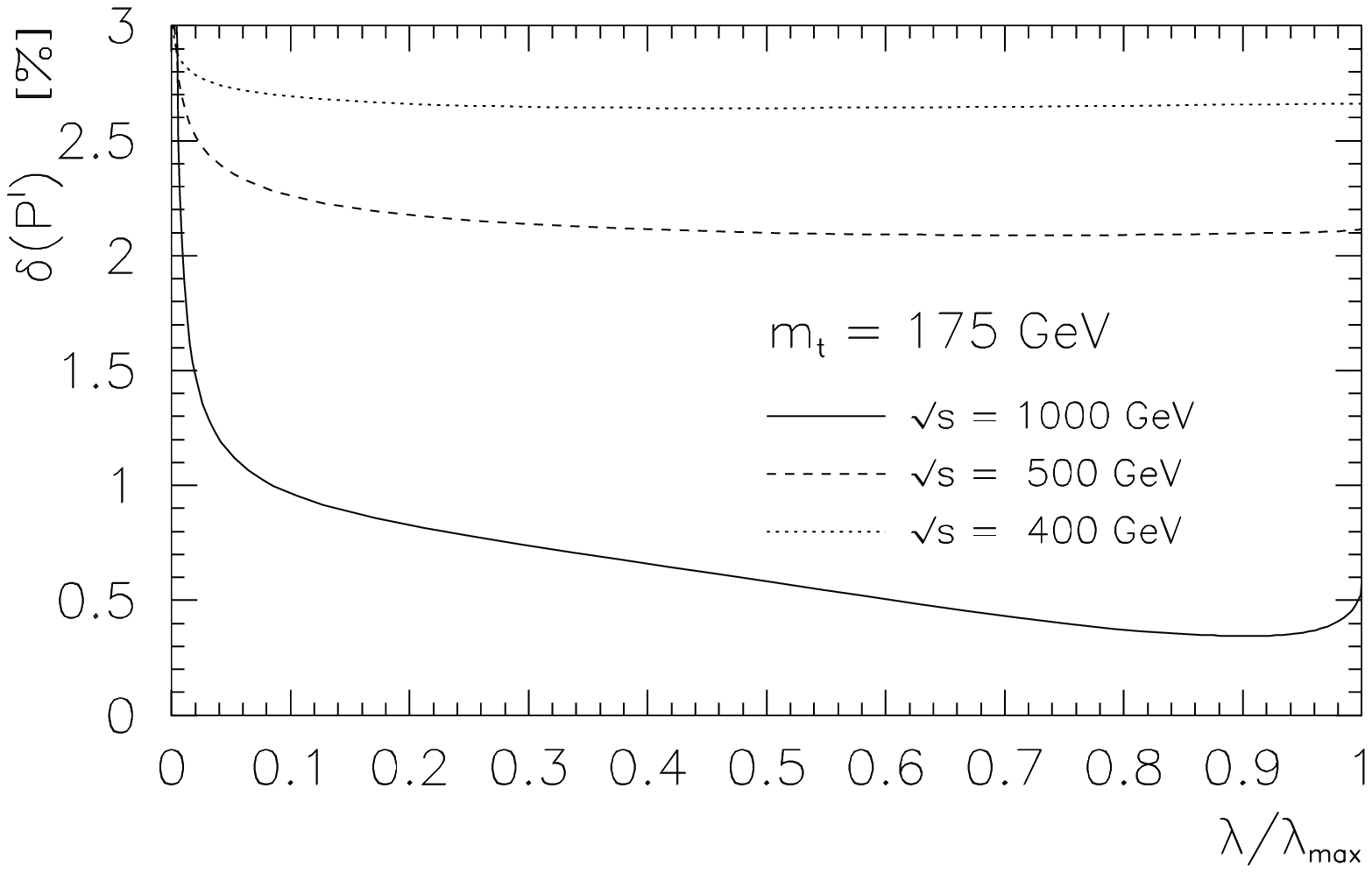, scale=0.7}\end{center}
\caption{\label{intewdol}
Dependence of the fractional deviation of the longitudinal polarization
$\delta(P^\ell)$ on the (upper) cut-off $\lambda/\lambda_{\rm max}$ in the
soft region. Curves are shown for the three center-of-mass energies
$\sqrt s=400$ (dotted line), $500$ (dashed line) and $1000\GeV$ (full line).}
\end{figure}

In order to highlight the size of the radiative corrections to $P^\ell$ we
define a fractional deviation of $P^\ell$ from its Born term value for
different cut-off values by writing
\begin{equation}
\delta(P^\ell)=\frac{P^\ell(\lambda)-P^\ell({\it Born\/})}
{P^\ell({\it Born\/})}
\end{equation}
where $P^\ell(\lambda)$ is the value of $P^\ell$ for the upper cut-off
parameter $\lambda$, i.e. in our above terminology $P^\ell(\lambda)$ refers
to the value of the observable in the soft region. Fig.~\ref{intewdol} shows
that close to $\lambda = 0$ the fractional deviations $\delta(P^\ell)$ tend
to infinity because the denominator in
$P^\ell(\lambda)=\sigma^\ell(\lambda)/\sigma(\lambda)$ go to zero, as
mentioned before. Away from $\lambda \approx 0$ the dependence of
$\delta(P^\ell)$ on the gluon cut $\lambda$ is not very pronounced
except for the highest energy value $\sqrt s=1000\GeV$. The fractional
deviation is largest for $\sqrt s=400\GeV$.

Also of interest are the values of a rate function in the hard gluon region.
To this end we define a lower scaled gluon energy cut-off
$\lambda_{\rm lower}$ and integrate from $\lambda_{\rm lower}$ to the upper
limit $\lambda_{\rm max}=(1-\xi)/2$. As before this is effectively done by
subtraction, i.e.\ $\sigma^{(m)}({\it hard\/})
=\sigma^{(m)}(\lambda_{\rm max})-\sigma^{(m)}(\lambda)$ since we have not
separately listed analytical formulas for the hard gluon rates. We then define
a forward-backward asymmetry $A_{FB}({\it hard\/})$ and a longitudinal
polarization $P^\ell({\it hard\/})$ in the hard region by writing
\begin{equation}\label{FB2}
A_{FB}({\it hard\/})=\frac{\sigma({\it forward\/})-\sigma({\it backward\/})}
{\sigma({\it forward\/})+\sigma({\it backward\/})} \,\,\bigg|_{{\it hard\/}}
\end{equation}
and
\begin{equation}\label{longlong1}
P^\ell({\it hard\/})=\frac{\sigma^\ell}{\sigma}\,\,\,\Big|_{{\it hard\/}}
\end{equation}

\begin{figure}\begin{center}
\epsfig{figure=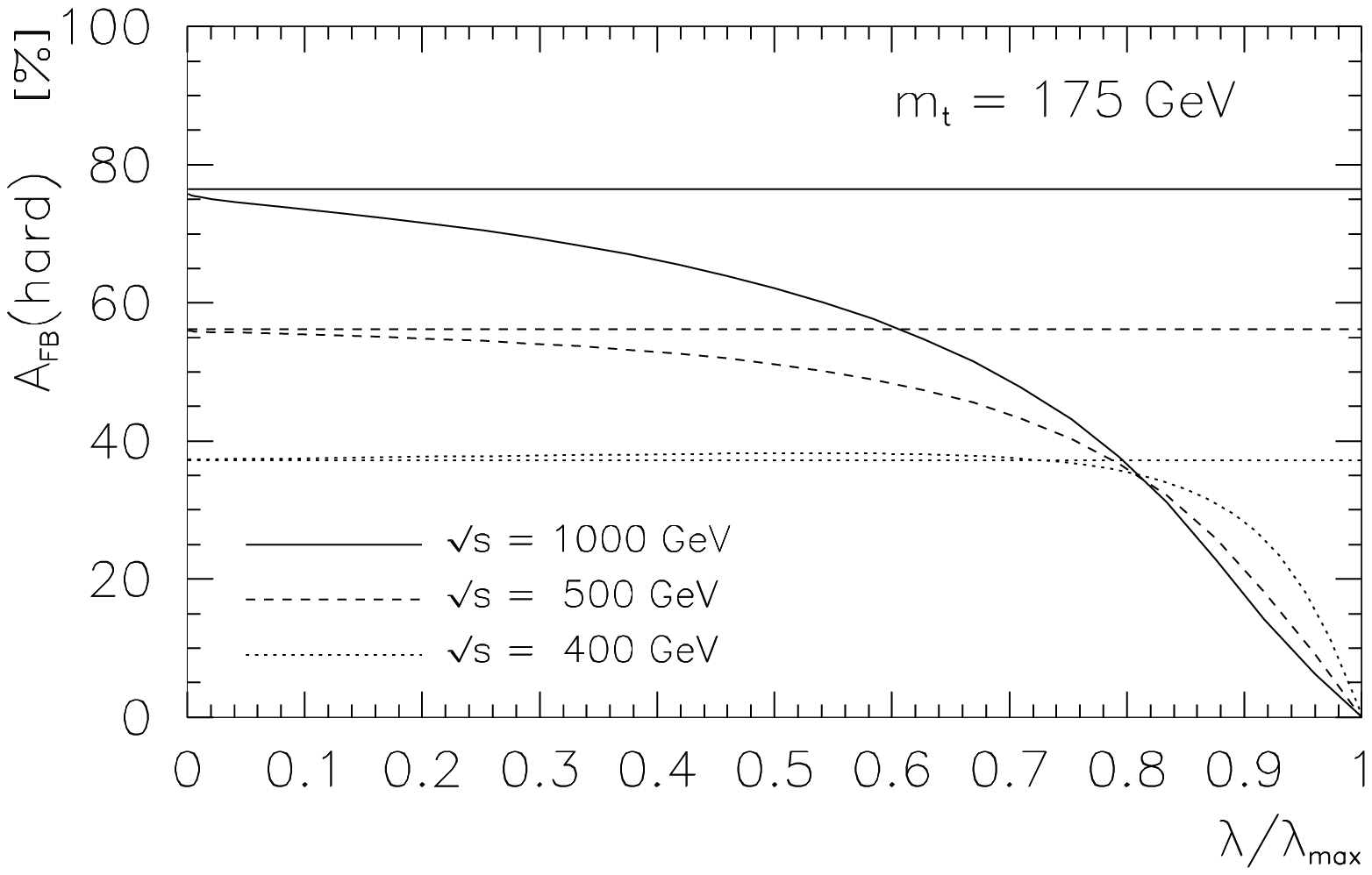, scale=0.7}\end{center}
\caption{\label{intewhfb}
Dependence of the forward-backward asymmetry $A_{FB}$ on
$\lambda/\lambda_{\rm max}$ in the hard region where $\lambda$ denotes a lower
cut-off. Curves are shown for the three center-of-mass energies
$\sqrt s=400$ (dotted line), $500$ (dashed line), and $1000\GeV$ (full line).
The straight lines indicate the Born term level results.}
\end{figure}

In Fig.~\ref{intewhfb} we show a plot of $A_{FB}({\it hard\/})$ as a function
of $\lambda/\lambda_{\rm max}$ again for the three c.m.\ energies
$\sqrt{s}=400$, $500$ and $1000 \GeV$. As the lower cut-off tends to zero
$A_{FB}({\it hard})$ reaches values very close to those of $A_{FB}({\it soft})$
in Fig.~\ref{intewafb} showing that the non-Born term structures in the 
$\alpha_s$-radiative corrections are not very significant. Only for larger 
cut-off values does one find significant deviation from the Born term values.
For example, for $\lambda/\lambda_{\rm max}=0.6$ and $\sqrt{s}=1000 \GeV$
one has a $30\%$ deviation from the Born term value. 
 
\begin{figure}\begin{center}
\epsfig{figure=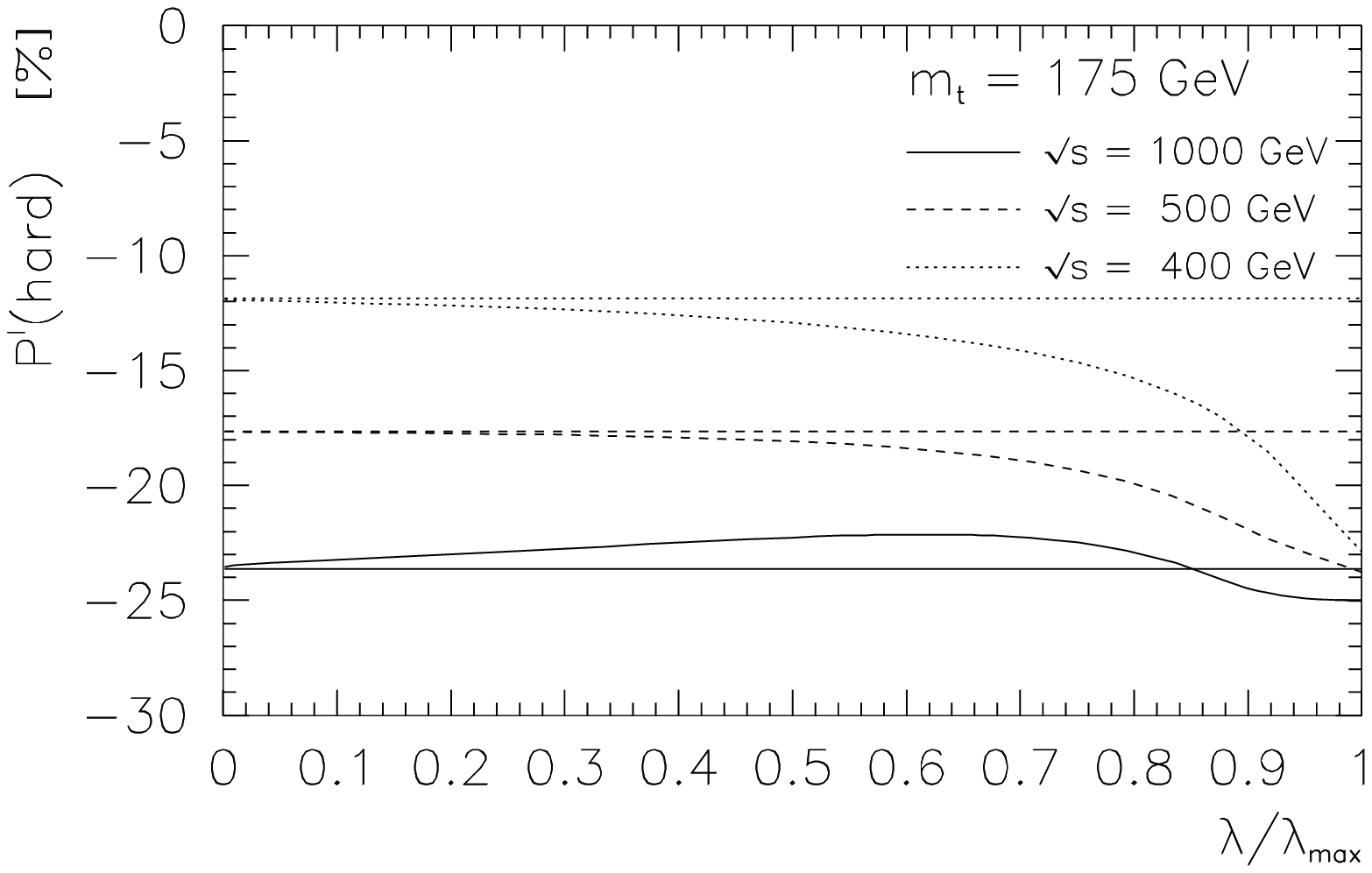, scale=0.7}
\end{center}
\caption{\label{intewhol}Dependence of the longitudinal polarization
$P^\ell$ on $\lambda/\lambda_{\rm max}$ in the hard region where $\lambda$
denotes a lower cut-off. Curves are shown for the three center-of-mass
energies $\sqrt s=400$ (dotted line), $500$ (dashed line), and $1000\GeV$ (full
line). The straight lines indicate the Born term level results.}
\end{figure}

Fig.~\ref{intewhol} shows the same plot for the longitudinal polarization
$P^\ell$. Similar remarks apply as in the discussion of $A_{FB}({\it hard\/})$
except that the dependence on the lower cut-off is not as pronounced as in
Fig.~\ref{intewhfb}. Marked deviations from the Born term values only set in at
larger values of $\lambda$. 

\subsection{NLO contributions to vanishing LO observables
  or structure functions}
It was pointed out already in Ref.~\cite{Groote:1995yc} that the longitudinal
polarization of the top quark produced from a longitudinally polarized gauge
boson ($\gamma$ and/or $Z$) denoted by $P_L^\ell$ vanishes at the Born term
level. $P_L^\ell$ vanishes at the Born term level and also for the one-loop
contribution due to the two facts that there are no second-class currents in
the SM and that one is dealing with a two-body final state in these two cases.
Technically this comes about since the contractions of the first class axial
currents $\bar u\gamma_\mu\gamma_5v$ and $\bar uq_\mu\gamma_5v$ with the
longitudinal projector $e_3^\mu$ (see Eq.~(\ref{project1})) vanish in the
two-body case. In the Standard Model a nonvanishing value of the polarization
$P_L^\ell$ is generated only at NLO (or higher orders) from real gluon
bremsstrahlung. This NLO effect is quite small as can be seen from Fig.~2a in
Ref.~\cite{Groote:1995yc} which shows that $P_L^\ell$ rises from zero at
threshold to $-0.21\%$ at $\sqrt{s}=1000\GeV$.  

\begin{figure}\begin{center}
\epsfig{figure=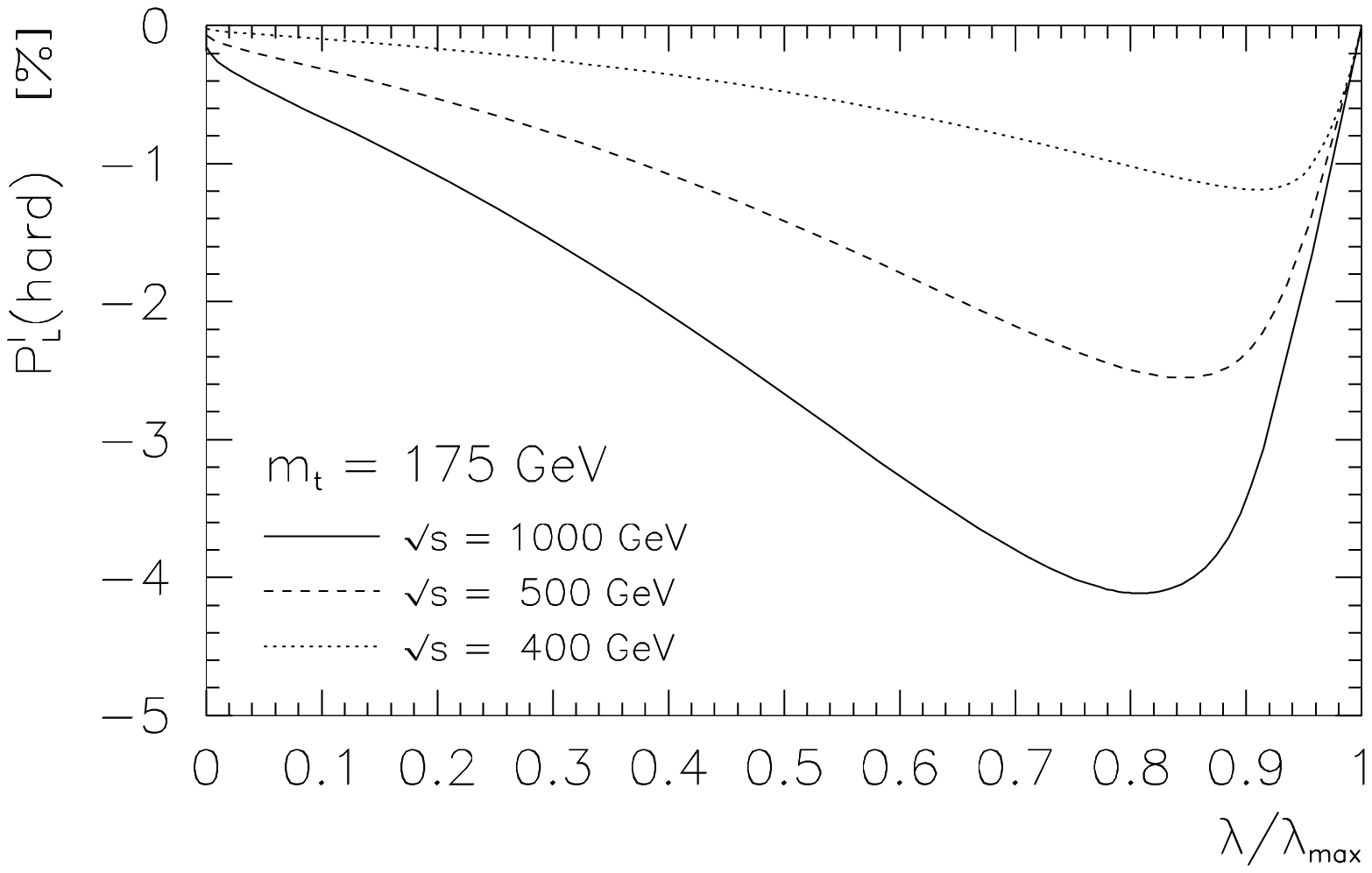, scale=0.7}\end{center}
\caption{\label{intewhll}
Dependence of the longitudinal polarization from a longitudinal polarized
gauge boson $P_L^\ell$ on $\lambda/\lambda_{\rm max}$ in the hard gluon region
where $\lambda$ denotes a lower cut-off. Curves are shown for the three
center-of-mass energies $\sqrt s=400$ (dotted line), $500$ (dashed line), and
$1000\GeV$ (full line).}
\end{figure}

A larger absolute value of $P_L^\ell$ is obtained in the hard gluon region
since $P_L^\ell$ is an $O(\as)$ effect. To this end we define the ratio
\begin{equation}
P_L^\ell({\it hard\/})=\frac{\sigma_L^\ell}{\sigma}\
  \Big|_{{\it hard\/}}
\end{equation}
where the hard gluon region is defined as in the beginning of this section.
In Fig.~\ref{intewhll} we show a plot of $P_L^\ell({\it hard \/})$ as a
function of the scaled gluon energy cut-off where the cut-off parameter
$\lambda$ now refers to a lower cut-off. It goes without saying that
$P_L^\ell({\it hard \/})=0$ in the soft-gluon or eikonal approximation
since then $\sigma_L^\ell(\as) \propto \sigma_L^\ell(Born)=0$ in the
soft-gluon or eikonal approximation. Fig.~\ref{intewhll} shows that
$P_L^\ell({\it hard \/})$ can become as large as $-4\%$ for
$\sqrt s=1000\GeV$ and $\lambda/\lambda_{\rm max}=0.8$.
$P_L^\ell({\it hard \/})$ increases when the energy increases.
$P_L^\ell({\it hard \/})$ goes to zero as $\lambda \to 0$ since in this
limit $\sigma_L^\ell$ is finite whereas $\sigma$ diverges.

We mention that a nonvanishing contribution to $P_L^\ell$ can also be obtained
by adding an anomalous axial current to the usual SM first class top quark
current structure. This will be discussed later on.

There are two classes of relations among the structure functions $H_a^{j(m)}$
at the two-body level. The first class of relations depends solely on the fact
that one is dealing with a two-body final state at the Born term and one-loop
level. There are four relations of this kind
\begin{eqnarray}\label{twobody1}
\mbox{\it real part:} &&H_U^1\ =\ H_F^{1\ell}\qquad H_U^2\ =\ H_F^{2\ell}
\qquad H_F^4\ =\ H_U^{4\ell} \\[7pt]
\mbox{\it imaginary part:}&&H_F^3\ =\ H_U^{3\ell} 
\end{eqnarray}
The second class of relations depends on the two-body dynamics and on the 
fact that one has only first class currents in the SM. There are six 
relations of this kind. These are 
\begin{eqnarray}\label{twobody2}
\mbox{\it real part:}&&H_L^1=H_L^2\qquad H_L^{4\ell}\ =\ 0
\qquad H_A^{1T}=H_A^{2T}\qquad H_A^{4T}\ =\ H_I^{3N}\\[7pt] 
\mbox{\it imaginary part:}&&H_A^{4N}\ =\ H_I^{3T}\qquad 
H_I^{1N}\ =\ H_I^{2N}
\end{eqnarray}
One can explicitly check with the Born term and one-loop expressions listed
in Sec.~3 that these relations are in fact satisfied.

Note that the class 1 relation $H_F^3=H_U^{3\ell}$, and the class 2 relations
$H_A^{4N}=H_I^{3T}$ and $H_I^{1N}=H_I^{2N}$ will not be affected by the
$O(\as)$ tree graph contributions since they result from the imaginary parts
of the (two-body) one-loop contributions. As mentioned before, the relation
$H_U^2=H_F^{2\ell}$ interestingly also holds at the $O(\as)$ tree graph level.
In the following we shall numerically investigate how the remaining relations
in (\ref{twobody1}) and (\ref{twobody2}) are affected by the $O(\as)$ tree
graph contributions. It goes without saying that the relevant remaining
relations in (\ref{twobody1}) and (\ref{twobody2}) still hold at NLO if one
uses the soft-gluon or eikonal approximations rather than the exact form of
the radiative corrections.

\begin{figure}\begin{center}
\epsfig{figure=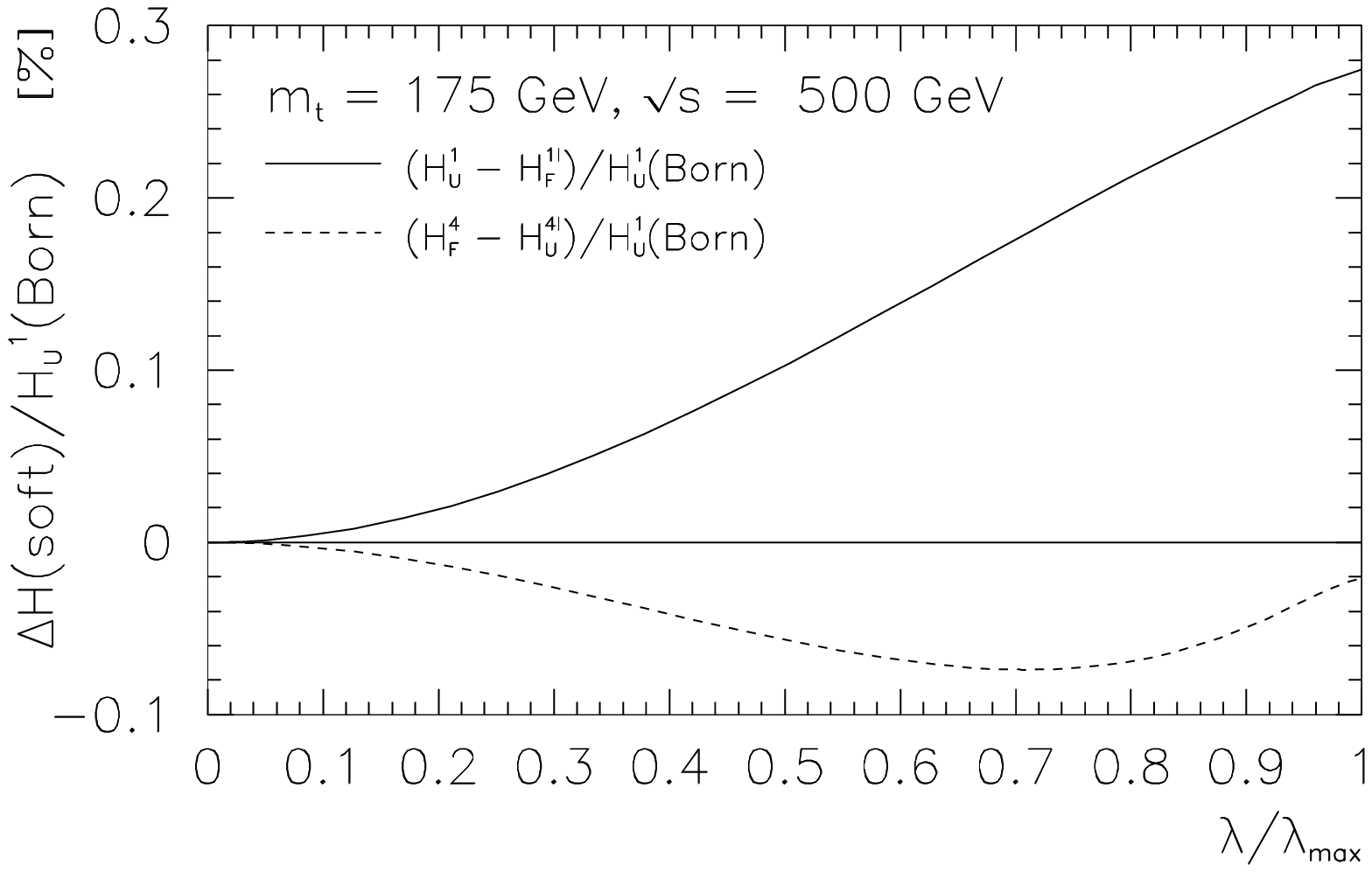, scale=0.7}\end{center}
\caption{\label{intewre1}Goodness of the class 1 relations against radiative
corrections using an upper gluon energy cut. Dependence of the ratios 
$(H_U^1-H_F^{1\ell})/H_U^1({\it Born\/})$ (solid line), and
$(H_F^4-H_U^{4\ell})/H_U^1({\it Born\/})$ (dashed line) on
$\lambda/\lambda_{\rm max}$ where $\lambda$ denotes an upper cut-off. Curves
are shown for the center-of-mass energy $\sqrt s=500\GeV$ in the soft region.}
\begin{center}\epsfig{figure=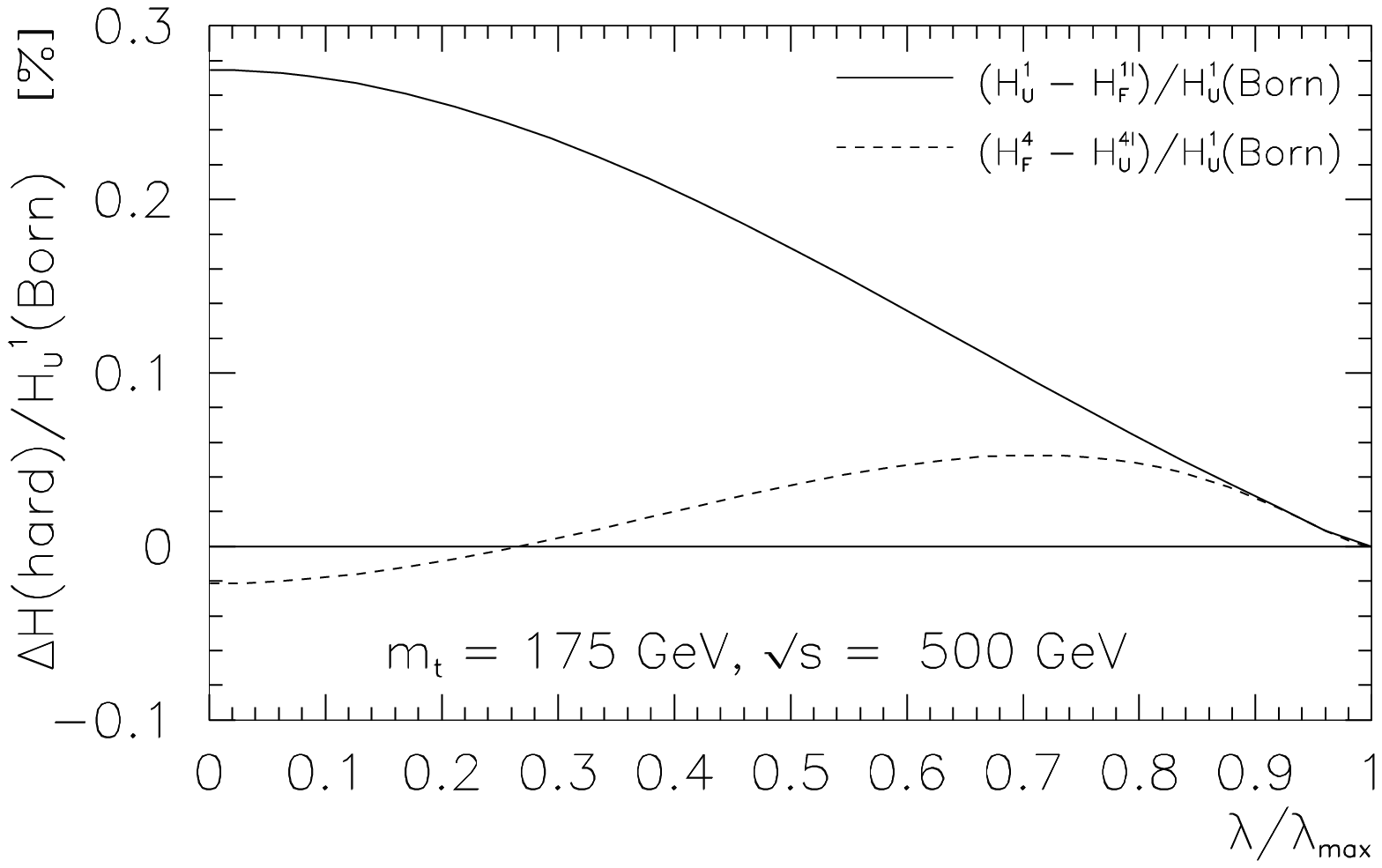, scale=0.7}\end{center}
\caption{\label{intewre3}
The same as in Fig.~\ref{intewre1} for the hard region.}
\end{figure}

We start our numerical discussion with the first class of relations in
Eq.~(\ref{twobody1}). In order to obtain a quantitative handle on how the tree
graph contributions affect the first class relations $H_U^1=H_F^{1\ell}$ and
$H_F^4=H_U^{4\ell}$ in Eq.~(\ref{twobody1}) we consider differences of the
relevant structure functions and (arbitrarily) normalize them to
$H_U^1({\it Born\/})$. In Fig.~\ref{intewre1} we show a plot of the ratios
$(H_U^1-H_F^{1\ell})/H_U^1({\it Born\/})$ and
$(H_F^4-H_U^{4\ell})/H_U^1({\it Born\/})$ as functions of the upper cut-off
(``soft region'') in terms of the scaled gluon energy cut
$\lambda/\lambda_{\rm max}$ for $\sqrt{s}=500\GeV$. The violation of the
class 1 relations slowly rises from zero at the soft-gluon point and reaches
values of $0.27$ and $-0.02\%$, respectively, for the two above ratios at
$\lambda_{\rm max}$ where one integrates over the full gluon phase-space. In
Fig.~\ref{intewre3} we consider the hard region where
$\lambda/\lambda_{\rm max}$ now refers to a lower cut-off in the gluon energy.
Now $\lambda=0$ corresponds to a full phase-space integration and one
therefore recovers the $\lambda/\lambda_{\rm max}=1$ limiting values of
Fig.~\ref{intewre1} remembering that there are no loop contributions to the
above four quantities. The relevant ratios go to zero for
$\lambda=\lambda_{\rm max}$ in Fig.~\ref{intewre3} since phase-space goes to
zero.  

\begin{figure}\begin{center}
\epsfig{figure=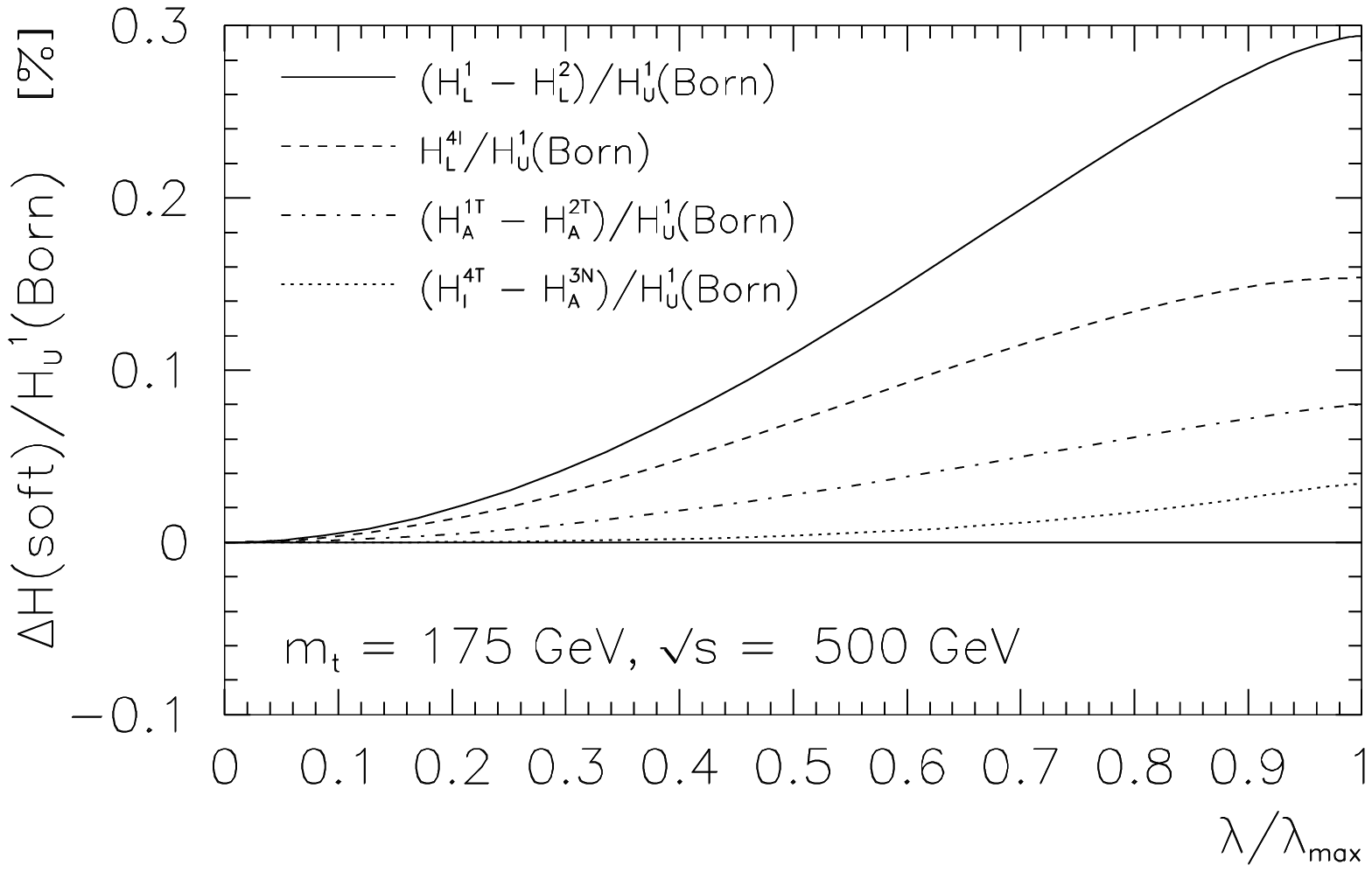, scale=0.7}\end{center}
\caption{\label{intewre2}Goodness of the class 2 relations against radiative
corrections using an upper gluon energy cut. Dependence of the ratios
$(H_L^1-H_L^2)/H_U^1({\it Born\/})$ (solid line) 
$H_L^{4\ell}/H_U^1({\it Born\/})$ (dashed line)
$(H_A^{1T}-H_A^{2T})/H_U^1({\it Born\/})$ (dash-dotted line),
and $(H_I^{4T}-H_A^{3N})/H_U^1({\it Born\/})$ (dotted line) on
$\lambda/\lambda_{\rm max}$ where $\lambda$ denotes an upper cut-off. Curves
are shown for the center-of-mass energy $\sqrt s=500\GeV$.}
\begin{center}\epsfig{figure=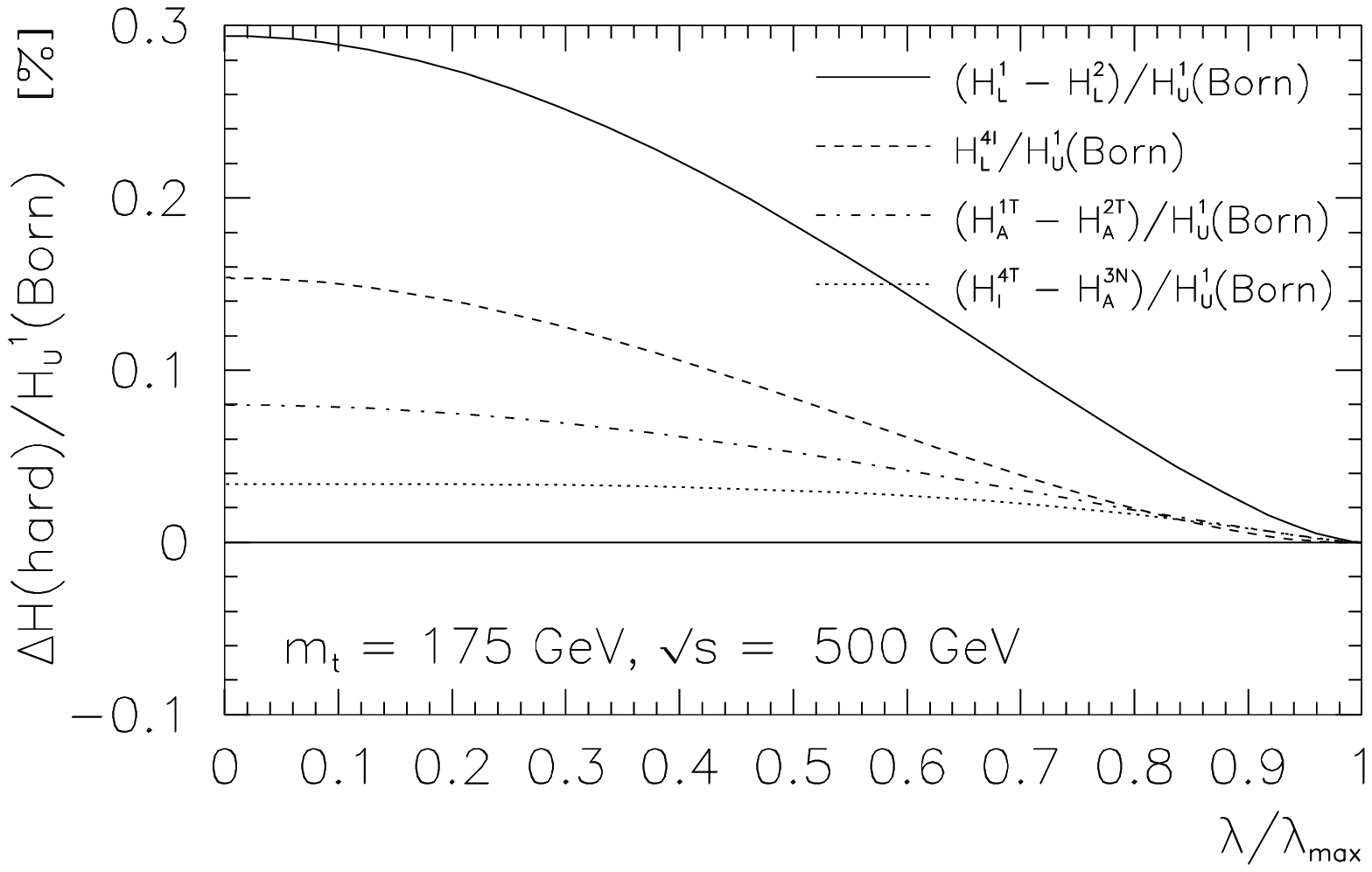, scale=0.7}\end{center}
\caption{\label{intewre4}
The same as in Fig.~\ref{intewre2} for the hard region.}
\end{figure}

The influence of the tree graph contributions on the second class of relations 
Eq.~(\ref{twobody2}) is tested in a similar manner. We consider again
differences of the relevant structure functions (or structure functions
themselves) normalized to $H_U^1({\it Born\/})$. In Figs.~\ref{intewre2} 
and~\ref{intewre4} we show plots of the ratios
$(H_L^1-H_L^2)/H_U^1({\it Born\/})$, $H_L^{4\ell}/H_U^1({\it Born\/})$, 
$(H_A^{1T}-H_A^{2T})/H_U^1({\it Born\/})$, and
$(H_I^{4T}-H_A^{3N})/H_U^1({\it Born\/})$ for upper and lower cut-off values
of the gluon energy, respectively. In Fig.~\ref{intewre2} (``soft region'')
the violations rise from zero at the soft-gluon point to the values $0.29$,
$0.15$, $0.08$, and $0.03\%$ for $\lambda=\lambda_{\rm max}$ where one 
integrates over the full gluon phase-space. Fig.~\ref{intewre4} shows the
same four ratios in the hard gluon region. As before the right-most values in
Fig.~\ref{intewre2} agree with their left-most pendants in Fig.~\ref{intewre4}.
The violations of the class 1 and class 2 relations due to hard gluon
radiation can be seen to be generally quite small.
 
The effect of the radiative corrections to the class 2 relations
(\ref{twobody2}) can be mimicked by adding an anomalous axial current to the
SM currents. The anomalous axial current to be added reads
(see e.g.~\cite{Bernreuther:1988jr,Korner:1990zk})
\begin{equation}\label{anocurrent}
j^\mu({\it anomalous\/})
  =g_a\bar\psi_t\frac{i\sigma^{\mu\nu}q_\nu}{2m_t}\gamma_5\psi_t.
\end{equation}
In general $g_a$ can be complex, $g_a=\real g_a+i\,\imag g_a$. Note that the 
current in Eq.~(\ref{anocurrent}) is a so-called second-class current with
$J^{PC}=1^{+-}$ quantum numbers. In particular, the contraction of the
anomalous current with the longitudinal projector $e_3^\mu$ (see
Eq.~(\ref{project1})) no longer vanishes, i.e.\ one now has
$e_3^\mu\bar u\sigma_{\mu\nu}q^\nu v\neq 0$, and therefore $H_L^{4\ell}\neq 0$.
It should be clear that the addition of the anomalous axial current does not
affect the class 1 two-body relations in Eq.~(\ref{twobody1}) but, in general,
violates the class 2 relations. We assume that the coupling strength $g_a$ is
small and we therefore only consider the interference contribution of
Eq.~(\ref{anocurrent}) with the SM $(t\bar t)$ current, i.e.\ terms that are
linear in $g_a$. 

The interference contribution of the anomalous axial-vector current can be
calculated using the projection formulas written down in Sec.~3. One finds
\begin{eqnarray}\label{anomal}
H_L^1-H_L^2&=&O(g_a^2)\nonumber\\[7pt]
H_L^{4\ell}&=&-2N_{c}q^2v \,\real g_a\nonumber \\[3pt]
H_A^{1T}-H_A^{2T}&=&N_c\frac{\sqrt{q^2}}{\sqrt2m}q^2v^2\,\real g_a\nonumber\\
H_I^{4T}-H_A^{3N}&=&N_c\frac{\sqrt{q^2}}{\sqrt2m}q^2v\,\real g_a\,.
\end{eqnarray}
It is noteworthy that only the real part of $g_{a}$ contributes to the
relations (\ref{anomal}).
In order to obtain a quantitative handle on the coupling parameter $g_a$ we
determine the values of the anomalous parameter $g_a$ that would reproduce the
fully integrated quantities $H_L^{4\ell}/H_U^1({\it Born\/})$,
$(H_A^{1T}-H_A^{2T})/H_U^1({\it Born\/})$ and
$(H_I^{4T}-H_A^{3N})/H_U^1({\it Born\/})$, i.e.\ the values that these
quantities take at the right-hand side of Fig.~\ref{intewre2} at
$\lambda/\lambda_{\rm max}=1$. One finds $g_a=-0.0032$, $0.0023$ and
$0.0007$ for $H_L^{4\ell}$, $H_A^{1T}-H_A^{2T}$ and
$H_I^{4T}-H_A^{3N}$, respectively. Values substantially larger than these
combinations of structure functions would signal contributions from a
second-class current with coupling strength exceeding the above values of
$g_a$.

\section{Summary and outlook}
We have presented analytical results for the $O(\as)$ radiative corrections to
polarized top quark pair production in $e^+e^-$ annihilation with a specific
gluon energy cut. When the gluon energy cut is taken to its maximal value we
recover previously known results ~\cite{Korner:1993dy,Ravindran:2000rz}. The
size of the radiative corrections to polarization-type observables involving
the top quark is generally quite small in the soft-gluon region but can become
substantial in the hard gluon region. This in turn implies that the dependence
of the polarization-type observables on the gluon energy cut is generally
quite small in the soft-gluon region but can become large in the hard gluon
region. We have calculated the contributions of a $CP$-odd non-SM coupling to
some linear combinations of structure functions that vanish in the two-body SM
case. These were compared to SM contributions resulting from radiative
corrections.  

We have not considered beam polarization effects in our analysis. However, in
as much as we have calculated the complete set of single spin structure
functions, beam polarization effects can be easily incorporated into our
analysis as described e.g.\ in more detail in Ref.~\cite{Groote:1995yc}. 
 
We have decomposed the top spin vector in the helicity basis, i.e.\ the
$z$ direction of our spin basis is determined by the momentum of the top
quark. In addition to the helicity basis the authors of
Refs.~\cite{Kodaira:1998gt,Parke:1996pr} have also considered a beamline and
an off-diagonal basis. A discussion of how these bases are related to the
helicity basis in the context of the NLO corrections can be found in
Ref.~\cite{Ravindran:2000rz}. 

All the results in this paper refer to the polarization of the top quark. In
order to obtain the SM and anomalous coupling predictions for the 
polarization of the antitop quark let us first set up an orthonormal spin 
basis for the antitop quark by replacing the momenta in Eq.~(\ref{basistop}) 
by their charge conjugate partners, i.e.\ $\vec p_1\to\vec p_2$ and
$\vec p_{e^-}\to\vec p_{e^+}$. The three orthonormal basis vectors
$(\vec e_T,\vec e_N,\vec e_\ell)$ are now given by
\begin{equation}\label{basisantitop}
\vec e_T=\frac{(\vec p_{e^+}\times\vec p_2)\times\vec p_2}
{|(\vec p_{e^+}\times\vec p_2)\times\vec p_2|},\qquad
\vec e_N=\frac{\vec p_{e^+}\times\vec p_2}
{|\vec p_{e^+}\times\vec p_2|},\qquad 
\vec e_\ell=\frac{\vec p_2}{|\vec p_2|}.
\end{equation}
In the polar angle distribution Eq.~(\ref{polar}) the polar angle now
refers to $\theta_{\bar te^-}$ and {\it not} to $\theta=\theta_{te^-}$ as in
the top quark case discussed in the main part of this paper. Since the lepton 
pair is back-to-back, one has 
$\theta_{\bar te^-}=\pi - \theta_{\bar t e^+}$, i.e.
the two terms in Eq.~(\ref{polar}) proportional to $\cos\theta$ change
sign if written in terms of $\cos\theta_{\bar t e^+}$. 

Let us list the SM Born term and the anomalous contributions in the
antitop quark case given by Eq.~(\ref{anocurrent}) together with the relevant
contributions in the top quark case. One finds 
\begin{eqnarray}
\label{bornplusanom}
H_U^1=2N_Cq^2(1+v^{2}),&& H_U^2=2N_Cq^2(1-v^{2}),\nonumber\\[7pt]
H_U^{3\ell}=0,&&H_U^{4\ell}=\pm4N_Cq^2v,\nonumber\\[7pt]
H_L^1=N_Cq^2(1-v^{2})+N_Cq^2v^2\frac{|g_a|^2}\xi,&&
H_L^2=N_Cq^2(1-v^{2})-N_Cq^2v^2\frac{|g_a|^2}\xi,\nonumber\\[7pt]
H_L^{3\ell}=-2N_Cq^2v\imag g_a,&&
H_L^{4\ell}=-2N_Cq^2v\real g_a,\nonumber\\[12pt]
H_F^{1\ell}=2N_Cq^2(1+v^{2}),&&
H_F^{2\ell}=2N_Cq^2(1-v^{2}),\nonumber\\[7pt]
H_F^3=0,&&H_F^4=\pm4N_cq^2v,\nonumber\\[7pt]
H_I^{3T}=\frac{N_Cq^2}{\sqrt{2\xi}}v\imag g_a,&&
H_I^{4T}=\frac{N_Cq^2}{\sqrt{2\xi}}v(\pm\xi+\real g_a),\nonumber\\
H_A^{1T}=\frac{N_Cq^2}{\sqrt{2\xi}}(\xi\pm v^2\real g_a),&&
H_A^{2T}=\frac{N_Cq^2}{\sqrt{2\xi}}(\xi \mp v^2\real g_a),\nonumber\\
H_I^{1N}=\pm\frac{N_Cq^2}{\sqrt{2\xi}}v^2\imag g_a,&&
H_I^{2N}=\mp \frac{N_Cq^2}{\sqrt{2\xi}}v^2\imag g_a,\nonumber\\
H_A^{3N}=\frac{N_Cq^2}{\sqrt{2\xi}}v(\pm \xi-\real g_a),&&
H_A^{4N}=\frac{N_Cq^2}{\sqrt{2\xi}}v\imag g_a,
\end{eqnarray}
where the upper and lower signs refer to the top quark and antitop quark 
cases, respectively. As concerns the SM Born term contributions one finds
\begin{eqnarray}
\label{smantitop}
\sigma_t(\cos\theta_{te^-})&=&
\sigma_{\bar t}(\cos\theta_{\bar te^+}) \nonumber \\
P^{\ell,N}_t(\cos\theta_{te^-})&=&
-P^{\ell,N}_{\bar t}(\cos\theta_{\bar te^+}) \nonumber \\ 
P^T_t(\cos\theta_{te^-})&=&
P^T_{\bar t}(\cos\theta_{\bar te^+}).
\end{eqnarray}
In the three-body case one has to simultaneously exchange 
$(y \leftrightarrow z)$ in the SM part of Eqs.~(\ref{bornplusanom})
and~(\ref{smantitop}). For example, one has 
$H_L^{3,4\ell}({\it top\/};y,z)=-H_L^{3,4\ell}({\it antitop\/};z,y)$. 
If one performs an integration symmetric
in $y$ and $z$ as done in this paper the SM part of the relations 
(\ref{bornplusanom}) and (\ref{smantitop}) also
hold for the integrated three--body results.
  
The linear contributions of the anomalous coupling to the polarization
vector behave in the opposite way to those in Eq.~(\ref{smantitop}), i.e.
\begin{eqnarray}
P^{\ell,N}_t({\it anomalous\/};\cos\theta_{te^-})&=&
P^{\ell,N}_{\bar t}({\it anomalous\/};\cos\theta_{\bar te^+}) \nonumber \\ 
P^T_t({\it anomalous\/};\cos\theta_{te^-})&=&
-P^T_{\bar t}({\it anomalous\/};\cos\theta_{\bar te^+}).
\end{eqnarray}
It is clear that one can obtain an additional handle on the anomalous
contributions by taking sums and differences of the top quark and antitop 
quark polarizations. For example, ($P^{\ell,N}_t(\cos\theta_{te^-})+
P^{\ell,N}_{\bar t}(\cos\theta_{\bar t e^+})$)
and ($P^T_t(\cos\theta_{te^-})-P^T_{\bar t}(\cos\theta_{\bar te^+})$)
are contributed to only by the anomalous contributions.

In this paper we have not discussed how the spin of the top quark can be
analyzed. The top quark decays weakly and is therefore self-analysing. If one
assumes SM interactions in the cascade decay
$t\to b\,W^+(\to l^+\nu_l,\,q\bar q)$ the polarization of the top quark can be
reconstructed by measuring spin-momentum correlations either in the top quark
rest system (see e.g.\
Refs.~\cite{Czarnecki:1993gt,Czarnecki:1994pu,Korner:1998nc,Groote:2006kq}) or 
in the $W$ rest system as e.g.\ discussed in 
Refs.~\cite{Fischer:2001gp,Fischer:1998gsa,Do:2002ky}. We mention that there
exists a large body of literature of how non-SM interactions in the production 
(see e.g.\ Ref.~\cite{Gounaris:1996vn}) (such as the anomalous coupling 
Eq.~(\ref{anocurrent})), and/or in the decay affect such spin--momentum
correlations (see e.g.\ Ref.~\cite{Antipin:2008zx} and references therein).

Gluons can be emitted from the original production process 
$e^+e^-\rightarrow t\bar{t}(G)$ as well as from the follow-up decay process 
$t \to b + W^{+}(G)$ and $\bar{t} \to \bar{b} +W^{-}(G)$ where we take the
$W$'s to decay leptonically. Interference effects between the two processes 
are expected to be quite small since they are suppressed by a factor of 
$\approx \Gamma_{t}/m_{t}\sim 1\%$. In order to identify the gluons of the 
original production process (which are the subject of this paper) one has to 
demand that the gluon's four-momentum 
satisfies $q=p_{t}+p_{\bar{t}}+p_{G}$. Gluons that satisfy 
$p_{t}=p_{b}+ p_{W} +p_{G}$ or $p_{\bar{t}}=p_{\bar{b}}+ p_{W} +p_{G}$ clearly
originate from the follow-up processes and can thus be vetoed. How effectively
gluons not originating from the original production process can be removed from
the data sample has to be carefully studied in detailed Monte Carlo simulation
runs.

With the appropriate modifications our results can also be applied to the
$(b\bar b)$ case. While the $\imag\cz$ contributions resulting from the
imaginary part of the Breit--Wigner line shape are negligibly small in the
$(t\bar t)$ case (since $(t\bar t)$ threshold is far away from the $Z$ pole)
the $\imag\cz$ contribution is more pronounced in the $(b\bar b)$ case in
particular in the vicinity of the $Z$ pole. However, close to the $Z$ pole the
transverse and normal polarization of the bottom quark are severely suppressed
due to the overall helicity suppression factor $2m/\sqrt{s}$. In this sense
the phenomenology of the top quark spin above $(t\bar t)$ threshold is richer
than that of the bottom quark in the high energy realm.

\vspace{1truecm}\noindent
{\bf Acknowledgements:} We would like to thank V.~Kleinschmidt for
participating in the early stages of this calculation. We are also grateful
for illuminating discussions with G.J.~Gounaris and F.M.~Renard. This work is
supported in part by the Estonian target financed project No.~0182647s04 and
by the Estonian Science Foundation under Grant No.~6216. S.G.\ also
acknowledges support from a grant of the Deutsche Forschungsgemeinschaft (DFG)
for staying at Mainz University as a guest scientist for a couple of months.

\newpage

\begin{appendix}

\section{SM values of the electroweak coupling coefficients}
\setcounter{equation}{0}\def\theequation{A\arabic{equation}}
The electroweak coupling matrix elements $g_{ij}(q^2)$ are given by
\begin{eqnarray}
g_{11}&=&Q_f^2-2Q_fv_ev_f\real\cz+(v_e^2+a_e^2)(v_f^2+a_f^2)|\cz|^2,\nonumber\\
g_{12}&=&Q_f^2-2Q_fv_ev_f\real\cz+(v_e^2+a_e^2)(v_f^2-a_f^2)|\cz|^2,\nonumber\\
g_{13}&=&-2Q_fv_ea_f\imag\cz,\\
g_{14}&=&2Q_fv_ea_f\real\cz-2(v_e^2+a_e^2)v_fa_f|\cz|^2,\nonumber\\
\nonumber\\
g_{21}&=&Q_f^2-2Q_fv_ev_f\real\cz+(v_e^2-a_e^2)(v_f^2+a_f^2)|\cz|^2,\nonumber\\
g_{22}&=&Q_f^2-2Q_fv_ev_f\real\cz+(v_e^2-a_e^2)(v_f^2-a_f^2)|\cz|^2,\nonumber\\
g_{23}&=&-2Q_fv_ea_f\imag\cz,\nonumber\\
g_{24}&=&2Q_fv_ea_f\real\cz-2(v_e^2-a_e^2)v_fa_f|\cz|^2,\nonumber\\
\nonumber\\
g_{31}&=&-2Q_fa_ev_f\imag\cz,\nonumber\\
g_{32}&=&-2Q_fa_ev_f\imag\cz,\nonumber\\
g_{33}&=&2Q_fa_ea_f\real\cz,\addtocounter{equation}{-1}\\
g_{34}&=&2Q_fa_ea_f\imag\cz,\nonumber\\
\nonumber\\
g_{41}&=&2Q_fa_ev_f\real\cz-2v_ea_e(v_f^2+a_f^2)|\cz|^2,\nonumber\\
g_{42}&=&2Q_fa_ev_f\real\cz-2v_ea_e(v_f^2-a_f^2)|\cz|^2,\nonumber\\
g_{43}&=&2Q_fa_ea_f\imag\cz,\nonumber\\
g_{44}&=&-2Q_fa_ea_f\real\cz+4v_ea_ev_fa_f|\cz|^2\nonumber
\end{eqnarray}
where $\cz(q^2)=gM_Z^2q^2/(q^2-M_Z^2+iM_Z\Gamma_Z)$, with $M_Z$ and 
$\Gamma_Z$ the mass and width of the $Z^0$ and
$g=G_F(8\sqrt 2\pi\alpha)^{-1}\approx 4.49\cdot 10^{-5}\mbox{\rm GeV}^{-2}$.
$Q_f$ are the charges of the final state quarks to which the electroweak 
currents directly couple; $v_e$ and $a_e$, $v_f$ and $a_f$ are the 
electroweak vector and axial-vector coupling constants. For example, in 
the Weinberg-Salam model, one has $v_e=-1+4\sin^2\theta_W$, $a_e=-1$ for 
leptons, $v_f=1-\frac83\sin^2\theta_W$, $a_f=1$ for up-type quarks 
($Q_f=\frac23$), and $v_f=-1+\frac43\sin^2\theta_W$, $a_f=-1$ for down-type 
quarks ($Q_f=-\frac13$). The left- and right-handed coupling constants are 
then given by $g_L=v+a$ and $g_R=v-a$, respectively. In the purely 
electromagnetic case one has $g_{11}=g_{12}=g_{21}=g_{22}=Q_f^2$ and all 
other $g_{r'r}=0$. The terms linear in $\real\cz$ and $\imag\cz$ come from 
$\gamma-Z^0$ interference, whereas the terms proportional to $|\cz|^2$ 
originate from $Z$ exchange.
    
Contributions coming from the imaginary part of the Breit--Wigner resonance
shape are of order $O(\imag\cz(q^2)/\real\cz(q^2))$ and can thus safely be
neglected for top quark pair production. For example, in the threshold region
of top quark pair production $\imag\cz/\real\cz$ is approximately $0.1\%$ and
decreases further with a $1/q^{2}$ power fall-off behaviour.

\section{Decay rate terms $t_i$}
\setcounter{equation}{0}\def\theequation{B\arabic{equation}}
It is convenient to define the mass dependent variables $a:=2+\sxi$, 
$b:=2-\sxi$ and $w:=\sqrt{(1-\sxi)/(1+\sxi)}$. The rate 
functions $t_1,\ldots,t_{12}$ appearing in the main text are then given by
\begin{eqnarray}
t_1&:=&\ln\pfrac{2\xi\sxi}{b^2(1+\sxi)},\quad
t_2\ :=\ \ln\pfrac{2\sxi}{1+\sxi}\quad
  \Rightarrow\quad t_1-t_2\ =\ \ln\pfrac\xi{b^2}\\
t_3&:=&\ln\pfrac{1+v}{1-v}\\
t_4&:=&\Li_2(w)-\Li_2(-w)+\Li_2(\frac abw)-\Li_2(-\frac abw)\\
t_5&:=&\frac12\ln\pfrac{a\sxi}{4(1+\sxi)}
  \ln\pfrac{1+v}{1-v}
  +\Li_2\pfrac{2\sxi}{a(1+w)}
  -\Li_2\pfrac{2\sxi}{a(1-w)}\,+\nonumber\\&&
  +\Li_2\pfrac{1+w}2-\Li_2\pfrac{1-w}2
  +\Li_2\pfrac{a(1+w)}4-\Li_2\pfrac{a(1-w)}4\\
t_6&:=&\ln^2(1+w)+\ln^2(1-w)+\ln\pfrac a8\ln(1-w^2)\,+\nonumber\\&&
  +\Li_2\pfrac{2\sxi}{a(1+w)}+\Li_2\pfrac{2\sxi}{a(1-w)}
  -2\Li_2\pfrac{2\sxi}a\,+\nonumber\\&&
  +\Li_2\pfrac{1+w}2+\Li_2\pfrac{1-w}2
  -2\Li_2\pfrac12\,+\nonumber\\&&
  +\Li_2\pfrac{a(1+w)}4+\Li_2\pfrac{a(1-w)}4-2\Li_2\pfrac a4\\
t_7&:=&2\ln\pfrac{1-\xi}{2\xi}\ln\pfrac{1+v}{1-v}
  -\Li_2\pfrac{2v}{(1+v)^2}
  +\Li_2\left(-\frac{2v}{(1-v)^2}\right)\,+\nonumber\\&&
  -\frac12\Li_2\left(-\pfrac{1+v}{1-v}^2\right)
  +\frac12\Li_2\left(-\pfrac{1-v}{1+v}^2\right)\,+\\&&
  +\Li_2\pfrac{2w}{1+w}-\Li_2\left(-\frac{2w}{1-w}\right)
  -2\Li_2\pfrac{w}{1+w}+2\Li_2\left(-\frac{w}{1-w}\right)
  \,+\nonumber\\&&
  +\Li_2\pfrac{2aw}{b+aw}-\Li_2\left(-\frac{2aw}{b-aw}\right)
  -2\Li_2\pfrac{aw}{b+aw}+2\Li_2\left(-\frac{aw}{b-aw}\right)
  \nonumber\\
t_8&:=&\ln\pfrac\xi 4\ln\pfrac{1+v}{1-v}
  +\Li_2\pfrac{2v}{1+v}-\Li_2\left(-\frac{2v}{1-v}\right)-\pi^2\\
t_9&:=&2\ln\pfrac{2(1-\xi)}{\sxi}
  \ln\pfrac{1+v}{1-v}
  +2\left(\Li_2\pfrac{1+v}2-\Li_2\pfrac{1-v}2\right)\,+\nonumber\\&&
  +3\left(\Li_2\left(-\frac{2v}{1-v}\right)
  -\Li_2\pfrac{2v}{1+v}\right)\\
t_{10}&:=&\ln\pfrac4\xi,\quad
t_{11}\ :=\ \ln\pfrac{4(1-\sxi)^2}\xi,\quad
t_{12}\ :=\ \ln\pfrac{4(1-\xi)}\xi
\end{eqnarray}

\section{Decay rate terms $\ell_i$, $t_{0\pm}$, $t_{1\pm}$, and $t_w$}
\setcounter{equation}{0}\def\theequation{C\arabic{equation}}
The logarithmic  rate terms $\ell_i$ are given by
\begin{eqnarray}
\ell_1&=&\ln\pfrac{w_1^2-w_\lambda^2}{w_0^2-w_1^2}
  -\ln\pfrac{1+w_1}{b-aw_1}-\ln\pfrac{(1+\sxi)\sxi}{1-2\lambda+\sxi}\\
\ell_2&=&\ln\pfrac{w_2^2-w_\lambda^2}{w_0^2-w_2^2}
  +\ln\pfrac{b+aw_2}{1-w_2}-\ln\pfrac{(1+\sxi)\sxi}{1-2\lambda+\sxi}\\
\ell_3&=&\ln\pfrac{w_2}{w_1}\\
\ell_{4+}&=&-\frac{\lambda\xi}{y_1}+\frac{\lambda\xi}{y_2}
  +2v\Bigg[4-2\ln\pfrac{4w_0y_1}{\sxi}
    +\ln\pfrac{w_0+w_1}{w_0-w_1}+\ln\pfrac{w_0+w_2}{w_0-w_2}\Bigg]\,+\\&&
  +\left(2v-(2-\xi)\ln\pfrac{1+v}{1-v}\right)
    \left[\ln\pfrac{\xi\Lambda}{v^2}
      +2\ln\pfrac{w_0^2-w_1^2}{1-w_1^2}-1\right]\nonumber\\
\ell_{4-}&=&2v\Bigg[2-2\ln\pfrac{2\sxi y_1}v
    +\ln\pfrac{(1+w_1)(b-aw_1)}{w_0^2-w_1^2}
    +\ln\pfrac{(b+aw_2)(1-w_2)}{w_0^2-w_2^2}\Bigg]\,+\nonumber\\&&
  +\left(2v-(2-\xi)\ln\pfrac{1+v}{1-v}\right)
    \left[\ln\pfrac{\xi\Lambda}{v^2}
      +2\ln\pfrac{w_0^2-w_1^2}{1-w_1^2}-1\right]\qquad\\
\ell_{5+}&=&\ln\pfrac{1-w_2}{1-w_0}-\ln\pfrac{1+w_1}{1+w_0},\qquad
\ell_{5-}\ =\ 2\ln\pfrac{1+v}{1-v}\\
\ell_{6+}&=&2\ln\pfrac{1+v}{1-v}
  -\ln\pfrac{1+w_1}{b-aw_1}-\ln\pfrac{b+aw_2}{1-w_2}\\
\ell_{6-}&=&\ln\xi+\ln\pfrac{1+w_1}{b-aw_1}-\ln\pfrac{b+aw_2}{1-w_2}\\
\ell_{7+}&=&\ln\pfrac{w_2^2-w_\lambda^2}{w_1^2-w_\lambda^2},\qquad
\ell_{7-}\ =\ \ln\pfrac{w_2-w_\lambda}{w_1-w_\lambda}
  -\ln\pfrac{w_2+w_\lambda}{w_1+w_\lambda}\\
\ell_{8+}&=&\ln\pfrac{w_0^2-w_2^2}{w_0^2-w_1^2},\qquad
\ell_{8-}\ =\ \ln\pfrac{w_0-w_2}{w_0-w_1}-\ln\pfrac{w_0+w_2}{w_0+w_1}\\
\ell_{9+}&=&\ln\pfrac{1-w_2^2}{1-w_1^2},\qquad
\ell_{9-}\ =\ \ln\pfrac{1-w_2}{1-w_1}-\ln\pfrac{1+w_2}{1+w_1}
\end{eqnarray}
while for the additional phase-space contribution we have to use
\begin{eqnarray}
\ell_2^c&=&\ln\pfrac{1+w_2}{1-w_2}+\ln\pfrac{b+aw_2}{b-aw_2},\nonumber\\
\ell_{4-}^c&=&\ln(1-w_2^2)+\ln(b^2-a^2w_2^2),\qquad
\ell_{4+}^c\ =\ \ln\pfrac{w_0+w_2}{w_0-w_2},\nonumber\\
\ell_{5-}^c&=&\ln b\ =\ \ln(2-\sxi),\qquad
\ell_{5+}^c\ =\ \ln\pfrac{1+w_2}{1-w_2},\nonumber\\
\ell_{6-}^c&=&\ln(1-w_2^2)-\ln(b^2-a^2w_2^2),\qquad
\ell_{7-}^c\ =\ \ln\pfrac{w_0^2}{w_0^2-w_2^2}.\kern128pt
\end{eqnarray}
For the double and dilogarithmic decay rate terms we obtain
\begin{eqnarray}
t_w&=&\frac12(2t_w^{ba}(w_0)-t_w^{ba}(w_1)-t_w^{ba}(w_2))
   +(t_w^z(w_2)-t_w^z(w_1))\,+\nonumber\\&&
  -\frac12(t_w^{ab}(w_2)-t_w^{ab}(w_1))
  -(t_w^\lambda(w_2)-t_w^\lambda(w_1))
  +\ln\pfrac{(1+\sxi)\sxi}{1-2\lambda+\sxi}\ln\pfrac{w_2}{w_1}\\
t_{0\pm}&=&\frac12(2t_{0\pm}^{ba}(w_0)-t_{0\pm}^{ba}(w_1)-t_{0\pm}^{ba}(w_2))
   +(t_{0\pm}^z(w_2)-t_{0\pm}^z(w_1))\,+\\&&
  -\frac12(t_{0\pm}^{ab}(w_2)-t_{0\pm}^{ab}(w_1))
  -(t_{0\pm}^\lambda(w_2)-t_{0\pm}^\lambda(w_1))
  \pm\ln\pfrac{(1+\sxi)\sxi}{1-2\lambda+\sxi}
  \ln\pfrac{w_0\pm w_2}{w_0\pm w_1}\nonumber\\
t_{1\pm}&=&\frac12(2t_{1\pm}^{ba}(w_0)-t_{1\pm}^{ba}(w_1)-t_{1\pm}^{ba}(w_2))
   +(t_{1\pm}^z(w_2)-t_{1\pm}^z(w_1))\,+\\&&
  -\frac12(t_{1\pm}^{ab}(w_2)-t_{1\pm}^{ab}(w_1))
  -(t_{1\pm}^\lambda(w_2)-t_{1\pm}^\lambda(w_1))
  \pm\ln\pfrac{(1+\sxi)\sxi}{1-2\lambda+\sxi}\ln\pfrac{1\pm w_2}{1\pm w_1}
  \nonumber
\end{eqnarray}
while for the additional phase-space contribution we take
\begin{equation}
t_w^c=t_w^{ba}(w_2)-t_w^{ba}(0),\qquad
t_{0\pm}^c=t_{0\pm}^{ba}(w_2)-t_{0\pm}^{ba}(0),\qquad
t_{1\pm}^c=t_{1\pm}^{ba}(w_2)-t_{1\pm}^{ba}(0)
\end{equation}
where
\begin{eqnarray}
t_w^{ba}(w)&=&\Li_2(w)-\Li_2(-w)+\Li_2\pfrac{aw}b-\Li_2\pfrac{-aw}b,\nonumber\\
t_w^z(w)&=&2\ln(w_0)\ln(w)+\Li_2(w)-\Li_2(-w)
  -\Li_2\pfrac{w}{w_0}-\Li_2\pfrac{-w}{w_0},\nonumber\\
t_w^{ab}(w)&=&2\ln(b)\ln(w)+\Li_2(w)-\Li_2(-w)
  -\Li_2\pfrac{aw}b-\Li_2\pfrac{-aw}b,\nonumber\\
t_w^\lambda(w)&=&\ln^2(w)+\Li_2(w)+\Li_2(-w)
  +\Li_2\pfrac{w_\lambda}w+\Li_2\pfrac{-w_\lambda}w,\\[12pt]
t_{0-}^{ba}(w)&=&-2\ln\pfrac{1+v}{1-v}\ln(w_0-w)\,+\nonumber\\&&
  +\Li_2\pfrac{w_0-w}{w_0+1}-\Li_2\pfrac{w_0-w}{w_0-1}
  +\Li_2\pfrac{a(w_0-w)}{aw_0+b}-\Li_2\pfrac{a(w_0-w)}{aw_0-b},\nonumber\\[7pt]
t_{0-}^{ba}(w_0)&=&2\ln\pfrac{y_1}{\sxi}\ln\pfrac{1+v}{1-v}
  -\Li_2\pfrac{2v}{(1+v)^2}+\Li_2\pfrac{-2v}{(1-v)^2}\,+\nonumber\\&&
  +\frac12\Li_2\left(-\frac{(1-v)^2}{(1+v)^2}\right)
  -\frac12\Li_2\left(-\frac{(1+v)^2}{(1-v)^2}\right),\nonumber\\[7pt]
t_{0+}^{ba}(w)&=&-2\ln\pfrac{1+v}{1-v}\ln(w_0+w)\,+\nonumber\\&&
  +\Li_2\pfrac{w_0+w}{w_0+1}-\Li_2\pfrac{w_0+w}{w_0-1}
  +\Li_2\pfrac{a(w_0+w)}{aw_0+b}-\Li_2\pfrac{a(w_0+w)}{aw_0-b},\nonumber\\[7pt]
t_{0-}^z(w)&=&\frac12\ln\pfrac\xi{1-\xi}\ln(w_0-w)
  -\frac12\ln^2(w_0-w)\,+\nonumber\\&&
  +\Li_2\pfrac{w_0-w}{2w_0}-\Li_2\pfrac{w_0-w}{w_0-1}
  -\Li_2\pfrac{w_0-w}{w_0+1},\nonumber\\[7pt]
t_{0+}^z(w)&=&-\frac12\ln\pfrac\xi{1-\xi}\ln(w_0+w)+\frac12\ln^2(w_0+w)
  \,+\nonumber\\&&
  -\Li_2\pfrac{w_0+w}{2w_0}+\Li_2\pfrac{w_0+w}{w_0-1}
  +\Li_2\pfrac{w_0+w}{w_0+1},\nonumber\\[7pt]
t_{0-}^{ab}(w)&=&-\ln\xi\ln(w_0-w)\,+\nonumber\\&&
  +\Li_2\pfrac{a(w_0-w)}{aw_0-b}+\Li_2\pfrac{a(w_0-w)}{aw_0+b}
  -\Li_2\pfrac{w_0-w}{w_0-1}-\Li_2\pfrac{w_0-w}{w_0+1},\nonumber\\[7pt]
t_{0+}^{ab}(w)&=&\ln\xi\ln(w_0+w)\,+\nonumber\\&&
  -\Li_2\pfrac{a(w_0+w)}{aw_0-b}-\Li_2\pfrac{a(w_0+w)}{aw_0+b}
  +\Li_2\pfrac{w_0+w}{w_0-1}+\Li_2\pfrac{w_0+w}{w_0+1},\nonumber\\[7pt]
t_{0-}^\lambda(w)&=&-\ln\pfrac{2\lambda}{1-2\lambda+\sxi}\ln(w_0-w)
  \,+\nonumber\\&&
  +\Li_2\pfrac{w_0-w}{w_0-w_\lambda}+\Li_2\pfrac{w_0-w}{w_0+w_\lambda}
  -\Li_2\pfrac{w_0-w}{w_0-1}-\Li_2\pfrac{w_0-w}{w_0+1},\nonumber\\[7pt]
t_{0+}^\lambda(w)&=&\ln^2(w_0+w)-\ln(1-w_0^2)\ln(w_0+w)\,+\nonumber\\[5pt]&&
  +\Li_2\pfrac{w_0-w_\lambda}{w_0+w}+\Li_2\pfrac{w_0+w_\lambda}{w_0+w}
  +\Li_2\pfrac{w_0+w}{w_0-1}+\Li_2\pfrac{w_0+w}{w_0+1},\\[12pt]
t_{1-}^{ba}(w)&=&\ln^2(1-w)+\ln\pfrac a8\ln(1-w)\,+\nonumber\\&&
  +\Li_2\pfrac{2\sxi}{a(1-w)}+\Li_2\pfrac{a(1-w)}4
  +\Li_2\pfrac{1-w}2,\nonumber\\[7pt]
t_{1+}^{ba}(w)&=&\ln^2(1+w)+\ln\pfrac a8\ln(1+w)\,+\nonumber\\&&
  +\Li_2\pfrac{2\sxi}{a(1+w)}+\Li_2\pfrac{a(1+w)}4
  +\Li_2\pfrac{1+w}2,\nonumber\\[7pt]
t_{1-}^z(w)&=&-\ln\pfrac{1+w_0}2\ln(1-w)
  -\Li_2\pfrac{1-w_0}{1-w}+\Li_2\pfrac{1-w}{1+w_0}
  -\Li_2\pfrac{1-w}2,\nonumber\\[7pt]
t_{1+}^z(w)&=&\ln\pfrac{1+w_0}2\ln(1+w)
  -\Li_2\pfrac{1+w}{1-w_0}+\Li_2\pfrac{1+w_0}{1+w}
  +\Li_2\pfrac{1+w}2,\nonumber\\[7pt]
t_{1-}^{ab}(w)&=&-\ln(2a)\ln(1-w)
  -\Li_2\pfrac{2\sxi}{a(1-w)}+\Li_2\pfrac{a(1-w)}4
  -\Li_2\pfrac{1-w}2,\nonumber\\[7pt]
t_{1+}^{ab}(w)&=&\ln(2a)\ln(1+w)
  +\Li_2\pfrac{2\sxi}{a(1+w)}-\Li_2\pfrac{a(1+w)}4
  +\Li_2\pfrac{1+w}2,\nonumber\\[7pt]
t_{1-}^\lambda(w)&=&\frac12\ln^2(1-w)+\ln 2\ln(1-w)
  -\ln(1-w_\lambda^2)\ln(1-w)\,+\nonumber\\&&
  +\Li_2\pfrac{1-w}{1-w_\lambda}+\Li_2\pfrac{1-w}{1+w_\lambda}
  -\Li_2\pfrac{1-w}2,\nonumber\\[7pt]
t_{1+}^\lambda(w)&=&\frac12\ln^2(1+w)-\ln 2\ln(1+w)\,+\nonumber\\&&
  +\Li_2\pfrac{1-w_\lambda}{1+w}+\Li_2\pfrac{1+w_\lambda}{1+w}
  +\Li_2\pfrac{1+w}2.
\end{eqnarray}

\end{appendix}

\end{document}